\documentclass[twocolumn]{aastex631}

\usepackage{xspace}

\newcommand{\onef}{$1/f$}
\newcommand{\pipename}{{\tt pjpipe}\xspace}

\graphicspath{{./}{figs/}{figs/early_science/}{figs/mast_comparison/}{figs/mosaics/}}

\begin{document}

\title{PHANGS-{\it JWST}: Data Processing Pipeline and First Full Public Data Release}

\correspondingauthor{Thomas G. Williams}
\email{thomas.williams@physics.ox.ac.uk}


\newcommand{\Ox}{Sub-department of Astrophysics, Department of Physics, University of Oxford, Keble Road, Oxford OX1 3RH, UK}

\newcommand{\UGent}{Sterrenkundig Observatorium, Universiteit Gent, Krijgslaan 281 S9, B-9000 Gent, Belgium}

\newcommand{\STScI}{Space Telescope Science Institute, 3700 San Martin Drive, Baltimore, MD 21218, USA}

\newcommand{\MPIA}{Max-Planck-Institut f\"{u}r Astronomie, K\"{o}nigstuhl 17, D-69117, Heidelberg, Germany}

\newcommand{\AURA}{AURA for the European Space Agency (ESA), Space Telescope Science Institute, 3700 San Martin Drive, Baltimore, MD 21218, USA}

\newcommand{\UCSD}{Department of Astronomy \& Astrophysics, University of California, San Diego, 9500 Gilman Dr., La Jolla, CA 92093, USA}

\newcommand{\JHU}{Department of Physics and Astronomy, The Johns Hopkins University, Baltimore, MD 21218, USA}

\newcommand{\OSU}{Department of Astronomy, The Ohio State University, 140 West 18th Avenue, Columbus, OH 43210, USA}

\newcommand{\CCAPP}{Center for Cosmology and Astroparticle Physics (CCAPP), 191 West Woodruff Avenue, Columbus, OH 43210, USA}

\newcommand{\ARI}{Astronomisches Rechen-Institut, Zentrum f\"{u}r Astronomie der Universit\"{a}t Heidelberg, M\"{o}nchhofstr. 12-14, D-69120 Heidelberg, Germany}

\newcommand{\UConn}{Department of Physics, University of Connecticut, 196A Auditorium Road, Storrs, CT 06269, USA}

\newcommand{\UHawaii}{Institute for Astronomy, University of Hawaii, 2680 Woodlawn Drive, Honolulu, HI 96822, USA}

\newcommand{\UniCA}{Universit\'{e} C\^{o}te d'Azur, Observatoire de la C\^{o}te d'Azur, CNRS, Laboratoire Lagrange, 06000, Nice, France}

\newcommand{\UAlberta}{Dept. of Physics, University of Alberta, 4-183 CCIS, Edmonton, Alberta, T6G 2E1, Canada}

\newcommand{\Arcetri}{INAF — Osservatorio Astrofisico di Arcetri, Largo E. Fermi 5, I-50125, Florence, Italy}

\newcommand{\UWyoming}{Department of Physics and Astronomy, University of Wyoming, Laramie, WY 82071, USA}

\newcommand{\LJMU}{Astrophysics Research Institute, Liverpool John Moores University, 146 Brownlow Hill, Liverpool L3 5RF, UK}

\newcommand{\ITA}{Universit\"{a}t Heidelberg, Zentrum f\"{u}r Astronomie, Institut f\"{u}r Theoretische Astrophysik, Albert-Ueberle-Str 2, D-69120 Heidelberg, Germany}

\newcommand{\CfA}{Center for Astrophysics $\mid$ Harvard \& Smithsonian, 60 Garden St., 02138 Cambridge, MA, USA}

\newcommand{\MPE}{Max-Planck-Institut f\"{u}r Extraterrestrische Physik (MPE), Giessenbachstr. 1, D-85748 Garching, Germany}

\newcommand{\Surrey}{Department of Physics, University of Surrey, Guildford GU2 7XH, UK}

\newcommand{\ESO}{European Southern Observatory, Karl-Schwarzschild Stra{\ss}e 2, D-85748 Garching bei M\"{u}nchen, Germany}

\newcommand{\IWR}{Universit\"{a}t Heidelberg, Interdisziplin\"{a}res Zentrum f\"{u}r Wissenschaftliches Rechnen, Im Neuenheimer Feld 205, D-69120 Heidelberg, Germany}

\newcommand{\ulyon}{Univ Lyon, Univ Lyon1, ENS de Lyon, CNRS, Centre de Recherche Astrophysique de Lyon UMR5574, F-69230 Saint-Genis-Laval France}

\newcommand{\COOL}{Cosmic Origins Of Life (COOL) Research DAO, coolresearch.io}

\newcommand{\OAN}{Observatorio Astron{\'o}mico Nacional (IGN), C/ Alfonso XII 3, E-28014 Madrid, Spain}

\newcommand{\UBonn}{Argelander-Institut f\"{u}r Astronomie, Universit\"{a}t Bonn, Auf dem H\"{u}gel 71, 53121 Bonn, Germany}

\newcommand{\kipac}{Kavli Institute for Particle Astrophysics \& Cosmology (KIPAC), Stanford University, CA 94305, USA}

\newcommand{\Umanc}{Jodrell Bank Centre for Astrophysics, Department of Physics and Astronomy, University of Manchester, Oxford Road, Manchester M13 9PL, UK}

\newcommand{\NRAO}{National Radio Astronomy Observatory, 520 Edgemont Road, Charlottesville, VA 22903, USA}

\newcommand{\ANU}{Research School of Astronomy and Astrophysics, Australian National University, Canberra, ACT 2611, Australia}

\newcommand{\AThreeD}{ARC Centre of Excellence for All Sky Astrophysics in 3 Dimensions (ASTRO 3D), Australia}

\newcommand{\Whitman}{Whitman College, 345 Boyer Avenue, Walla Walla, WA 99362, USA}

\newcommand{\IAC}{Instituto de Astrof\'isica de Canarias, C/ V\'ia L\'actea s/n, E-38205, La Laguna, Spain}

\newcommand{\ULL}{Departamento de Astrof\'isica, Universidad de La Laguna, Av. del Astrof\'isico Francisco S\'anchez s/n, E-38206, La Laguna, Spain}

\newcommand{\Princeton}{Department of Astrophysical Sciences, Princeton University, 4 Ivy Lane, Princeton, NJ 08544, USA}

\newcommand{\IRAM}{IRAM, 300 rue de la Piscine, 38400 Saint Martin d'H\'{e}res, France}

\newcommand{\LERMA}{LERMA, Observatoire de Paris, PSL Research University, CNRS, Sorbonne Universit\'{e}s, 75014 Paris}

\newcommand{\YB}{Centro de Desarrollos Tecnol\'ogicos, Observatorio de Yebes (IGN), 19141 Yebes, Guadalajara, Spain}

\newcommand{\CamIoAKavli}{Institute of Astronomy and Kavli Institute for Cosmology, University of Cambridge, Madingley Road, Cambridge CB3 0HA, UK}
\author[0000-0002-0012-2142]{Thomas~G.~Williams}
\affiliation{\Ox}


\author[0000-0002-2278-9407]{Janice~C.~Lee}
\affiliation{\STScI}

\author[0000-0003-3917-6460]{Kirsten~L.~Larson}
\affiliation{\AURA}

\author[0000-0002-2545-1700]{Adam~K.~Leroy}
\affiliation{\OSU}
\affiliation{\CCAPP}

\author[0000-0002-4378-8534]{Karin Sandstrom}
\affiliation{\UCSD}

\author[0000-0002-3933-7677]{Eva~Schinnerer}
\affiliation{\MPIA}

\author[0000-0002-8528-7340]{David~A.~Thilker}
\affiliation{\JHU}


\author[0000-0002-2545-5752]{Francesco~Belfiore}
\affiliation{\Arcetri}

\author[0000-0002-4755-118X]{Oleg~V.~Egorov}
\affiliation{\ARI}

\author[0000-0002-5204-2259]{Erik Rosolowsky}
\affiliation{\UAlberta}

\author[0000-0002-9183-8102]{Jessica Sutter}
\affiliation{\UCSD}
\affiliation{\Whitman}


\author{Joseph DePasquale}
\affiliation{\STScI}

\author{Alyssa Pagan}
\affiliation{\STScI}

\author{Travis A. Berger}
\affiliation{\STScI}


\author[0000-0002-5259-2314]{Gagandeep S. Anand}
\affiliation{\STScI}

\author[0000-0003-0410-4504]{Ashley T. Barnes}
\affiliation{\ESO}

\author[0000-0003-0166-9745]{Frank Bigiel}
\affiliation{\UBonn}

\author[0000-0003-0946-6176]{M\'ed\'eric Boquien}
\affiliation{\UniCA}

\author[0000-0001-5301-1326]{Yixian Cao}
\affiliation{Max-Planck-Institut f\"ur Extraterrestrische Physik (MPE), Giessenbachstr. 1, D-85748 Garching, Germany}

\author[0000-0002-5235-5589]{J\'{e}r\'{e}my Chastenet}
\affiliation{\UGent}

\author[0000-0002-5635-5180]{M\'{e}lanie Chevance}
\affiliation{\ITA}
\affiliation{\COOL}

\author[0000-0001-8241-7704]{Ryan Chown}
\affiliation{\OSU}

\author[0000-0002-5782-9093]{Daniel~A.~Dale}
\affiliation{\UWyoming}

\author[0000-0003-1943-723X]{Sinan Deger}
\affiliation{\CamIoAKavli}

\author[0000-0002-1185-2810]{Cosima~Eibensteiner}
\affiliation{\NRAO}\thanks{Jansky Fellow of the National Radio Astronomy Observatory}

\author[0000-0002-6155-7166]{Eric Emsellem}
\affiliation{\ESO}
\affiliation{\ulyon}

\author[0000-0001-5310-467X]{Christopher M. Faesi}
\affiliation{\UConn}

\author[0000-0001-6708-1317]{Simon~C.~O.~Glover}
\affiliation{\ITA}

\author[0000-0002-3247-5321]{Kathryn~Grasha}
\altaffiliation{ARC DECRA Fellow}
\affiliation{\ANU}   
\affiliation{\AThreeD}   

\author[0000-0001-9628-8958]{Stephen Hannon}
\affiliation{\MPIA}

\author[0000-0002-8806-6308]{Hamid Hassani}
\affiliation{\UAlberta}

\author[0000-0001-9656-7682]{Jonathan D. Henshaw}
\affiliation{\LJMU}
\affiliation{\MPIA}

\author[0000-0002-9165-8080]{Mar\'ia~J.~Jim\'enez-Donaire}
\affiliation{\OAN}
\affiliation{\YB}

\author[0000-0002-0432-6847]{Jaeyeon Kim}
\affiliation{\kipac}

\author[0000-0002-0560-3172]{Ralf S. Klessen}
\affiliation{\ITA}
\affiliation{\IWR}

\author[0000-0001-9605-780X]{Eric W. Koch}
\affiliation{\CfA}

\author[0000-0002-4825-9367]{Jing Li}
\affiliation{\ARI}

\author[0000-0001-9773-7479]{Daizhong Liu}
\affiliation{\MPE}

\author[0000-0002-6118-4048]{Sharon E. Meidt}
\affiliation{\UGent}

\author[0000-0002-6972-6411]{J. Eduardo M\'{e}ndez-Delgado}
\affiliation{\ARI}

\author[0000-0001-7089-7325]{Eric J.\,Murphy}
\affiliation{\NRAO}

\author[0000-0002-3289-8914]{Justus Neumann}
\affiliation{\MPIA}

\author[0000-0001-9793-6400]{Lukas~Neumann}
\affiliation{\UBonn}

\author[0000-0002-6922-2598]{Nadine~Neumayer}
\affiliation{\MPIA}

\author[0000-0002-0119-1115]{Elias K. Oakes}
\affiliation{\UConn}

\author[0000-0003-2721-487X]{Debosmita Pathak}
\affiliation{\OSU}

\author[0000-0003-3061-6546]{J\'{e}r\^{o}me Pety}
\affiliation{\IRAM}
\affiliation{\LERMA}

\author[0000-0001-5965-3530]{Francesca Pinna}
\affiliation{\IAC}
\affiliation{\ULL}
\affiliation{\MPIA}

\author[0000-0002-0472-1011]{Miguel Querejeta}
\affiliation{\OAN}

\author[0000-0002-9190-9986]{Lise Ramambason}
\affiliation{\ITA}

\author{Andrea Romanelli}
\affiliation{\ITA}

\author[0000-0001-6113-6241]{Mattia C. Sormani}
\affiliation{\Surrey}

\author[0000-0002-9333-387X]{Sophia K. Stuber}
\affiliation{\MPIA}

\author[0000-0003-0378-4667]{Jiayi~Sun}
\altaffiliation{NASA Hubble Fellow}
\affiliation{\Princeton}

\author[0000-0003-4209-1599]{Yu-Hsuan Teng}
\affiliation{\UCSD}

\author[0000-0003-1242-505X]{Antonio Usero}
\affiliation{\OAN}

\author[0000-0002-7365-5791]{Elizabeth J. Watkins}
\affiliation{\Umanc}

\author[0009-0005-8923-558X]{Tony D. Weinbeck}
\affiliation{\UWyoming}

\begin{abstract}
The exquisite angular resolution and sensitivity of {\it JWST} is opening a new window for our understanding of the Universe. In nearby galaxies, {\it JWST} observations are revolutionizing our understanding of the first phases of star formation and the dusty interstellar medium. Nineteen local galaxies spanning a range of properties and morphologies across the star-forming main sequence have been observed as part of the PHANGS-{\it JWST} Cycle 1 Treasury program at spatial scales of $\sim$5--50~pc.  Here, we describe \pipename, an image processing pipeline developed for the PHANGS-{\it JWST} program that wraps around and extends the official {\it JWST} pipeline. We release this pipeline to the community as it contains a number of tools generally useful for {\it JWST} NIRCam and MIRI observations. Particularly for extended sources, \pipename\ products provide significant improvements over mosaics from the MAST archive in terms of removing instrumental noise in NIRCam data, background flux matching, and calibration of relative and absolute astrometry. We show that slightly smoothing F2100W MIRI data to 0\farcs9 (degrading the resolution by about 30~percent) reduces the noise by a factor of $\approx$3. We also present the first public release (DR1.1.0) of the \pipename processed eight-band 2-21 $\mu$m imaging for all nineteen galaxies in the PHANGS-{\it JWST} Cycle 1 Treasury program. An additional 55 galaxies will soon follow from a new PHANGS-{\it JWST} Cycle 2 Treasury program. 
\end{abstract}


\section{Introduction}\label{sec:intro}

\begin{figure*}[ht]
    \includegraphics[width=\textwidth]{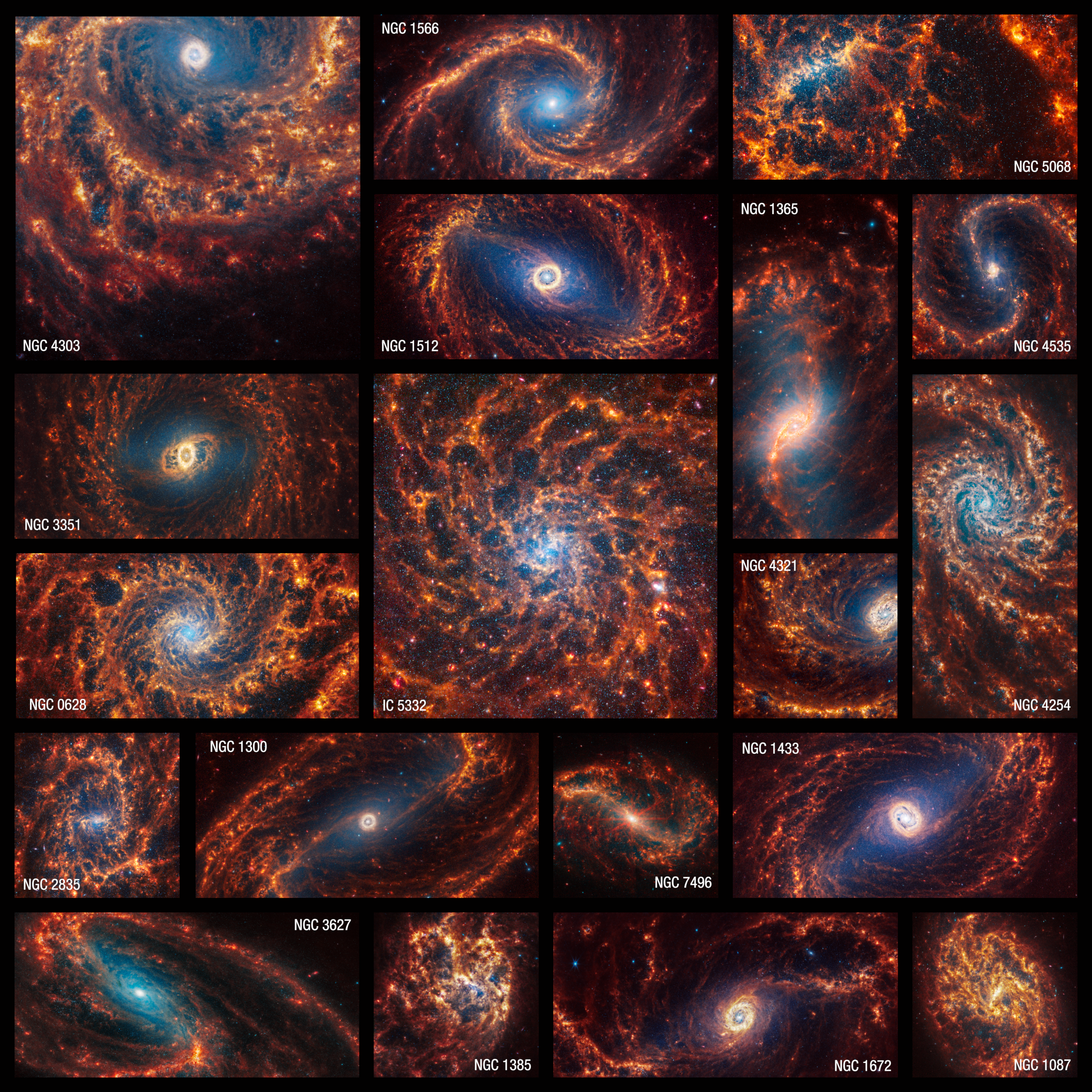}
    \caption{Color mosaics showing the {\it JWST} images for our full 19 galaxy sample developed to support a 2024 January press release, \url{https://webbtelescope.org/contents/news-releases/2024/news-2024-105}. Red is a combination of the MIRI filters (F770W, F1000W, F1130W, and F2100W), green is a combination of NIRCam and MIRI (F360M and F770W), and blue is purely NIRCam (F300M and F335M). F200W imaging has also been obtained, but is not included in the color composites. Note that in many cases, these mosaics are zoomed, and so do not show the full extent of coverage of the galaxy. For the full coverage of each observation, see Appendix \ref{app:atlas}.    
    }
    \label{fig:all_gals}
\end{figure*}

Near- and mid-infrared (N/MIR) wavelengths provide
 vital windows for studying star formation and the interstellar medium (ISM) in galaxies. Stars are born enshrouded deep within dusty molecular clouds and due to this, infrared observations are much more effective at detecting these earliest phases than optical \citep[e.g.,][]{2004Allen,2008Robitaille}. Emission from polycyclic aromatic hydrocarbons (PAHs) is also present at NIR and MIR wavelengths \citep[e.g.,][]{2008Tielens, 2023HensleyDraine}. This PAH emission gives crucial information on the size and charge distributions of the PAHs, and characterizing the evolution of PAHs through the ISM is central to understanding their life cycle \citep[e.g.][]{2005Engelbracht, 2019Chastenet, 2022Wolfire}. However, localizing these young stars and the clouds in which they are born \emph{requires} high spatial resolution observations, 10s of parsecs or better. For previous MIR telescopes including {\it Spitzer} such resolution could only be achieved in the nearest, $D \lesssim 5~{\rm Mpc}$, galaxies. {\it JWST} is finally allowing us to study galaxies at these spatial resolutions at distances up to $\sim20$~Mpc. This will enable us to determine both the galaxy-wide and local processes that affect how stars are born, exert feedback on the surrounding ISM, and then die, enriching the ISM for future stellar generations (the baryon cycle).

Studying this baryon cycle that surrounds star formation and feedback has been the goal of the Physics at High Angular resolution in Nearby GalaxieS \citep[PHANGS;][]{2021Leroy}\footnote{\url{http://phangs.org/}} collaboration through a series of multi-observatory surveys. The latest of these is PHANGS-{\it JWST} \citep{2023Lee}, a Cycle 1 Treasury Program to observe 19 galaxies in eight {\it JWST} filters from 2~$\mu$m to 21~$\mu$m. These observations reveal a complex web of dust and gas as well as dust-embedded stellar clusters that are hidden from similar resolution {\it HST} observations. The goals of this program are wide-ranging, and a number of results have already been published on, for example, PAH properties \citep{2023aChastenet,2023bChastenet,2023Dale,2023Egorov,2023Sandstrom}, feedback-driven bubbles \citep{2023Barnes,2023Watkins}, and embedded stellar clusters \citep{2023Hoyer,2023Rodriguez}\footnote{For the full set of these ``PHANGS-{\it JWST} first results'', we direct interested readers to \url{https://iopscience.iop.org/collections/2041-8205_PHANGS-JWST-First-Results}.}. 

Most of the PHANGS-{\it JWST} first results used data from the first four galaxies observed (NGC~628, NGC~7793, NGC~1365, and IC~5332). Observations of all 19 targets are now complete, and studies are beginning to exploit the full sample \citep{2023Belfiore, 2023Pathak}. This work presents a full public data release (DR1.1.0\footnote{Some issues were found with the astrometry in NIRCam observations of NGC~3627 and have been updated to a DR1.1.1. Other data remains unchanged.}) for the PHANGS-{\it JWST} Cycle 1 program and the associated pipeline (\pipename\footnote{\url{https://pjpipe.readthedocs.io/}, \url{https://doi.org/10.5281/zenodo.10458746}}). It also details the lessons we have learned during data acquisition for this Treasury program. The survey and much of the technical information have already been described in \cite{2023Lee}, and we present here the significantly upgraded view of the reduction approach, emphasizing the (occasionally dramatic) applied changes.

We have undertaken the task of building a pipeline to robustly deal with shortcomings we have discovered using the official {\it JWST} pipeline that are general to imaging observations, or in some cases more specific to observations where emission fills the field of view. These are \onef\ noise in NIRCam data, astrometric alignment, background matching between tiles in a mosaic\footnote{We will refer to these throughout as ``mosaic tiles,'' as opposed to the individual single exposure ``tiles'' (in the case of the NIRCam short images, these are the individual detector images, rather than the combination of the multiple detectors). These mosaic tiles are formed by the combination of all (four) of the dithers in a single pointing (for the NIRCam short images, this combines the four individual detectors as well, for a total of 16 input exposures to a mosaic tile).}, and calibrating flux levels in the absence of a true background. The pipeline also addresses the reality of widely varying resolution across the JWST bands, as well as the opportunity for a small smoothing operation to dramatically improve the signal to noise at the long wavelength MIRI bands. \pipename\ offers robust implementations to address these issues, and the main focus of this paper is to describe our algorithms, illustrate their application to PHANGS-{\it JWST} data, and explain how these link with the official {\it JWST} pipeline. We note that most of the issues addressed by \pipename\ are not specific to PHANGS-{\it JWST}. Indeed, other programs have already made use of this pipeline to improve their data products \citep{2023Peltonen,2024Chastenet}.

The layout of this paper is as follows. We briefly introduce the PHANGS-{\it JWST} survey in Section \ref{sec:data}, before detailing our data processing steps in Section \ref{sec:data_processing}, noting differences between this and our early science efforts \citep{2023Lee}. We provide details of our quality assurance checks in Section \ref{sec:qa}, highlighting remaining known issues in Section \ref{sec:outstanding_issues}. We compare to our early science mosaics and those obtained from MAST in Section \ref{sec:mosaic_comparison}. A description of the output files for the data release is given in Section \ref{sec:data_products}. We summarise in Section \ref{sec:summary}, and present an atlas of our 19 galaxies in Appendix \ref{app:atlas}. Our data (both combined mosaics and individual tiles for PSF photometry) are available publicly at \url{https://archive.stsci.edu/hlsp/phangs/phangs-jwst}.

\section{Data Overview}\label{sec:data}

PHANGS-{\it JWST} is a {\it JWST} Cycle 1 Treasury Program, building upon existing PHANGS efforts to map nearby ($D$~\textless~20~Mpc) galaxies with Atacama Large Millimeter/submillimeter Array \citep[ALMA;][]{2021Leroy} CO observations, Multi Unit Spectroscopic Explorer \citep[MUSE;][]{2022Emsellem} optical integral field spectroscopy, and high-resolution {\it Hubble Space Telescope} \citep[{\it HST};][]{2022Lee} optical imaging. Specifically, this Cycle 1 Treasury (program 2107, PI: Lee) has observed the 19 main-sequence galaxies that form the PHANGS-MUSE subset of the full sample, which have been observed with {\it JWST} between 2022--06--06 and 2024--02--06\footnote{An up-to-date observing log can be found at \url{https://www.stsci.edu/cgi-bin/get-visit-status?id=2107&markupFormat=html&observatory=JWST&pi=1}}. 

\begin{table*}[ht]
\begin{tabular}{c|c|c|c|c}
Filter & Total exposure time\tablenotemark{a} & Avg. tile overlap & 5$\sigma$ point source sensitivity & 1$\sigma$ surface brightness sensitivity
\\
& s & & $\mu$Jy & MJy~sr$^{-1}$\\
\hline
\multicolumn{5}{c}{{\bf NIRCam}}\\
\hline
F200W & 1202.5 & 16 & 0.039 & 0.071 \\
F300M & 386.5 & 4 & 0.084 & 0.052 \\
F335M & 386.5 & 4 & 0.092 & 0.050 \\
F360M & 429.5 & 8 & 0.107 & 0.054 \\
\hline
\multicolumn{5}{c}{{\bf MIRI}}\\
\hline
F770W & 88.8 & 4 & 1.028 & 0.13 \\
F1000W & 122.1 & 4 & 1.253 & 0.13 \\
F1130W & 310.8 & 4 & 2.637 & 0.24 \\
F2100W & 321.9 & 4 & 5.152 & 0.27 \\
\hline
\end{tabular}
\tablenotetext{a}{This is the total exposure time per mosaic tile.}
\caption{Overview of the observational setup for our {\it JWST} filters. We use  a four-point dither pattern in all cases, so the average tile overlap is an indication of how many repeats we perform in the same area. Exposure times and overlaps are from \citet{2023Lee}, but sensitivities have been updated using the same method presented in \citet{2023Lee}, reflecting the new data processing.}\label{tab:data_overview}
\end{table*}

For full details of the survey, we refer the reader to \cite{2023Lee} and we briefly summarize details relevant to this paper here. The observations cover eight bands, spanning 2--21~\micron, using both NIRCam (F200W, F300M, F335M, F360M) and MIRI (F770W, F1000W, F1130W, F2100W). Our observations are primarily mosaics, with a maximum 2$\times$2 tiling pattern. Because our targets are large on the sky, even these mosaics are often not sufficient to cover the full galaxy, though the coverage always captures most of the infrared emission and star formation activity from the galaxy (Figure \ref{fig:all_gals}), and we place the {\it JWST} footprint to overlap existing data. Details of the observations are given in Table \ref{tab:data_overview}.
We use a minimum of four dithers per integration to fully sample the PSF, up to a maximum of sixteen overlapping dithers in the case of F200W. To maximize the observing efficiency, we observed off positions for the MIRI data in parallel with our NIRCam science observations. 
A collage of our mosaics is shown in Figure \ref{fig:all_gals}. A rich tapestry of structure is clearly visible, from extended dust emission to point-like, embedded stellar clusters. 

\section{Data Processing}\label{sec:data_processing}

\begin{figure}[ht]
    \includegraphics[width=\columnwidth]{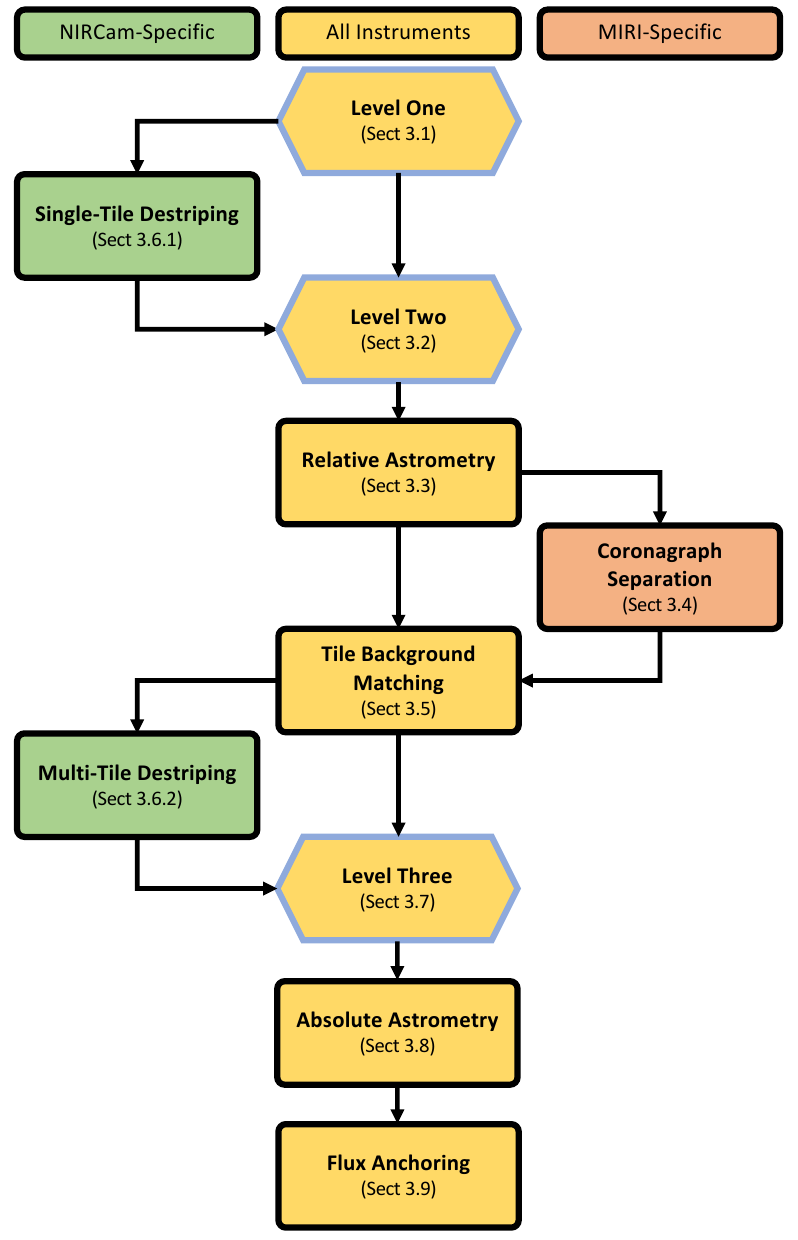}
    \caption{Schematic overview of our data reduction. Whilst most stages are instrument-agnostic, there are some specific to either NIRCam or MIRI. Steps that are present in the official {\it JWST} pipeline are highlighted as blue-outlined hexagons.}
    \label{fig:flowchart}
\end{figure}

In \cite{2023Lee}, we presented our processing methodology for the four galaxies observed during the first month and a half of our program. Since then, we have seen significant improvements in a number of directions. Firstly, the reference files (for this release, we use reference files {\tt jwst\_1201.pmap}\footnote{\url{https://jwst-crds.stsci.edu/}}) have improved as the in-flight performance of {\it JWST} and its instruments have been better understood. Secondly, improvements in the official {\it JWST} pipeline\footnote{\url{https://github.com/spacetelescope/jwst}} \citep{bushouse_howard_2023_8157276} have fixed a number of bugs and improved many of the underlying reduction steps (for our DR1.1.0 reprocessing, we use pipeline version 1.13.4). Finally, we have identified and understood several issues associated with observations of extended, bright sources that fill the field of view and developed robust strategies for many of these. This paper will focus on this last point, but we note our latest data processing represents a large improvement both over our original data release (e.g., improved flux calibration, background matching and astrometry), as well as the mosaics that are currently available in the MAST archive (see Sect. \ref{sec:mosaic_comparison}).


Our pipeline (\pipename) wraps around the official {\it JWST} pipeline \citep{bushouse_howard_2023_8157276}, adding a number of bespoke steps not present in the official pipeline, which we show schematically in Figure \ref{fig:flowchart}. We make \pipename\ publicly available at \url{https://github.com/phangsTeam/pjpipe}, which also includes the configuration required to reproduce our reduction. We show the main output of \pipename, the mosaics, in Appendix \ref{app:atlas}. For users who wish to use \pipename\ for their science, documentation and examples of configuration files are available at \url{https://pjpipe.readthedocs.io/en/}. As an open-source project, we also encourage users to contribute bugfixes and new features. By making our code fully available, we hope that some of the general techniques we have developed for our image processing will also feed back into improvements in the official {\it JWST} pipeline\footnote{Many of the \pipename\ algorithms have been presented at the ``Improving {\it JWST} Data Workshop,'' and the materials and discussions from that workshop are available at \url{https://outerspace.stsci.edu/display/JEA/Improving+JWST+Data+Products+Workshop}.}. A current list of known issues with the {\it JWST} pipeline is maintained at \url{https://jwst-docs.stsci.edu/known-issues-with-jwst-data} (some of which are dealt with by \pipename).

\subsection{Level One Processing}\label{sec:lv1}

The Level One stage of the pipeline transforms uncalibrated images to count rate images via a number of steps (e.g., flagging bad pixels, subtracting reference pixels, and fitting slopes to ramps). The details for each step are given at \url{https://jwst-pipeline.readthedocs.io/en/latest/jwst/pipeline/calwebb_detector1.html}\footnote{For the specific documentation for this version, see \url{https://doi.org/10.5281/zenodo.8436689}}. We essentially keep this step at the observatory-recommended defaults, although we process individual exposures (tiles) from each dither group in observation sequence rather than simultaneously. Processing the observations sequentially allows us to use the output files from the previous observation to identify persistence, i.e., faint residual images caused by previous exposure to bright sources appearing in subsequent integrations. Then data affected by persistence can be flagged in the subsequent persistence step. In our observations, we have not noticed persistence occurring from previous observations, but this could be a potential issue for observations immediately following, for example, observations of solar system planets. We also attempt to mitigate saturation by setting {\tt suppress\_one\_group} to {\tt False} in the ramp fitting step so that a flux can be obtained for any pixels where the first group is not saturated.

\subsection{Level Two Processing}\label{sec:lv2}

Level Two processing applies calibrations and corrections to count rate images on a per-exposure level. This stage performs background subtraction (for MIRI data), initial world coordinate assignment, flat-fielding, and photometric calibration. The details for each step here are given at \url{https://jwst-pipeline.readthedocs.io/en/latest/jwst/pipeline/calwebb_image2.html}. As with Level One, we keep most of these at the observatory-recommended defaults. 

In this stage, we combine into a single master background frame the MIRI background observations in image space. A background estimate is calculated per-pixel after applying sigma-clipping and combining all dithers executed during the off observation. In the case of PHANGS-{\it JWST}, the MIRI background observations are obtained in parallel during the NIRCam science observations and as a result, the dither used is also the one executed to obtain the NIRCam images. When constructing the background image, we use a relatively strict sigma-clip of $1.5\sigma$. Our background fields occasionally contain clear foreground stars or background galaxies and this strict tolerance limits the number of negative artifacts seen in our calibrated images, although in a few cases unavoidable artifacts persist. These are extended background galaxies that the dither pattern fails to remove, and appear as negative spots in our background images. Work is ongoing to mitigate for this in this step (e.g.\ by masking sources and interpolating over missing data), but our current data occasionally contain these artifacts. For our larger mosaics, we expect the overall noise level to be slightly lower because we acquire background observations for every NIRCam position, meaning we generate a background from more observations.

This background estimate is then subtracted from the science image. Following this step, we obtain fully calibrated individual tiles (in units of MJy/sr), to which we perform a number of bespoke steps before mosaicking. We note that for this data release, the flux calibration does not take into account the decrease in sensitivity at long MIRI wavelengths. This will be updated in a future version of the data. 

\begin{figure*}[ht]
    \includegraphics[width=\textwidth]{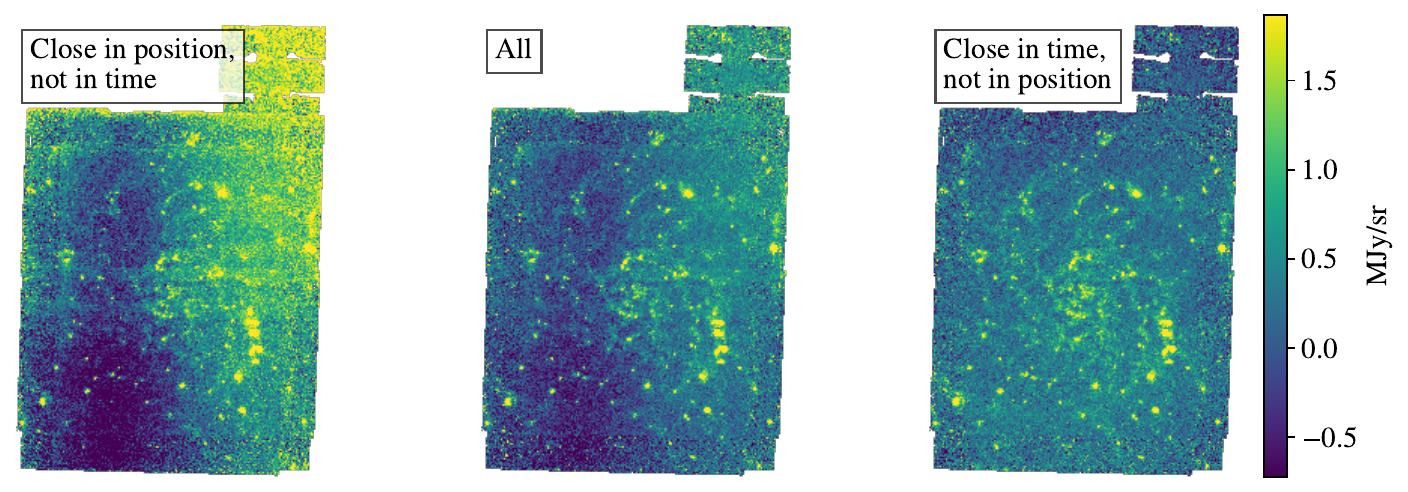}
    \caption{Illustration of how the temporal variability of MIRI backgrounds affects the final science images, using F2100W IC~5332 observations. {\it Left}: MIRI background using later NIRCam parallel observation, as the initial observation failed. This background is therefore taken close in position to IC~5332, but not close in time. {\it Middle}: Combination of IC~5332 background (close in position), and NGC~7496 background (close in time) observations. {\it Right}: Using only the NGC~7496 (close in time) background observation. Clearly, using backgrounds taken closer in time leads to a better, and overall flatter background.}
    \label{fig:ic5332_bg}
\end{figure*}

When constructing MIRI background images, proximity of the background to the on-source image in \textit{time} appears to be a critical factor. For PHANGS-{\it JWST}, the initial NIRCam observations, and hence the MIRI background images, failed for IC~5332, and the successful NIRCam science and MIRI background observations were only obtained months later. Another PHANGS-{\it JWST} target, NGC~7496 was observed immediately prior to IC~5332. Figure \ref{fig:ic5332_bg} compares the results subtracting backgrounds obtained close in position but not in time (i.e., using the later IC~5332 for the background) and close in time but not in position (i.e., using NGC~7496). We perform a sigma-clipped RMS measurement on these three images, using a source mask created from the ``close in time, not in position'' backgrounds. We obtain RMS=[0.80, 0.46, 0.36]~MJy~sr$^{-1}$ in these three cases. This shows that using the MIRI backgrounds obtained months after the MIRI science observations led to large-scale artifacts in the final mosaics, highlighting the temporal variability of the MIRI backgrounds. In the current release, we use the NGC~7496 ``off'' observations that ended 14 minutes before the IC~5332 science observations started to construct the background for IC~5332. For pointings where the background observations fail, we would therefore recommend re-observing both science and background given the clear time variability of the MIRI background.

\subsection{Relative Astrometric Alignment}\label{sec:relative_astrometry}

As noted in \cite{2023Rigby}, the relative pointing accuracy after guide star acquisition is excellent, to within a few milli-arcseconds. However, given that many of our observations are mosaics that use multiple tiles, we found that the absolute alignment between these guide stars is insufficient for our purposes. This could be due to a number of reasons -- the star tracker locking onto an incorrect target, that target being extended at the {\it JWST} resolution, or outstanding uncertainties in the location of the chips relative to the guide star tracker, for instance. We expect that as the guide star catalogs improve through the lifecycle of {\it JWST} and as the instruments are better understood, these issues will become less common and severe, but for these observations we require correction to get a good common astrometric alignment between the various mosaic tiles. 

Using {\tt tweakreg}, we align the mosaic tiles using a single shift for each set of dithers. {\tt tweakreg} uses an iterative rejection algorithm to match sources between point source catalogs, in order to align a set of input images. For each instrument and individual mosaic tile, the tile is observed in filter sequence so that all observations of a specific tile with an individual instrument use the same guide star. Therefore, because of the excellent relative pointing accuracy, initial testing found that the astrometric correction is almost identical for the four NIRCam images, while the four MIRI filters also share a single correction that is different from the NIRCam correction. We therefore base our astrometric solutions on two bands, F300M for the NIRCam observations and F1000W for the MIRI observations. The F1000W band is chosen instead of the shorter-wavelength F770W band because it is not expected to be as heavily dominated by PAH emission, making stellar point sources within the galaxies clearer. It also has improved angular resolution compared to the F1130W band, allowing for better centroiding of point sources. For NIRCam, we work with F300M because we found the sheer density of point sources in F200W to be a detriment to solving for a good astrometric correction, so that using F200W as the reference band often led to a poor initial alignment. 

For each visit (i.e., each mosaic tile), we calculate a single $x$, $y$ shift and apply it to all individual exposures targeting that mosaic tile. The offsets derived during this step are typically small, on the order of 0\farcs1. We experimented with including rotations but found this degraded the astrometric solution, without measuring any significant rotation.  Note that we apply the same shift to the NIRCam short and long filters, although given some uncertainties in the exact chip position we will allow for a further shift during our Level Three processing (Sect. \ref{sec:lv3}). 

\subsection{MIRI Coronagraph Separation}\label{sec:corona_sep}

The MIRI data contain Lyot coronagraph observations. The coronagraph is separate from the primary imager but can still be used to obtain science-quality data because it shares the same optical path as the primary MIRI imager. Our testing showed that the background levels for the coronagraph observations were often significantly different compared to those on the primary imager. To account for these different background levels, we separate out the coronagraph from the primary imager observations by editing the data quality array in the Level Two images. We produce two images, one image flags the primary imager as `non-science' and the other flags the coronagraph as `non-science.' This means we can perform the background matching (Sect.~\ref{sec:bg_match}) separately for these two parts of the CCD and include the coronagraph in our final mosaics after accounting for their different background levels.

\subsection{Background Matching}\label{sec:bg_match}

Particularly for MIRI data, the background levels can vary substantially, even after subtracting the background calculated in the Level Two processing. This is mainly driven by instrumental backgrounds (primarily the telescope thermal background at longer wavelengths, and scattered light at shorter wavelengths), which we observe to vary at the $\sim$0.1~MJy~sr$^{-1}$ level between dithers even within a single mosaic tile. These must be corrected for in order to provide consistent level matching between individual observations, and tiles in a mosaic. 

The {\it JWST} pipeline uses the {\tt skymatch} algorithm to provide consistent levels between tiles. It does this by calculating a single sigma-clipped average within overlap regions, and then minimizing the difference among the background values for the various overlapping observations (i.e., minimizing the difference of the medians). However, we found no combination of parameters that allowed this algorithm to produce good results across our full dataset. Instead, ``jumps'' between adjacent tiles always remained visible. This likely reflects the fact that our overlap regions are full of bright extended emission, with no real ``background'' to sigma-clip to. 

We therefore take a different approach. For each pixel in the overlap region between a pair of images, including those that contain signal from the source, we calculate the difference between values at that pixel in the different images. Then we take the median of the pixel-by-pixel differences to calculate the average offset between these two images. To do this, we first reproject all images to a common pixel grid and then calculate the median difference of all overlapping pixels. When calculating these median differences between pairs of tiles, we use sigma-clipping to reject outlying differences. We then adjust the background level of the images being considered, in order to minimize the differences among all image pairs being considered. This is done using least-squares minimization, weighting each image pair by the root mean square (RMS) spread of the difference histograms for each tile overlap. To avoid including noisy and untrustworthy overlaps between images, we do not include any image pairs with a total pixel overlap less than 0.2~percent the average number of pixels in an overlap region. We also do not include image pairs that do not overlap by at least 25~percent along at least one axis of the image, or image pairs that have RMS values that are more than $2\sigma$ greater than the average RMS value of all the other image pair differences (i.e.\ very noisy difference histograms, which often occur when there are few overlapping pixels). These cuts typically remove the small diagonal overlaps that occur in the centers of the $2\times2$ mosaics or similar (or the centers of the four NIRCam short chips). These choices mirror those used by {\tt Montage} \citep{2010Jacob} for their background matching step. 

In practice, we perform this background matching in two stages. First, we match the level between dithers within a visit group. Then we stack these and match the tiles in the overall mosaic. We find that this process produces excellent background matching both for NIRCam and MIRI tiles, significantly better than what we could achieve with the {\tt skymatch} step in the {\it JWST} pipeline.

\subsection{NIRCam Destriping}\label{sec:destripe}

NIRCam data are well-documented to suffer from \onef\ (pink) noise \citep{2020Schlawin}. For example, this is clearly visible in the left panel of Figure \ref{fig:single_tile_destripe} as horizontal striping is pervasive across the entire image. This noise is due to the way NIRCam data are read out
and has power at a number of spatial scales. This \onef\ noise is particularly challenging to remove from our data because diffuse emission also pervades our observations with power on overlapping spatial scales. 

Reference pixels are present as a four-pixel border around the NIRCam chips, and the Level One processing provides a first-order correction to remove this noise, but the striping is still persistent even following this {\tt refpix} step. Fortunately, algorithms have been developed for mitigating this \citep[e.g., {\tt remstriping};][]{2023Bagley}. Unfortunately, these have mostly been designed for blank fields and high redshift observations.

To mitigate this noise, we adopt a two-stage approach. First, we employ a single-tile destriping technique that uses a single exposure. Second, we use temporal information from overlapping exposures to remove remaining large-scale stripes that the single-tile destriping does not correct. Because the \onef\ noise is instrumental we apply this destriping to the data {\it before} our Level Two processing and flat fields are applied. 

Note that for our initial reduction of the first four galaxies \citep[presented in][and associated papers]{2023Lee}, we used a robust principal component analysis (PCA) method. Further testing has revealed that this method is extremely sensitive to source masking, and using a median filter yielded more robust results. Particularly, our testing with the PCA method often showed imprints of the source in the noise model, indicating that we are removing real emission. The same is not true in the case of the median filter. We do believe that PCA may ultimately represent the optimal strategy when there is limited extended emission in a tile, but our data contain too much emission for the PCA to work successfully.

\subsubsection{Single-Tile Destriping}

First, following \cite{2023Bagley}, we measure \onef -noise induced stripes on flat-fielded images, but subtract the stripes before the flats are applied.  We start by filtering our data using a \cite{1930Butterworth} filter, which removes large-scale structure (both emission and large-scale \onef\ noise). We opt to use this filter because its smooth frequency response suppresses the Fourier ``ringing'' that is often seen around discontinuous sources in many high-pass filters. 

Having filtered the data, we then remove the \onef\ noise using a median filter at a number of scales. The \onef\ noise is present in both the $x$ and $y$ directions, and we subtract striping from these independently. To avoid oversubtraction, we create a source mask and exclude any remaining bright sources after the filtering. Then, we calculate and subtract a row median in the $y$ direction (collapsing along columns), before calculating and subtracting a median in the $x$ direction (i.e.\ collapsing along the rows). In this step, we split the data per-amplifier because the noise properties are different in each amplifier. For rows or columns where more than 80~percent of the data are masked, we fall back to the full row or column median, to avoid spurious values. 

\begin{figure*}[ht]
    \includegraphics[width=\textwidth]{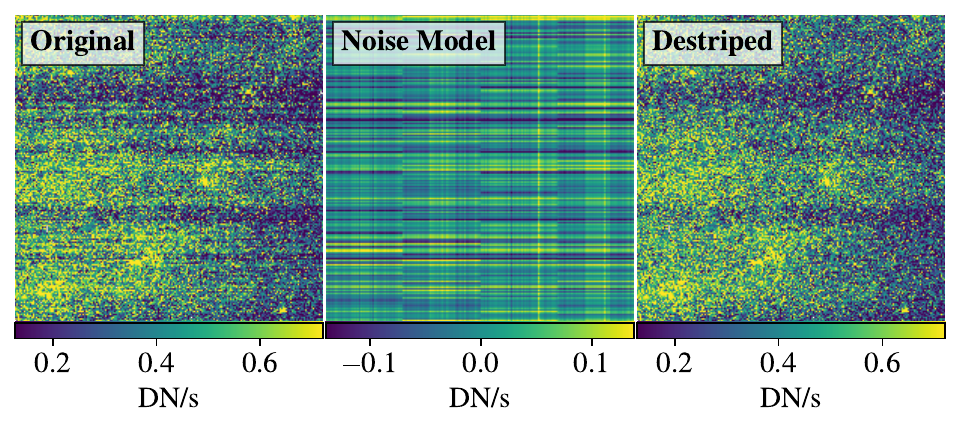}
    \caption{Single-tile destripe model for one F200W observation of NGC~628. Left shows the original data, the middle is our stripe model and the right the destriped data. Although effective at removing much of the striping, some low-level, large-scale stripes remain.}
    \label{fig:single_tile_destripe}
\end{figure*}

We show an example of this single-tile destriping in Figure \ref{fig:single_tile_destripe}. We find the amplitude of the noise removed in this step to be typically around the 0.05~DN~s$^{-1}$ level. This step effectively removes the \onef\ noise on small scales whilst retaining the large-scale structure present in the images. However, large-scale `ripples' remain, which we mitigate with our second destriping technique (Sect. \ref{sec:multi_tile_destripe}).

\begin{figure*}[ht]
    \includegraphics[width=\textwidth]{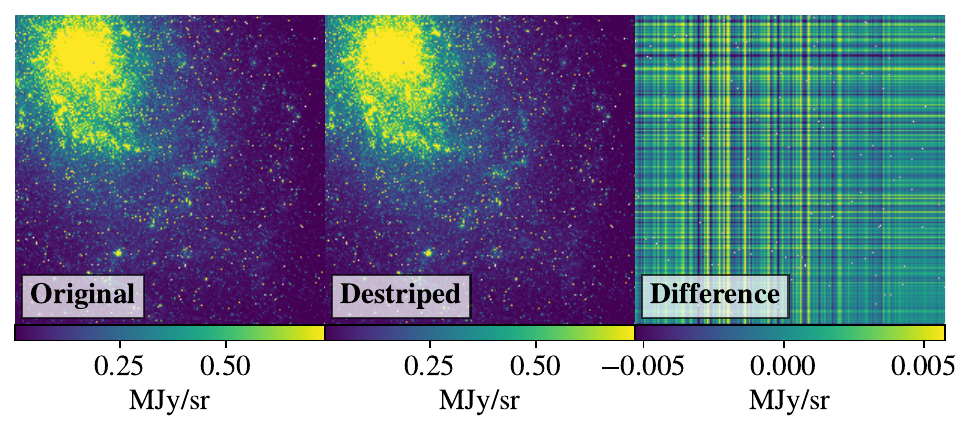}
    \caption{Applying the single-tile destriping to {\it Spitzer} observations of M33. The left and middle panels show the original and destriped data respectively, and are locked at the same colorscale. The right panel shows the difference image, on the order of 0.001~MJy/sr, essentially below the noise level of the original image. The white pixels in the difference image are due to pixels with no data; these are not present in the {\it JWST} noise models but are present here due to differences in the specifics of the data between {\it JWST} and {\it Spitzer}.}
    \label{fig:destripe_test}
\end{figure*}

We can see in the middle panel of Figure \ref{fig:single_tile_destripe} that there is no clear imprint of the galaxy in the noise model, indicating that this method does not remove flux in the image. To test this further, and ensure that this destriping procedure does not alter the true flux levels to a significant extent, we take {\it Spitzer} observations of M33 at a similar physical resolution. These data do not suffer from \onef\ noise, and so this test is to ensure that our destriping algorithm essentially does not subtract anything in the absence of \onef\ noise. We apply the destriping to this image after ensuring it looks like a calibrated {\it JWST} image to the pipeline, with the exception of the lack of \onef\ noise. The difference is shown in Fig. \ref{fig:destripe_test}, with the difference (i.e., the ``stripes'' measured here) being on the order of 0.001~MJy/sr. We are therefore confident that our destriping does not add or remove flux to any great degree.

\subsubsection{Multi-Tile Destriping}\label{sec:multi_tile_destripe}

Although our previous step effectively removes small-scale \onef\ noise, large-scale noise is still present because it survives the Butterworth filtering. This is particularly visible in the short NIRCam bands, but is prevalent throughout the NIRCam observations. The stripes are typically at around the $1\sigma$ level compared to the error map, but can be larger (particularly for the NIRCam short wavelengths). We take advantage of the fact that this noise is not correlated between different exposures to remove striping from each individual exposure whilst leaving real galactic structure intact. 

\begin{figure*}[ht]
    \includegraphics[width=\textwidth]{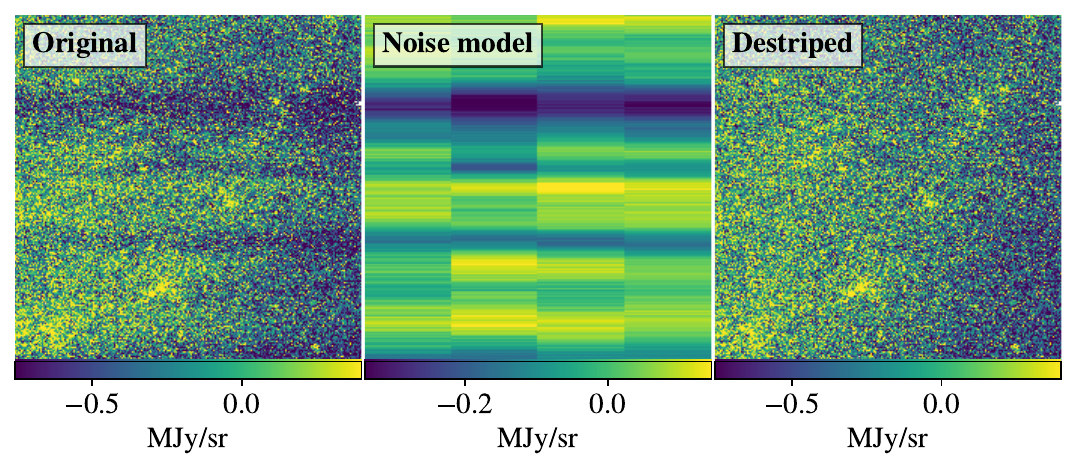}
    \caption{Multi-tile destriping for one F200W observation of NGC~628. {\it Left}: uncorrected data. {\it Middle}: noise model. {\it Right}: corrected data. This step successfully removes the majority of the remaining large-scale striping in the data, leaving a mostly flat background.}
    \label{fig:multi_tile_destripe}
\end{figure*}

In detail, for each exposure, we create a sigma-clipped median image of all overlapping exposures. For PHANGS-JWST, there are at least four overlapping exposures for all bands and up to 12 for F200W. When creating this image, we make sure that the overall flux level matches between tiles by performing this after our background matching step (Sect.~\ref{sec:bg_match}). For each individual exposure, we then create a difference image by comparing that exposure to this median image. Because it is present in both the median and the image, real structure will not be present in this difference image but it will include any large-scale ripples. Using this difference image we then calculate a sigma-clipped median along rows, per-amplifier as before. We subtract these medians from our image to effectively remove any large-scale stripes still left in the data. An example of this procedure is shown in Figure \ref{fig:multi_tile_destripe}. Typically, the amplitude of the noise removed here is at around the 0.05~MJy~sr$^{-1}$ level, and is worst for the F200W band. We also attempted to perform destriping using only this method, but found that the small-scale stripes remained; hence, both the single- and multi-tile destriping are necessary to remove the \onef\ noise on all scales.

\begin{figure*}[ht]
    \includegraphics[width=\textwidth]{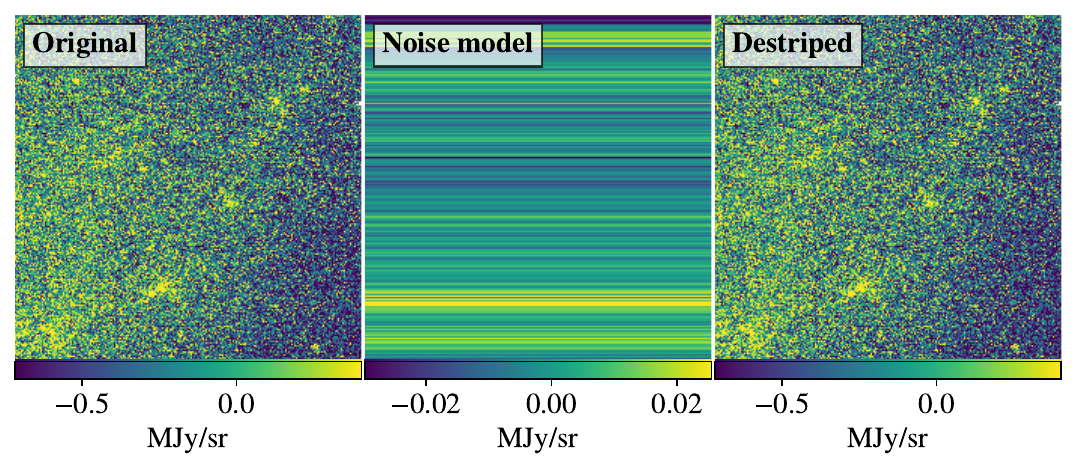}
    \caption{Multi-tile destriping for one F200W observation of NGC~628, having smoothed the stacked image perpendicular to the stripe axis. {\it Left}: uncorrected data. {\it Middle}: noise model. {\it Right}: corrected data. This step successfully removes the remaining striping, so having completed this step our backgrounds are flattened.}
    \label{fig:multi_tile_destripe_smooth}
\end{figure*}

Particularly in the short NIRCam observations, there remains some very low-level striping in the data following this first pass using overlapping tiles. This reflects the fact that whilst the \onef\ noise is essentially random, given its prevalence it can interfere constructively when creating an average image. We therefore carry out a second pass where we create a smoothed, stacked image, performing a large-scale (a tenth the size of the full array) median filter along the axis perpendicular to the direction of the stripes in the stacked image. This creates a new median image as in the previous step, but with any remaining large-scale noise filtered out. We then subtract the large-scale median filtered image from the individual tiles and then take a row-by-row median to construct a last noise estimate. We calculate the median across the full tile rather than per amplifier, as the median filter redistributes flux, potentially leading to a bias along rows completely full of emission. We show an example of this final step in Figure \ref{fig:multi_tile_destripe_smooth}, and note that these last remaining stripes have significantly smaller amplitude than those subtracted in the previous destriping steps, at around the 0.01~MJy~sr$^{-1}$ level. Whilst we have not performed detailed tests on how much our destriping lowers the noise in our imaging, initial tests in blank fields appear to lower the RMS noise level by around 10~percent, consistent with other algorithms \citep[e.g.][]{2023Bagley}. We leave a full exploration of optimal destriping methods and noise improvements to future work. Having carried out these destriping steps, the images appear stripe-free.

\subsection{Level Three Processing}\label{sec:lv3}

Level Three processing combines individually calibrated exposures into a final mosaic via drizzling \citep{2002FruchterHook}, along with astrometric alignment, background matching, and cosmic ray rejection. The details for each step here are given at \url{https://jwst-pipeline.readthedocs.io/en/latest/jwst/pipeline/calwebb_image3.html}. 

We disable astrometric alignment for the MIRI images, having already performed this step manually (Sect. \ref{sec:relative_astrometry}); we find that keeping this on tends to degrade the astrometric solutions, given the small number of point sources in each MIRI exposure. We do, however, leave this on with a strict tolerance for the NIRCam images. The derived shifts are significantly less than one pixel, highlighting that our initial alignment step is performing well. 

We also disable level matching for all bands, as we have implemented our bespoke solution. We do, however, subtract a constant value as the lowest sigma-clipped value in each group of tiles. This would get us to a ``true'' background level in the case of blank fields in our observations, although as we will see in Section \ref{sec:anchoring}, this is not generally the case. 

We perform outlier flagging with default parameters, although to speed this step up we keep all images in memory ({\tt in\_memory = True}; if this setting is False then this step takes on the order of hours, rather than minutes). F200W is the most memory-intensive band here, and can use up to 250GB of RAM. Following Level Three processing, we have final mosaics in all of our eight bands, which are aligned north-up. These mosaics are produced at the native pixel scale of the data (i.e.\ 0\farcs031 for NIRCam short, 0\farcs063 for NIRCam long, 0\farcs11 for MIRI). We use the default {\tt pixfrac} of 1.0, and note that as we perform another round of astrometric alignment following the mosaicking (Sect.\ \ref{sec:absolute_astrometry}), the images are not aligned to the same pixel grid. We expect that future versions of the data will feature products aligned to the same pixel grid.

\subsection{Absolute Astrometric Alignment}\label{sec:absolute_astrometry}

Following Level Three processing, our mosaics have good internal alignment but are not astrometrically locked to an external reference frame. Given the relatively small fields of view for our PHANGS-{\it JWST} mosaics, they often include few Gaia sources. The Gaia sources that are in the field are often extended H{\sc ii} regions or stellar clusters within the galaxy, and not suitable point sources on which to centroid a final astrometric solution. Therefore, we rely on an external catalog to supply our absolute astrometric reference. 

For PHANGS-JWST, these are luminous red stellar sources, a mix of AGB stars and red supergiants, that are bright in our {\it JWST} bands. These stars are identified from DOLPHOT \citep{2000Dolphin, 2016Dolphin} catalogs produced from the {\it HST} imaging, and their astrometric calibration descends from those data  \citep{2022Lee}. To select these sources, we use a combination of detection quality and color-magnitude cuts. To obtain bright, uncrowded, well-PSF-fitted point-like sources, we use the following DOLPHOT parameter criteria: ${\rm S/N_{F814W}} \ge 10$, ${\rm crowd_{F814W}} \le 0.01$, ${\rm PhotQualFlag_{F814W}} = 0$, ${\rm OBJTYPE}=1$, $-0.025 < {\rm sharp_{F814W}} \le 0.01$ and $-0.15 < {\rm round_{F814W}} \le 0.4$. The crowding parameter describes how much brighter a star would be (in magnitudes) if the flux from neighboring stars was not removed. By definition the value is zero for an isolated star.  The object type is a classification for the morphology of the sources, where 1 corresponds to a well-detected star.  The quality flag provides information on artifacts, such as bad or saturated pixels, and whether the extent of the source is not fully captured because it is at the edge of the chip.  A flag of zero indicates no such issues.  Sharpness is a morphological parameter that is computed relative to a point source: if a source is perfectly fit by the PSF model, then the value is zero.  Positive values indicate sources are more centrally peaked than the PSF (e.g., a cosmic ray), while negative values indicate those that are more extended.  Finally, roundness is a symmetry parameter, with a value of zero for perfectly circular sources. For further discussion, we refer the reader to \cite{2022Thilker} and the {\tt DOLPHOT} \citep[][]{2000Dolphin,2016Dolphin} documentation (\url{http://americano.dolphinsim.com/dolphot/}) Sharpness and roundness limits are slightly adjusted if necessary per galaxy. For the color-magnitude cuts, we require F814W in the range $3-5$ mag brighter than the foreground extinction corrected tip of the red giant branch (TRGB) and take only sources with $V$-$I$ color in the range $2 \le F555W-F814W < 4$ mag.  

{\tt tweakreg} is used to perform the alignment, but because our astrometric solutions can often be different from the internal astrometry by an arcsecond or more, we take a two-step approach. Firstly, we use a relatively wide search area with a loose tolerance to find matches, using a simple $x$/$y$ shift (i.e., {\tt shift} in {\tt tweakreg}). Then we use this initial guess with significantly stricter tolerances to improve the initial solution. In this second step, we also allow for rotation ({\tt rshift}). We generally find small rotations, although occasionally up to $\sim0.01^\circ$, leading to sub-pixel shifts across the full mosaics. 

We perform this procedure separately for each NIRCam band. For MIRI, there are a lack of true point sources at many of the longer wavelength bands (particularly 21~\micron). For this, we calculate a solution from the F1000W band (as the highest resolution MIRI band with many clear point sources), and apply this solution to all the MIRI bands as they have consistent internal astrometry. At this stage, we also back-propagate these astrometric solutions to the individual tiles, as these are used as the inputs to the stellar photometry package {\tt DOLPHOT} \citep{2000Dolphin, 2016Dolphin}.

\subsection{Anchoring the Background Level}\label{sec:anchoring}

\begin{figure}[ht]
    \includegraphics[width=\columnwidth]{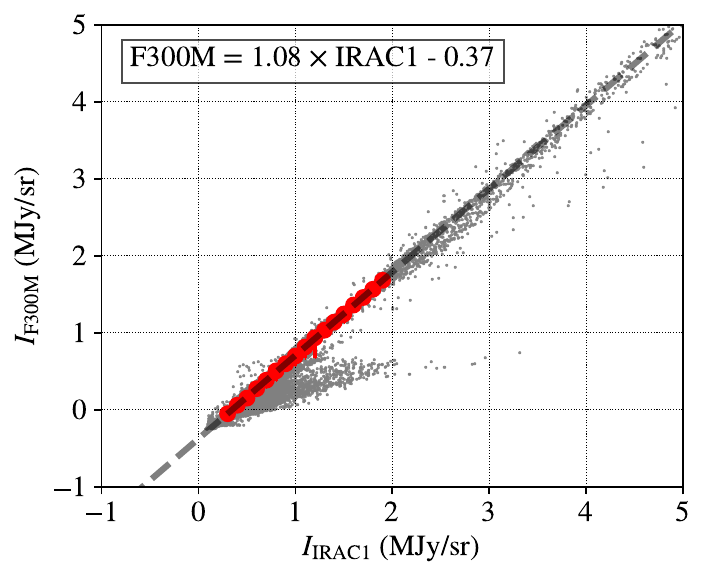}
    \caption{External anchoring for NGC~628 at F300M to IRAC1 (both convolved to 4\arcsec\ resolution). The gray dots indicate individual pixels, the red dots are binned medians for the region over which we fit to anchor. The best fit line is shown in dashed gray, with the fit parameters given in the top left.}
    \label{fig:anchor}
\end{figure}

After the tile background matching step above, our mosaic tiles have a consistent internal background (Sect.~\ref{sec:bg_match}), with little or no jump in intensity from tile to tile within a mosaic. However, there is still uncertainty in the overall intensity of the background level. That is, at this stage the intensity of empty sky near the galaxy is not necessarily zero, and our empirical tests suggest background level variations by several times $\sim 0.1$~MJy~sr$^{-1}$ from galaxy to galaxy and band to band to be common, even up to $\sim \pm 1$~MJy~sr$^{-1}$ in some cases. Given the small field of view of many PHANGS-JWST mosaic compared to the angular extent of the galaxy, it is often not trivial to identify a region of true galaxy-free sky to establish the background level.

In order to establish a consistent, correct background level across all bands we adopt a two step ``anchoring'' procedure that builds on the approach described in \citet{2023Leroy}. First, we solve for an internally consistent background level across all NIRCam bands and all MIRI bands. Then, we set the overall background level by comparing one MIRI and one NIRCam band to wide-area \textit{Spitzer} or WISE images as these include plenty of galaxy-free sky and therefore have well-established sky background levels.

All stages of this ``anchoring'' procedure assume that near-IR or mid-IR intensities of the same region at different bands exhibit a linear relation, on average, at low to moderate intensities. We solve for the slope and intercept of this relation. The slope represents the average band ratio, which may reflect underlying dust or stellar physics but does not matter to this exercise. The intercept represents the intensity offset between the current sky background level in the two images. Empirical testing \citep[e.g., Fig. \ref{fig:anchor} and see Appendix B of][]{2023Leroy} shows that tight linear relations hold among every near-IR and mid-IR band pair that we have examined. That is, the bands appear to approach empty sky in approximate lock-step.

We apply this procedure to solve for four sets of offsets for each source:

\begin{enumerate}
\item The sky background offset between the NIRCam F300M image and a wide-field reference \textit{Spitzer} IRAC1 $3.6\mu$m image (when available) or a wide-field WISE1 $3.4\mu$m image (when IRAC1 data are not available).
\item Internal sky background offsets among NIRCam images.
\item The sky background offset between the MIRI F770W image and a wide-field reference IRAC4 $8\mu$m image (when available) or a wide-field WISE3 $12\mu$m image (when IRAC4 data are not available).
\item Internal sky background offsets among the MIRI images.
\end{enumerate}

\noindent Then we use these to place the NIRCam and MIRI data on a common sky background level that is ``anchored'' to share the sky background level of the wide-field reference image.

An example of the procedure is shown in Figure \ref{fig:anchor}. In detail, we conduct the fit by constructing medians within bins defined to span a range of intensities close to zero (which we adjust somewhat from source to source and band to band to capture the behavior near the background level). The median binning approach down-weights band ratio variations in individual regions, e.g., see the outlying spur at low intensities in Fig. \ref{fig:anchor}. The focus on low intensities avoids, e.g., curvature in the band-to-band intensity relations due to the influence of changing stellar and dust colors in \textsc{Hii} regions or galaxy centers. Note that in this approach the solved slope is a nuisance term. Though the band ratios are of physical interest elsewhere, here we only care that we can solve for a linear relation and use this to adjust the background level of the image to a consistent, externally validated scale.


The procedure appears to work empirically. The bands show tight, clear linear relations with one another, almost always similar to Fig. \ref{fig:anchor}. The ratio images among bands in low intensity parts of galaxies, which tend to be sensitive to background-level discrepancies, also appear mostly free of background-related artifacts. Most directly, in a few cases we do have some regions that contain mostly empty sky (which we will also use to study the noise properties as a function of resolution in Sect. \ref{sec:noise_scaling}). After this anchoring procedure, the intensity histograms appear centered at intensities within a few times $\sim 0.01$~MJy~sr$^{-1}$ of zero.

We caution that the anchoring to a true, correct sky background will only be as good as the reference image. Our IRAC1 images come mostly from S$^{4}$G \citep{2010Sheth}, our IRAC4 images from SINGS \citep{2003Kennicutt} or LVL \citep{2009Dale}, and our WISE1 and WISE3 images from z0MGS \citep{2019Leroy}. In each case, we conducted an additional round of by-hand background fitting to the reference image, but we do note that this remains an area where additional work (as well as improvements to the algorithm) may further improve the images at the few times $0.01$~MJy~sr$^{-1}$ level.

\subsection{PSF Matching}\label{sec:psf_matching}

\begin{table*}[ht]
\begin{tabular}{cc|c|c|c|c}

Instrument & Filter & Source PSF & Very safe Gauss. & Safe Gauss. & Aggressive Gauss. \\
  &   & FWHM [$''$] & FWHM [$''$] & FWHM [$''$] & FWHM [$''$] \\
  \hline
NIRCam & F200W & 0.064 & 0.089 & 0.077 & 0.069 \\
NIRCam & F300M & 0.104 & 0.141 & 0.125 & 0.115 \\
NIRCam & F335M & 0.115 & 0.161 & 0.139 & 0.122 \\
NIRCam & F360M & 0.122 & 0.170 & 0.147 & 0.130 \\
MIRI & F770W & 0.238 & 0.320 & 0.286 & 0.258 \\
MIRI & F1000W & 0.314 & 0.415 & 0.375 & 0.339 \\
MIRI & F1130W & 0.360 & 0.478 & 0.436 & 0.400 \\
MIRI & F2100W & 0.651 & 0.863 & 0.777 & 0.699 \\
\end{tabular}
\caption{FWHM for very safe, safe, and aggressive convolution kernels to a Gaussian PSF for our {\it JWST} filters, calculated using the metrics of \citet{2011Aniano}.}\label{tab:kernels}
\end{table*}

Much of the PHANGS science relies on multi-wavelength photometry, for which performing an analysis at matched resolution is important. We therefore produce images convolved to a common resolution as a standard part of the pipeline. We use the method outlined in \cite{2011Aniano} to produce convolution kernels that can be convolved with the data to produce a common, Gaussian-resolution image. Our code, called \texttt{jwst\_kernels}, is available at \url{https://github.com/francbelf/jwst_kernels}. We assume the JWST PSFs computed using the {\tt WebbPSF} simulation tool \citep{Perrin2014} version 1.2.1 taking the detector effects into account.

Based on the $W_{-}$ metric from \cite{2011Aniano} (their equation 21, which quantifies the level of negative values in the kernel, and is related to the necessity of moving power from large to smaller scales), the minimum ``very safe'' common resolution across our {\it JWST} dataset is determined by the F2100W image and is 0\farcs86. In testing we find that this resolution occasionally produces ringing around bright point sources. This is likely due to the highly non-axisymmetric and non-Gaussian nature of the {\it JWST} PSFs. We therefore produce beam-matched images at 0\farcs85 (close to the ``very safe'' value), 0\farcs9 and 1\arcsec, with the latter two being very conservative choices, which may nonetheless be preferable for comparison with low-resolution data. 

Users interested in mitigating the impact of the JWST PSF may wish to convolve the data with a suitable kernel to obtain the smallest ``safe'' Gaussian PSF for a specific band, different from our default F2100W. We report in Table \ref{tab:kernels} the Gaussians PSFs corresponding to ``very safe'', ``safe'', and ``aggressive'' kernels according to \cite{2011Aniano}. Suitable kernels for all these bands can be generated using our \texttt{jwst\_kernels} code with the \texttt{make\_jwst\_kernel\_to\_Gauss} function. For matched PSF work, we recommend experimenting with the kernels presented here; if resolution is particularly important, then opting for an ``aggressive'' kernel may be the optimal choice. However, these more aggressive convolutions are more prone to artifacts, especially around very bright sources. We therefore strongly suggest using visual inspection to ensure any convolved images are artifact free to the level the science requires.

We also carry the convolution through to the error maps, by following standard error propagation (i.e., convolving the square of the error map with the square of the kernel, and taking the square root of each image). This assumes the noise is uncorrelated at the pixel level. This appears to be true to first order based on our tests of the scale-dependence of the noise (Sect. \ref{sec:noise_scaling}), but is unlikely to hold in all cases given that the PSF is often significantly oversampled. As the community continues to better understand the noise properties, including contribution of sky noise and inter-pixel correlations of noise within the {\it JWST} instruments, we expect these error maps to become more sophisticated.

\section{Data Quality}\label{sec:qa}

\begin{figure*}[ht]
    \includegraphics[width=\textwidth]{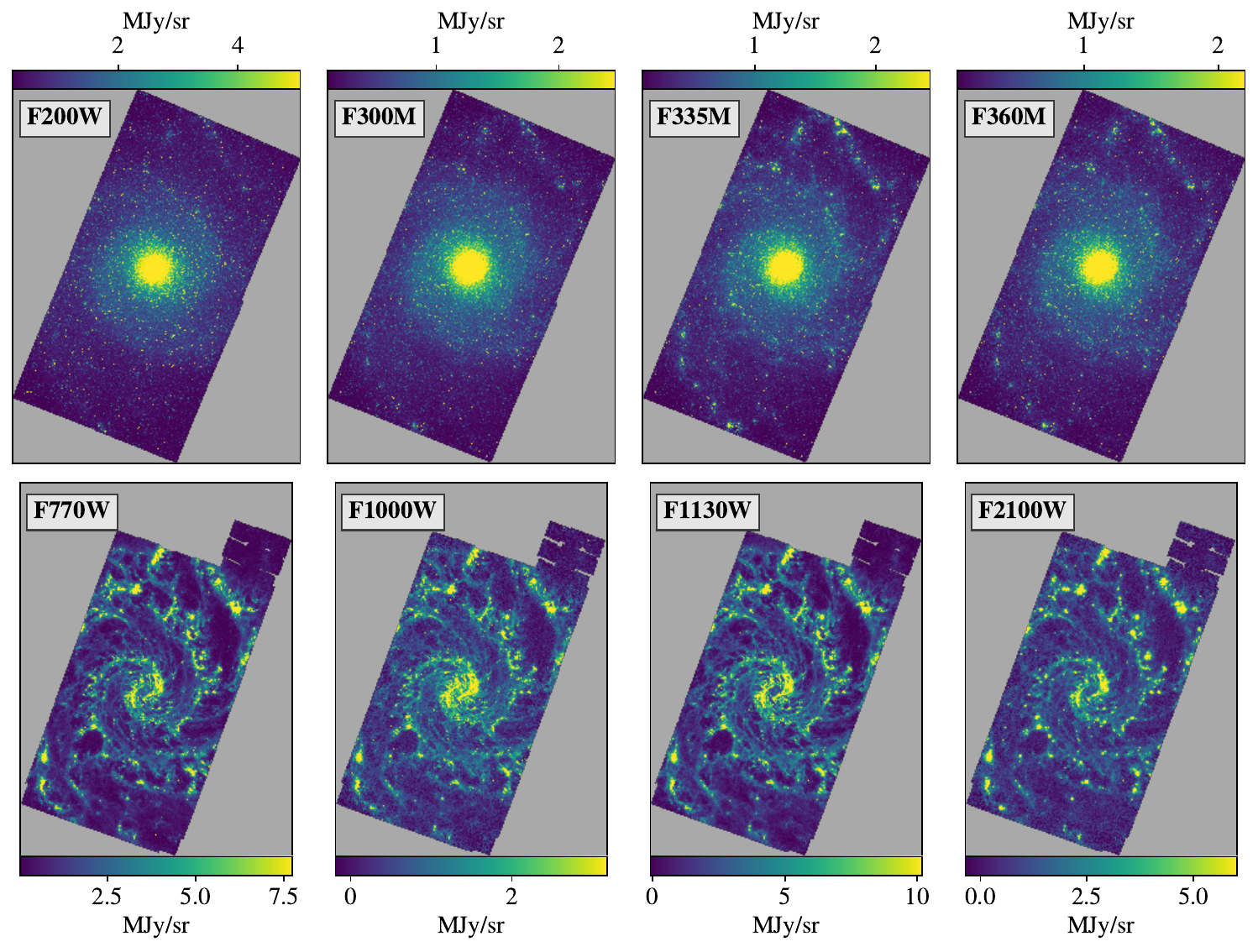}
    \caption{Final mosaics for each band for NGC~628. NIRCam filters are shown in the top row, and MIRI on the bottom. Each image is on a linear stretch between the $2^{\rm nd}$ and $98^{\rm th}$ percentiles of the pixel distribution, to highlight the fainter structure in the images. The complete figure set (19 images) is available in the online journal.}
    \label{fig:final_mosaics}
\end{figure*}

In this Section, we detail the quality of the data, with particular attention to how the noise properties scale as a function of resolution. We also describe our quality assurance process, and note outstanding issues with the data. For an example of the quality of our data, we show the mosaics for NGC~628 in Figure \ref{fig:final_mosaics}, with equivalent plots for the other galaxies in Appendix \ref{app:atlas}.

\subsection{Noise Scaling with Resolution}\label{sec:noise_scaling}

Of the bands used in this survey, the F2100W band has the largest native PSF (FWHM of 0\farcs674), much larger than the 0\farcs11 size of the pixels.  The F2100W band also has the highest noise level of any band in the PHANGS-JWST data set. As a result, any signal to noise cuts performed in an analysis that includes F2100W data will be dominated by the large number of pixels cut across these maps. The F2100W noise is instrumental in nature, and correlated at the pixel level. The pixel size is much smaller than the PSF (the F2100W PSF is oversampled by $\sim 10\times$ in area relative to Nyquist sampling). Therefore the noise in these images can be dramatically decreased by smoothing the images to a slightly worse resolution. Smaller gains can be also realized for the other bands.

\begin{figure*}[ht]
    \includegraphics[width=\textwidth]{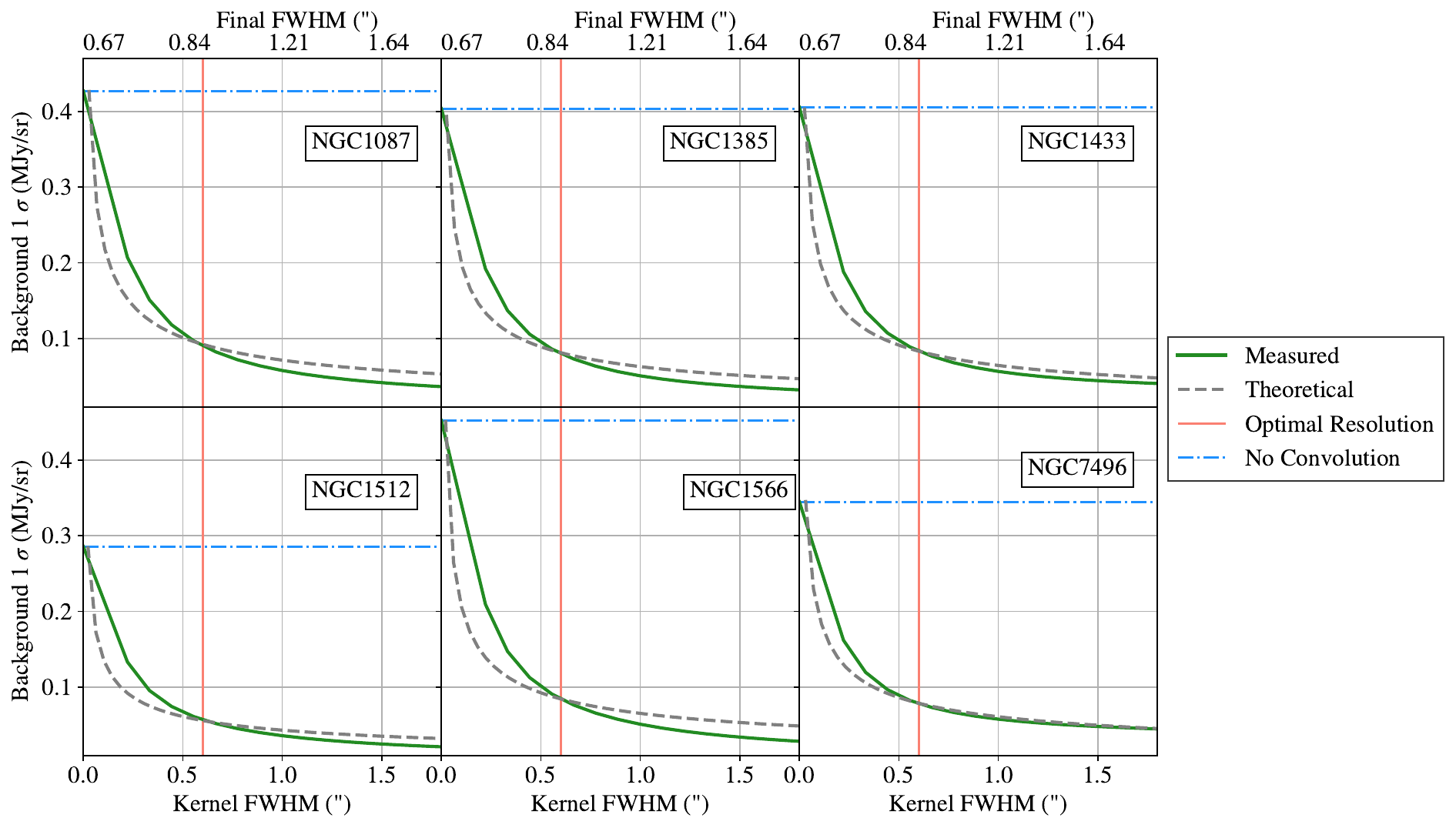}
 \caption{Variations of the standard deviation of an empty sky region after convolving the F2100W maps with a 2D-Gaussian with the FWHM value displayed on the lower $x$--axis.  The approximate final resolution is listed on the upper $x$--axis. Green lines show the measured standard deviations. Vertical red lines represent the Gaussian kernel that will approximately produce our chosen 0\farcs90 final resolution, a kernel FWHM of 0\farcs60. Horizontal blue dash-dot lines represent the background 1$\sigma$ of the regions at the native resolution, before any convolution. The gray dashed line represents the theoretical prediction for uncorrelated noise, normalized to the measured curve at our chosen final resolution. Each panel represents a different map that included an empty sky region in all of the MIRI bands.} 
    \label{fig:smoothing}
\end{figure*}

Figure \ref{fig:smoothing} demonstrates this effect, showing the empirically measured noise in several regions with no source emission in the MIRI bands. The Figure shows the empirically estimated noise in each map after convolution of the image with a series of 2D Gaussian kernels of increasing FWHM. Because the FWHM of the original F2100W PSF is $\approx 0.65\arcsec$, convolutions with kernels at or below this size will only modestly change the resolution of the image. Though the optimal resolution-noise tradeoff depends on the specific science application, these curves suggest a target resolution of $\approx 0\farcs90$\footnote{Assuming the F2100W beam is Gaussian, this is smoothing with a Gaussian of FWHM 0.62\arcsec.} offers a good tradeoff, degrading the FWHM of the PSF by only $\approx 30$~percent but lowering the noise by a factor of $\approx 3$. This is similar to the closest safe Gaussian discussed in Sect. \ref{sec:psf_matching}, so the Gaussian common resolution images will already have realized many of the key gains here.

\subsection{Noise Estimates Compared to Error Maps}

\begin{figure}[ht]
    \includegraphics[width=\columnwidth]{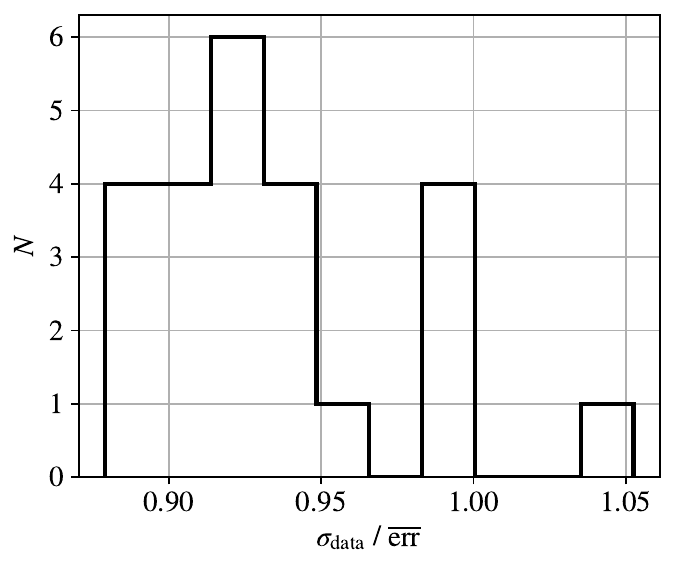}
 \caption{Comparison of the scatter measured within background regions of MIRI data to the mean of the pipeline-generated error map value within the same region (one region per galaxy).} 
    \label{fig:data_err_ratios}
\end{figure}

Using the background regions from the previous Subsection, we compare the error measured at the native resolution to the pipeline-calculated error map. We take the standard deviation of the pixels within the background region in the science image for the four MIRI bands (we do not use NIRCam as due to the source density, we are unable to find any background regions within our images, although given the good agreement between our calculated sensitivities and the exposure time calculator estimates, these error maps are likely also reasonable), and compare these to the mean of the pixels within the same region of the error map image. We note that this error map does not account for pixel-correlated noise. We have not included this effect at this point, but will revisit this at some point in the future. For an example of how this can be accounted for, we refer readers to \cite{2023Yang}. This comparison is shown in Figure \ref{fig:data_err_ratios} -- there is a good agreement here, with the error map value typically being higher by around 5 to 10~percent. We therefore deem these error maps suitable for providing estimates of the noise for observations where no background regions are present.

\subsection{Quality Assurance and Outstanding Issues}\label{sec:outstanding_issues}

Having created our final mosaics, we perform a round of internal quality assurance (QA). We have examined each individual image for issues such as poor astrometric alignment (which manifests as ``blurring'' between tiles), stepping between tiles which indicates poor level matching, and other defects in the data quality (e.g., remaining \onef\ noise). We find that each mosaic is free of these issues.

We then performed a round of inter-band quality checks, producing three-color images (of F200W, F300M, and F1000W) using SAOImageDS9 \citep{2003JoyeMandel} to assess the NIRCam short, NIRCam long, and MIRI overall alignment. Here, we are essentially checking that the astrometry between these mosaics is good, as large misalignments will lead to clear blurring between the color channels. We find that the NIRCam-to-NIRCam agreement is very good (with any misalignment significantly less than a NIRCam pixel), as is the NIRCam-to-MIRI agreement in most cases (with any misalignment less than a MIRI pixel), although there are some exceptions. The mosaics themselves generally look excellent and are ready for a wide range of science cases, but there are some outstanding issues that remain unresolved in this data release. We expect that these may improve as characterization of the instrumental features of {\it JWST} improve, and substantial improvements will be followed by new version releases. In particular, our QA has noted the following issues:

\subsubsection{PSF spikes and saturation in bright galaxy centers}

A few bright sources lead to large PSF spikes that can cover a significant portion of the observations. These spikes primarily affect MIRI images and they mostly originate from bright point sources, both star clusters and AGN, in galaxy centers. We have experimented with using {\tt WebbPSF} to subtract these and so recover the affected extended emission. However, we find that {\tt WebbPSF} models matched to the observed sources tend to oversubtract the spikes, particularly near the bright sources themselves. This could be due to the ``brighter-fatter'' effect \citep{2023Argyriou} whereby charge can leak into neighbouring pixels, leading to an increase in the PSF size of 10--25~percent in bright sources. Additionally, limitations in the current theoretical model for the PSF are likely prevalent in the outskirts of the PSF wings. Whatever the root cause, we currently do not have a robust method for removing these PSF spikes. Lacking the ability to correct for these, many of our analyses identify and mask regions affected by PSF spikes before analyzing the data.

Additionally, in the brightest galaxy centers, a small number of pixels are saturated by the first readout and so are unrecoverable. It may be possible to estimate the flux here via a curve-of-growth analysis \citep[e.g.][]{2023Liu} or PSF fitting. However, we have not yet developed a robust, automated solution for this and so in some cases, a number of central pixels do not have measured flux values. This tends to be more prevalent in the MIRI bands, although some NIRCam wavelengths are also affected. These bright centers also occasionally cause negative horizontal banding. As of now, we have not corrected for this and note that this may affect the level matching between tiles, as the centers typically fall in the overlap regions of our mosaics.

\subsubsection{Astrometry issues}

\begin{figure*}[ht]
    \includegraphics[width=\textwidth]{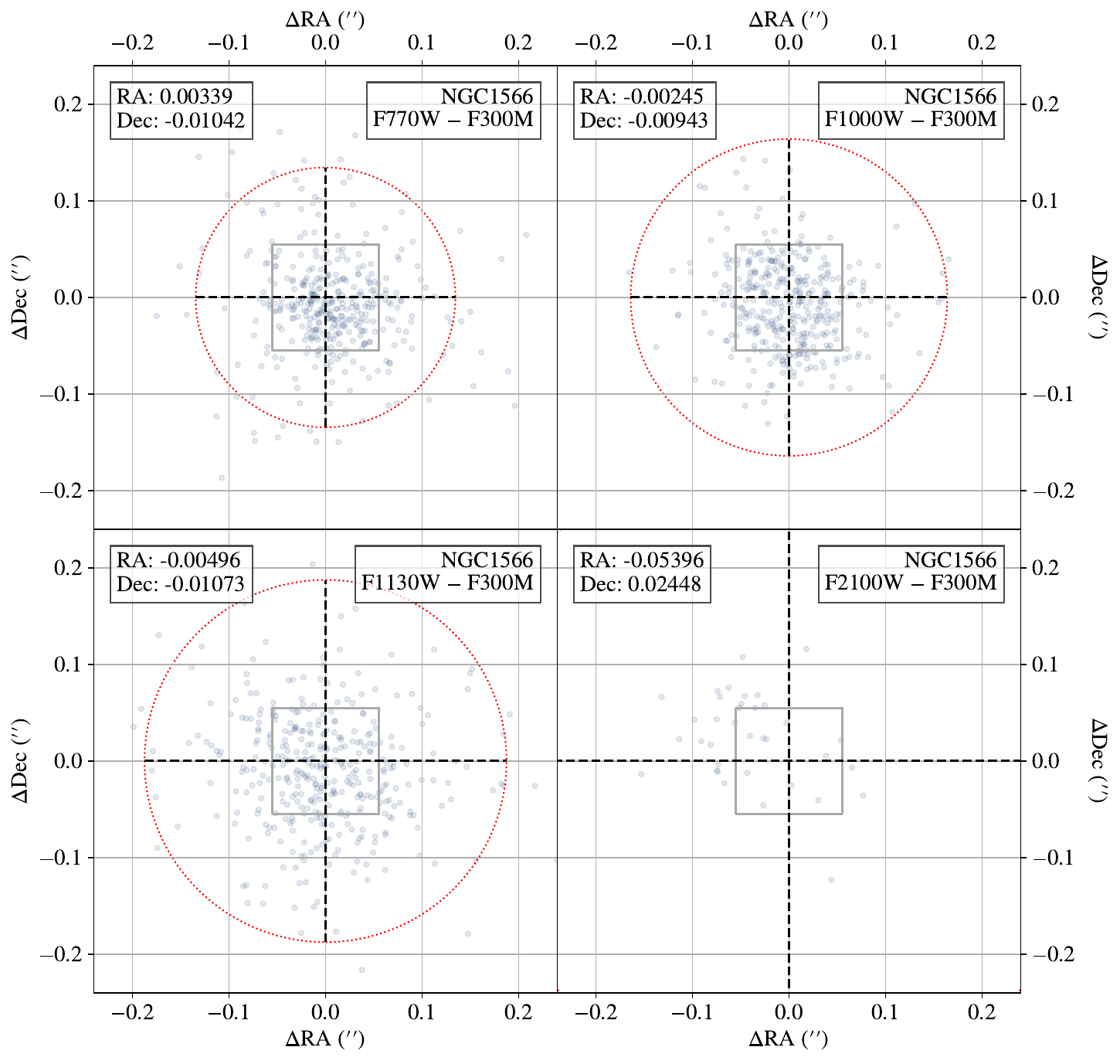}
 \caption{Diagnostic diagrams showing the MIRI-to-NIRCam relative astrometric alignment quality for example galaxy NGC~1566. Each of the panels displays the resulting RA/Dec `scatterball' for a different MIRI band using our final MIRI-measured sky coordinate minus NIRCam F300M RA/Dec.  The plotted points are plotted with transparency in order to allow assessment of peak location and distribution shape for each scatterball. The FWHM extent of the MIRI PSF in each band is drawn with a red dotted circle with a black dashed cross, and the pixel size on the MIRI detector is marked as a thick grey square.  With text in each panel, we provide the sigma-clipped mean residual RA/Dec offset from NIRCam F300M in arcseconds. This galaxy is in the top third of galaxies in ranked quality of MIRI-to-NIRCam alignment.}
    \label{fig:astrometry_diag_nircam_to_hst}
\end{figure*}

\begin{figure*}[ht]
    \includegraphics[width=\textwidth]{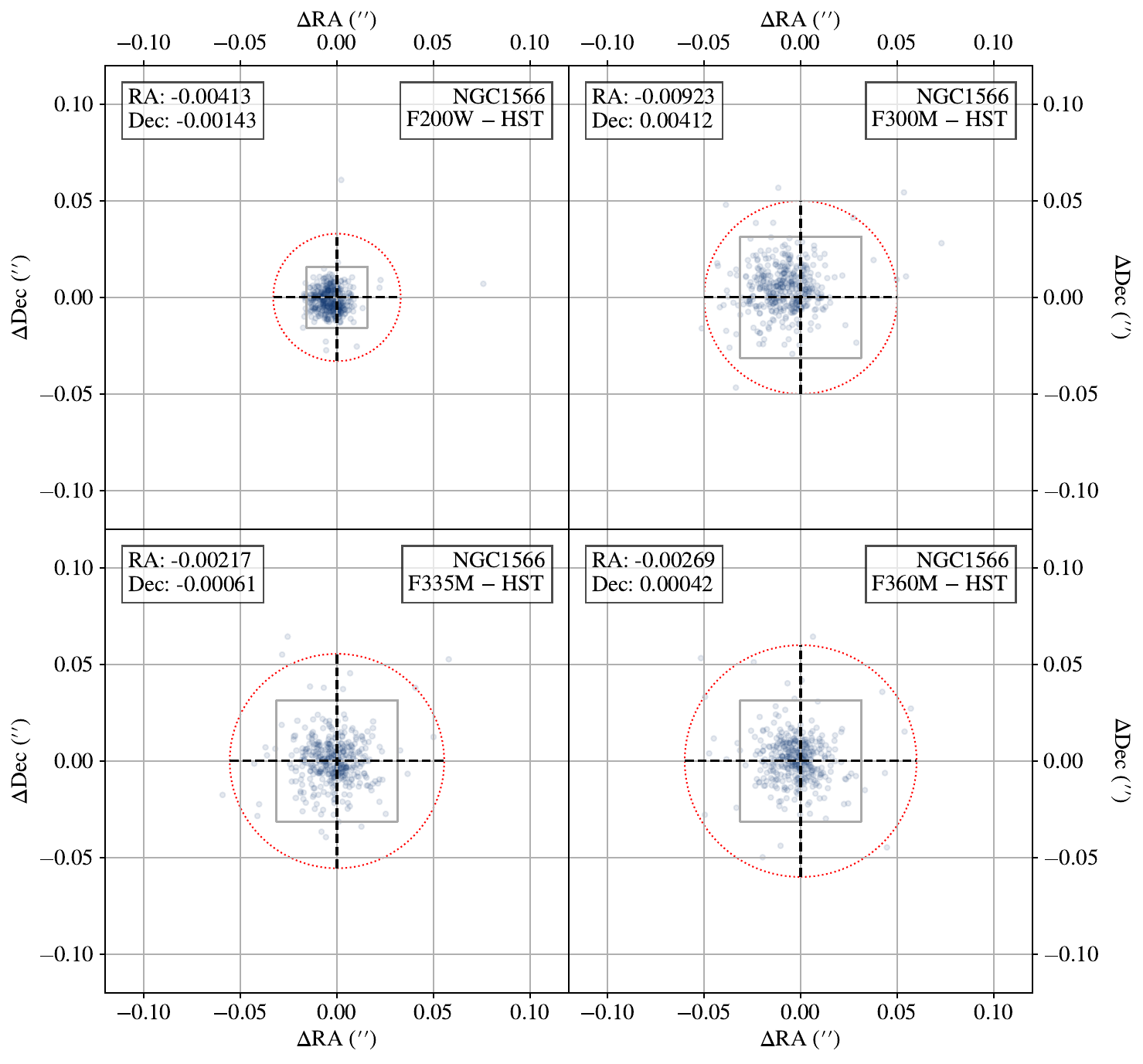}
 \caption{Diagnostic diagrams showing the NIRCam-to-HST (ICRS) astrometric alignment quality for example galaxy NGC~1566. Each of the panels displays the resulting RA/Dec `scatterball' for a different NIRCam band using our final NIRCam-measured sky coordinate minus HST RA/Dec.  The points are plotted with transparency in order to allow assessment of peak location and distribution shape for each scatterball. The FWHM extent of the NIRCam PSF in each band is drawn as a red dotted circle with a black dashed cross, and the pixel size on either the short (F200W) or long (F300M, F335M, F360M) NIRCam detectors is marked as a thick grey square. With text in each panel, we provide the sigma-clipped mean residual RA/Dec offset from HST in arcseconds. This galaxy is typical in quality of NIRCam alignment, though does not show evidence for tile-to-tile intramosaic alignment residuals as some galaxies do.}
    \label{fig:astrometry_diag_miri_to_nircam}
\end{figure*}

In Figures~\ref{fig:astrometry_diag_nircam_to_hst} and \ref{fig:astrometry_diag_miri_to_nircam}, we show the quality of the alignment using sources confidently matched by eye between {\it HST}, NIRCam, and MIRI observations. Whilst we can see that the NIRCam-to-{\it HST} alignment is generally very good -- the median offset across all NIRCam bands with respect to {\it HST} is 0\farcs012 (around 0.5 NIRCam short or 0.2 NIRCam long pixels) with a median RMS scatter of 0\farcs006. However, the agreement is less good between MIRI and NIRCam -- the median offset across all MIRI bands with respect to F300M is 0\farcs05 (around 0.5 MIRI pixels) with a median RMS scatter of 0\farcs03. This is mainly driven by the much lower density of point sources in the MIRI data, and the source extraction not being optimized for the {\it JWST} PSF shape. As well as this, in some cases, a point source at the MIRI resolution (particularly F2100W) can be an extended source at the higher NIRCam resolution (often a background galaxy), making this comparison more complicated. The shifts are typically less than a pixel but can be more severe in some cases; our worst case is NGC3351, where the shift between NIRCam and MIRI is around one MIRI pixel. In a few cases there are multiple peaks in the distribution of residual astrometric offsets, indicating some persistent unresolved issues with the tile-to-tile matching. 

We expect the astrometry to improve in future releases as we improve our alignment algorithms---fitting bespoke PSF models to sources may yield better centroiding (this is currently in development by the {\it JWST}-FEAST team, with their algorithm at \url{https://github.com/Vb2341/One-Pass-Fitting})---and as the overall astrometric properties of the {\it JWST} instruments and guide star catalog improve. There is also ongoing work to optimize the matching algorithm, which may improve the astrometric solution \citep[e.g., {\tt JHAT;}][]{Rest2023_jhat}. Our tests here hint that a more hands-on approach may be required, however for this release we prefer a repeatable, automated approach. We intend to revisit this topic in the future. We deem our current absolute astrometry to be acceptable for most science cases in our dataset, although for high-precision, multi-band aperture photometry some manual correction may still be required.

\section{Existing mosaic comparison}\label{sec:mosaic_comparison}

In this Section, we show the level of changes in our data processing relative to both our early science data release and the mosaics obtainable through the MAST service. Overall, our latest version (DR1.1.0) is a significant improvement over both of these products, and so we suggest users to prefer our latest versions over these other available versions.

\subsection{Comparison to early science release}\label{sec:early_sci_comparison}

\begin{figure*}[ht]
    \includegraphics[width=\textwidth]{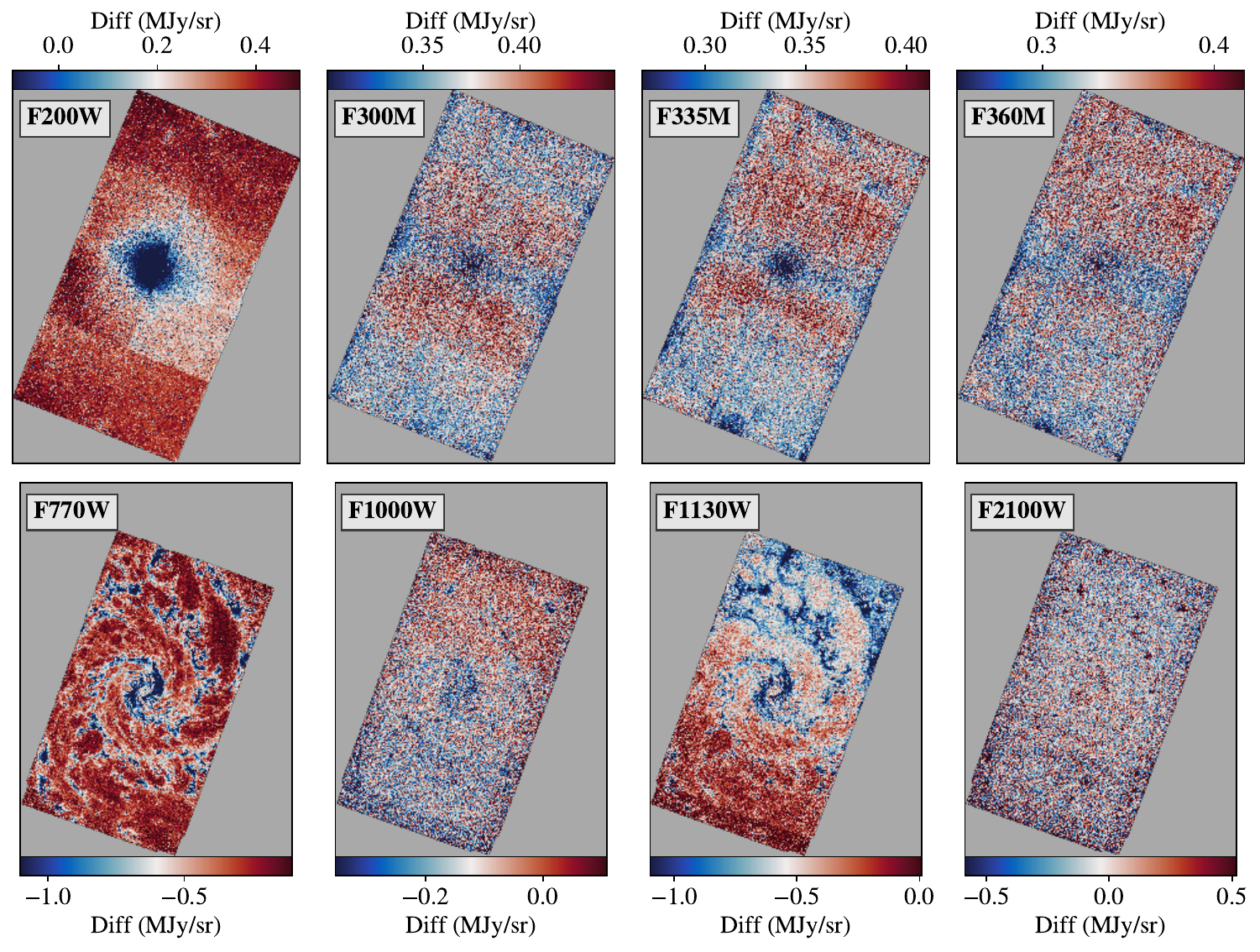}
    \caption{Comparison between our latest reprocessing and our early science mosaic for NGC~628. The difference here is the early science version subtracted from our latest reprocessed version, after reprojecting to the same WCS. There are significant differences in the short NIRCam band, caused by both background level matching and our new destriping method. At long NIRCam wavelengths, the differences are relatively minor. For MIRI, the improved calibration shows up as a galaxy imprint (most clear at F770W). Differences in astrometry show neighboring red/blue shifts and the background matching algorithm shows clear steps between mosaic tiles. The complete figure set (4 images) is available in the online journal.}
    \label{fig:early_sci_comparison}
\end{figure*}

Our first point of comparison is to the data we released as part of our early science release. This comprised four galaxies (IC~5332, NGC~628, NGC~1365, and NGC~7496) that were observed in the first few months of {\it JWST} science operations. A comparison between our latest version and the early science NGC~628 images is shown in Figure \ref{fig:early_sci_comparison}, and analogous figures for the other three galaxies are shown in Appendix \ref{app:early_sci}. The differences are stark in the short NIRCam band, which are caused both by our differences in destriping (PCA tended to subtract too much flux around bright sources that extended over much of the field of view) as well as the background matching (steps between tiles are clearly visible). These issues are significantly improved in our latest version. For the long-wavelength NIRCam bands, the differences are minor, a few tenths of a MJy/sr at most.

The MIRI images also see significant improvements, both in astrometry and background matching. The astrometry differences are most clear at F2100W, but indeed are present across the four MIRI filters. The background is also much more consistent between tiles in our latest version, highlighted as steps between mosaic tiles (most clearly seen in F1130W). Calibration updates have also substantially changed since our early science, by as much as 20~percent in the F770W, though less in other bands\footnote{\url{https://www.stsci.edu/contents/news/jwst/2023/updates-to-the-miri-imager-flux-calibration-reference-files}}. 

We will be revisiting many of our early science results with the full Cycle 1 sample in due course, as the increased number statistics and various changes to the data may affect results. For the most part, our First Results works took a conservative approach and many focused on source detection rather than detailed photometry. These results are unlikely to be affected by our improvements to the data processing. However, those that used flux measurements are likely to vary at the 10-20~percent level, especially those using long wavelength MIRI data, and any new publications will supersede the First Results \citep[e.g.][]{2024Sutter}. Initial testing has also shown that the improvements to astrometry allow for better measurements of the concentration index \citep[CI, see][]{2023Rodriguez}. This means that in the F200W band we can better discriminate star clusters from individual stars \citep{2024Rodriguez}.

\subsection{Comparison to MAST mosaics}\label{sec:mast_comparison}

\begin{figure*}[ht]
    \includegraphics[width=\textwidth]{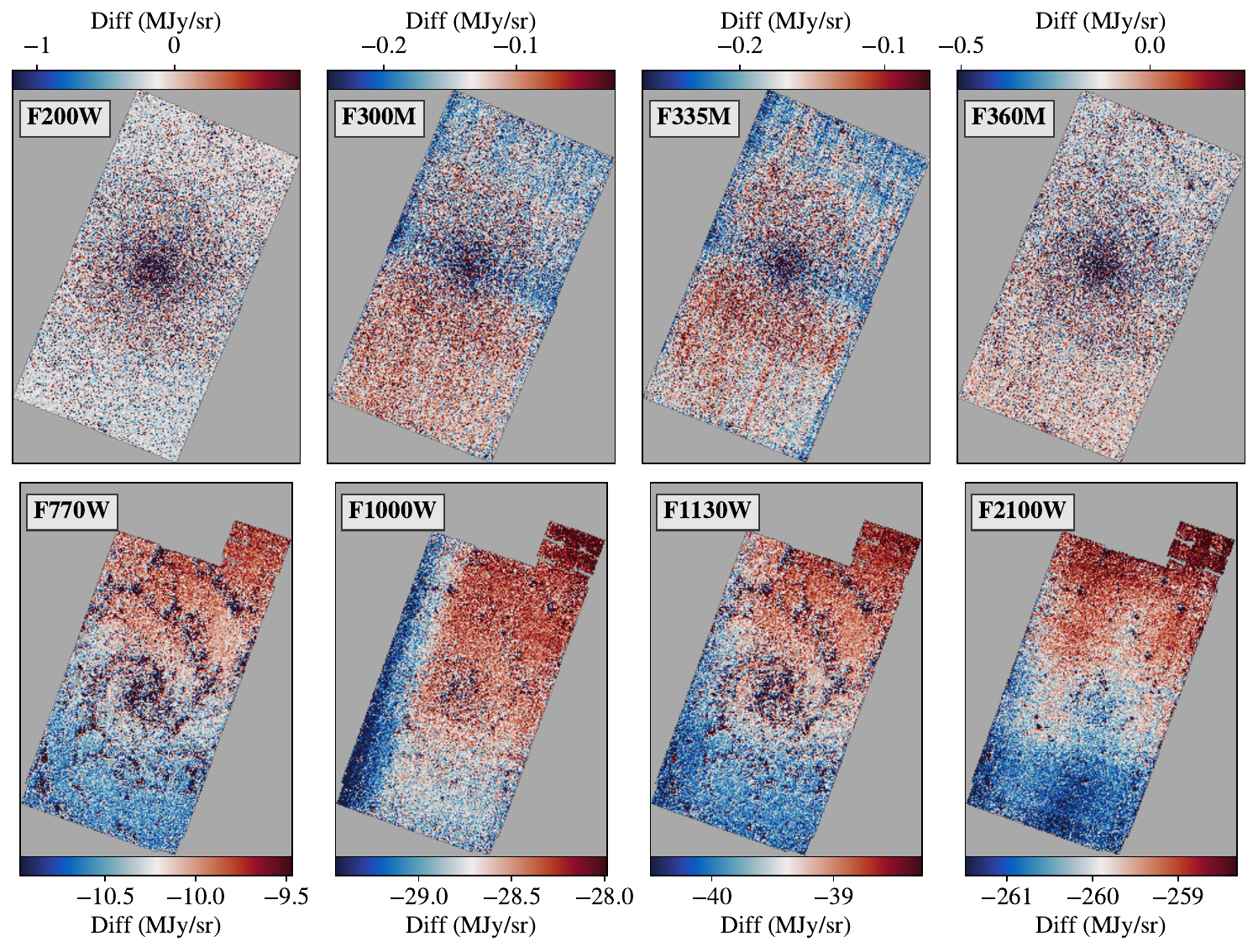}
    \caption{Comparison between our reprocessing and Level Three MAST-obtained mosaics for NGC~628. The difference here is the MAST version subtracted from our reprocessed version, after reprojecting to the same WCS. As can be seen, the astrometry is somewhat different in NIRCam, and the striping in the MAST mosaic is clear. The background levels are substantially different in the MIRI bands, due to a combination of our improved background matching algorithm and the fact that the MAST mosaics do not include our dedicated background observations. The complete figure set (19 images) is available in the online journal.}
    \label{fig:mast_comparison}
\end{figure*}

Here, we make a comparison to the Level Three mosaics that are available from MAST\footnote{As downloaded on 2024/04/02.} (\url{https://mast.stsci.edu/}). These images are automatically processed using observatory-recommended default settings per-band, and as such are designed to produce reasonable quality mosaics across the full breadth of science that {\it JWST} will cover. 

We show an example comparison in Figure~\ref{fig:mast_comparison} for NGC~628 for all eight bands, and for the other 18 galaxies in Appendix \ref{app:mast}. The differences are generally subtle in the NIRCam bands: we see the striping that is accounted for in our data processing showing up, as well as astrometric offsets between the two different versions. However, differences are small, generally on the order of a few tenths of MJy/sr. 

The difference is more substantial for the MIRI bands. We see the pronounced effects of our improved tile matching (Sect. \ref{sec:bg_match}) as clear steps between the tiles. Due to how we set up our observations, MAST does not associate the background observations with the science automatically. This leads to clear non-linear gradients across the tiles that have not been accounted for. For non-parallel observations where the backgrounds are directly linked to the science observations in the observing metadata, this will not be an issue. Astrometric offsets also appear in the difference images as galactic structures that have not been subtracted out. These trends are similar across all of our targets (see App.\ \ref{app:mast}). Particularly for the MIRI, we strongly recommend our processing over the MAST-obtainable mosaics. For NIRCam, the differences are more subtle but our reprocessing efforts still represent an improvement over the automated {\it JWST} pipeline.

\section{Released Data Products}\label{sec:data_products}

\begin{table*}[ht]
\begin{tabular}{p{0.15\textwidth} | p{0.75\textwidth}}

Extension name & Description\\
  \hline

SCI & 2-D data array containing the pixel values, in units of surface brightness (MJy/sr) \\
ERR & 2-D data array containing uncertainty estimates for each pixel. These values are based on the combined VAR\_POISSON and VAR\_RNOISE data, given as standard deviation \\
DQ & 2-D data array containing data quality flags for each pixel \\
VAR\_POISSON & 2-D data array containing the variance estimate for each pixel, based on Poisson noise only \\
VAR\_RNOISE & 2-D data array containing the variance estimate for each pixel, based on read noise only \\
VAR\_FLAT & 2-D data array containing the variance estimate for each pixel, based on uncertainty in the flat-field \\
CON$^1$ & 3-D context image, which encodes information about which input images contribute to a specific output pixel \\
WHT$^1$ & 2-D weight image giving the relative weight of the output pixels \\
\end{tabular}
\caption{Description of fits extensions for our released data products. $^1$Only present in {\tt img.fits} files.}\label{tab:data_products}
\end{table*}

Our data products are hosted at the MAST high-level science products (HLSP) webpages, at \url{https://archive.stsci.edu/hlsp/phangs/phangs-jwst}, as well as on CANFAR at \url{https://www.canfar.net/storage/vault/list/phangs/RELEASES/PHANGS-JWST}. Our primary data products are mosaics ({\tt img.fits}), although we also provide compressed directories containing individually corrected tiles ({\tt tweakback.tar.gz}), which are fully processed up to the final mosaic stage, and have absolute astrometric corrections back-propagated to them. The files are presented as multi-extension .fits files, where the extensions are standard outputs from the {\it JWST} pipeline. These are detailed in Table \ref{tab:data_products}. We note that \pipename\ does not currently propagate the effects of any convolution to any arrays except the science and error arrays, and so these other extensions will be invalid for PSF matched images.

\section{Summary}\label{sec:summary}

In this work, we have presented the overview of the steps we have taken in producing science-ready data for the PHANGS-{\it JWST} survey, implemented in our custom pipeline \pipename (our DR1.1.0 was produced using \pipename\ version 1.1.0). This pipeline provides:
\begin{enumerate}
    \item Wrappers around the official pipeline for Level One, Two, and Three processing (Sects. \ref{sec:lv1}, \ref{sec:lv2}, and \ref{sec:lv3});
    \item Optimized methods for both relative astrometry between mosaic tiles (Sect. \ref{sec:relative_astrometry}) and absolute astrometry to external references (Sect. \ref{sec:absolute_astrometry});
    \item A number of algorithms to remove \onef\ noise in NIRCam data (Sect. \ref{sec:destripe});
    \item Improved background matching between mosaic tiles (Sect. \ref{sec:bg_match}), and the ability to separate the MIRI coronagraph from the main science observations to improve this process (Sect. \ref{sec:corona_sep});
    \item Flux anchoring to external images, to provide absolute flux levels in images with no ``true'' background (Sect. \ref{sec:anchoring});
    \item Convolution algorithms to produce images at a matched angular resolution (Sect. \ref{sec:psf_matching}).
\end{enumerate}
Whilst these techniques have been tailored specifically for our survey, they have flexible implementations, and will be of general use in other observations of nearby galaxies.

We have extensively tested our processing methodology for 19 galaxies observed in four NIRCam and four MIRI bands, and our approach produces excellent mosaics across our sample. Further testing on the 55 galaxies that make up our Cycle 2 Treasury (program 3707, PI: Leroy) will help refine the data processing further. The automated pipeline enables rapid processing of these data so that we can make products available in a timely manner. We have developed a number of bespoke algorithms that will be of general use to the community, and hence our pipeline, \pipename, is made public at \url{https://github.com/phangsTeam/pjpipe}, and is pip-installable ({\tt pip install pjpipe}). We have designed \pipename\ to be flexible and easy to interface with, to maximize its usefulness. We make our first full public data release (DR1.1.0) mosaics available at \url{https://archive.stsci.edu/hlsp/phangs/phangs-jwst}, and stress that these represent a significant improvement on those we used in our early science papers, as well as those obtainable from the MAST database.

\clearpage

\begin{acknowledgments}
This work has been carried out as part of the PHANGS collaboration. This work is based on observations made with the NASA/ESA/CSA {\it JWST}. The data were obtained from the Mikulski Archive for Space Telescopes at the Space Telescope Science Institute, which is operated by the Association of Universities for Research in Astronomy, Inc., under NASA contract NAS 5-03127 for {\it JWST}. These observations are associated with program 2107. The specific observations analyzed can be accessed via doi:10.17909/ew88-jt15. The authors would like to thank the anonymous referee for their constructive comments, which have improved this work. We also thank M. Garc\'{i}a Mar\'{i}n for comments on the manuscript. TGW would like to personally thank P. Woolford for technical assistance throughout the preparation of this work, as well as Beryl Williams for everything throughout the years. KS, JS acknowledge funding from JWST-GO-2107.006-A. R.S.K.\ and S.C.O.G.\ acknowledge funding from the European Research Council via the Synergy Grant ``ECOGAL'' (project ID 855130), from the German Excellence Strategy via the Heidelberg Cluster of Excellence (EXC 2181 - 390900948) ``STRUCTURES'', and from the German Ministry for Economic Affairs and Climate Action in project ``MAINN'' (funding ID 50OO2206). AKL, DP, and RC gratefully acknowledges support by grants 1653300 and 2205628 from the National Science Foundation, by award JWST-GO-02107.009-A, JWST-GO-03707.001-A, JWST-GO-04256.001-A,  and by a Humboldt Research Award from the Alexander von Humboldt Foundation. MC gratefully acknowledges funding from the DFG through an Emmy Noether Research Group (grant number CH2137/1-1). COOL Research DAO is a Decentralized Autonomous Organization supporting research in astrophysics aimed at uncovering our cosmic origins. J.K. is supported by a Kavli Fellowship at the Kavli Institute for Particle Astrophysics and Cosmology (KIPAC).
JPe acknowledges support by the French Agence Nationale de la Recherche through the DAOISM grant ANR-21-CE31-0010 and by the Programme National ``Physique et Chimie du Milieu Interstellaire'' (PCMI) of CNRS/INSU with INC/INP, co-funded by CEA and CNES. OE acknowledges funding from the Deutsche Forschungsgemeinschaft (DFG, German Research Foundation) in the form of an Emmy Noether Research Group (grant number KR4598/2-1, PI Kreckel). JDH gratefully acknowledges financial support from the Royal Society (University Research Fellowship; URF/R1/221620).
ES acknowledges funding from the European Research Council (ERC) under the European Union's Horizon 2020 research and innovation programme (grant agreement No. 694343). 
SD is supported by funding from the European Research Council (ERC) under the European Union’s Horizon 2020 research and innovation programme (grant agreement no. 101018897 CosmicExplorer).
\end{acknowledgments}

%

\vspace{5mm}
\facilities{{\it JWST}.}


\software{{\tt astropy} \citep{astropy:2013, astropy:2018, astropy:2022}, 
{\tt jwst} \citep{bushouse_howard_2023_8157276},
{\tt matplotlib} \citep{Hunter:2007},
{\tt numpy} \citep{harris2020array},
{\tt photutils} \citep{larry_bradley_2022_6825092},
SAOImageDS9 \citep{2003JoyeMandel},
{\tt scipy} \citep{2020SciPy-NMeth}.
          }



\newpage

\appendix

\section{Image Atlas}\label{app:atlas}

Here, we show per-filter mosaics for our full galaxy sample.

\begin{figure*}
    \includegraphics[width=\textwidth]{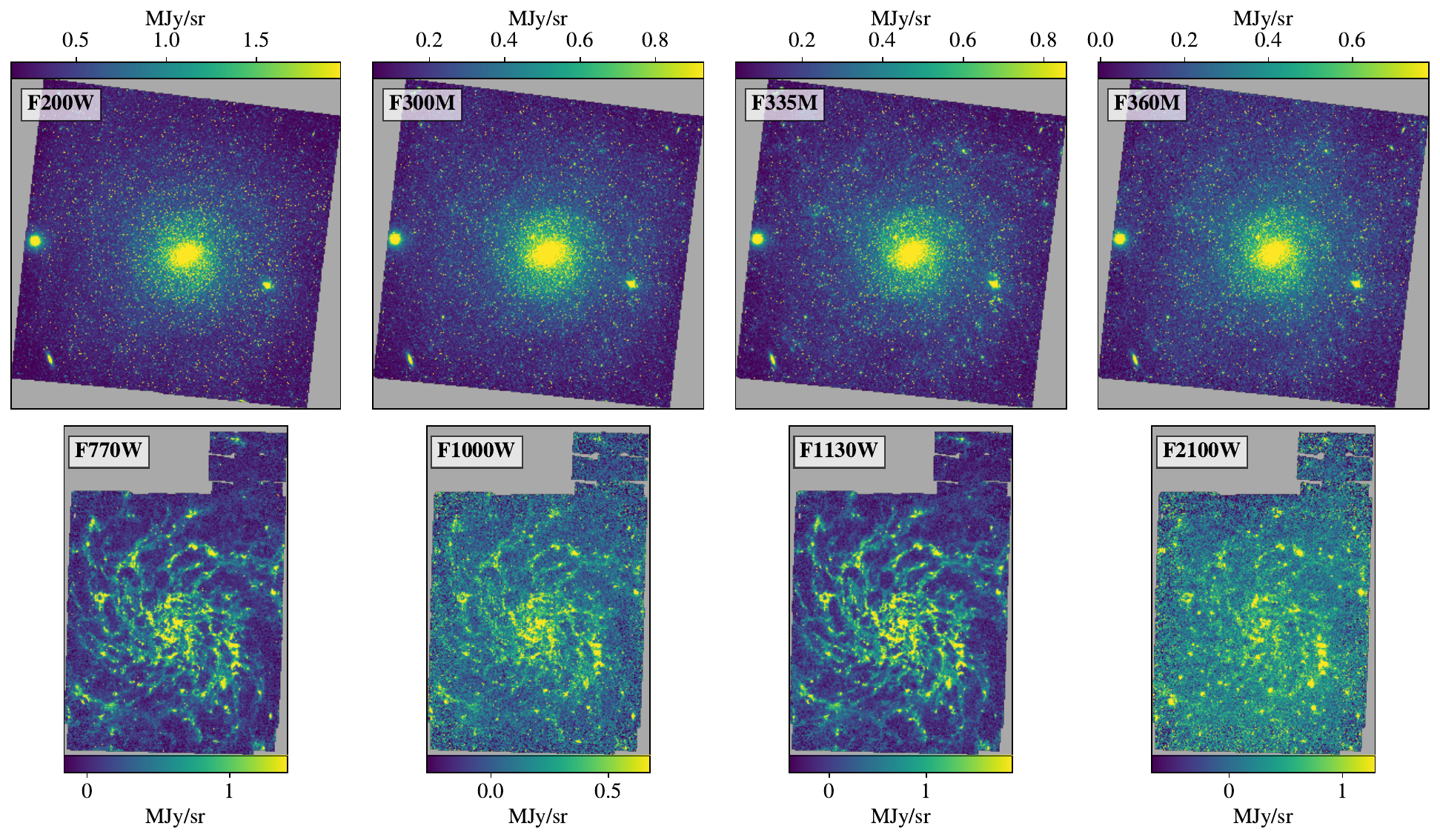}
    \caption{As Figure \ref{fig:final_mosaics}, but for IC~5332.}
    \label{fig:ic5332_mosaics}
\end{figure*}

\begin{figure*}
    \includegraphics[width=\textwidth]{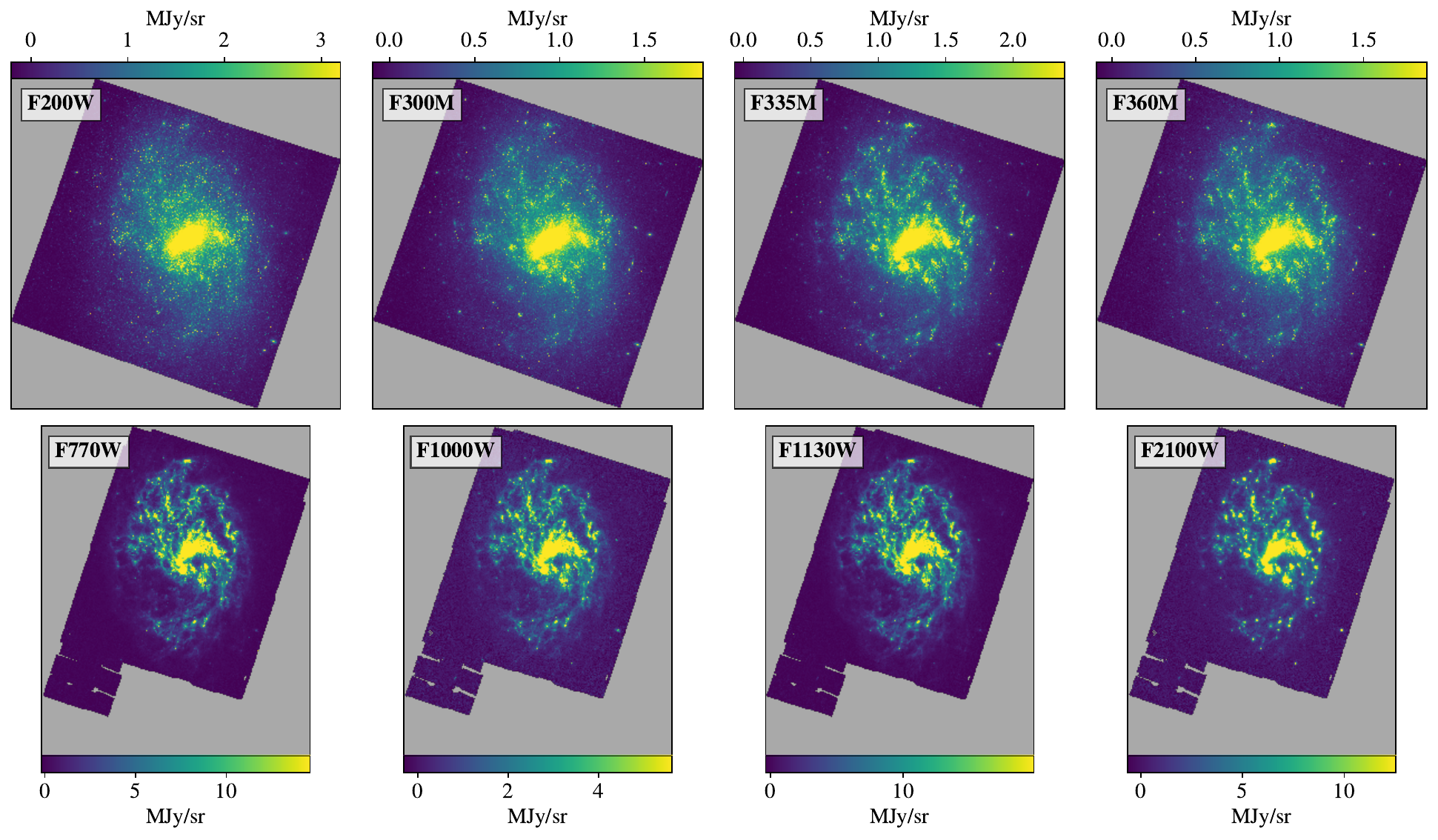}
    \caption{As Figure \ref{fig:final_mosaics}, but for NGC~1087.}
    \label{fig:ngc1087_mosaics}
\end{figure*}

\begin{figure*}
    \includegraphics[width=\textwidth]{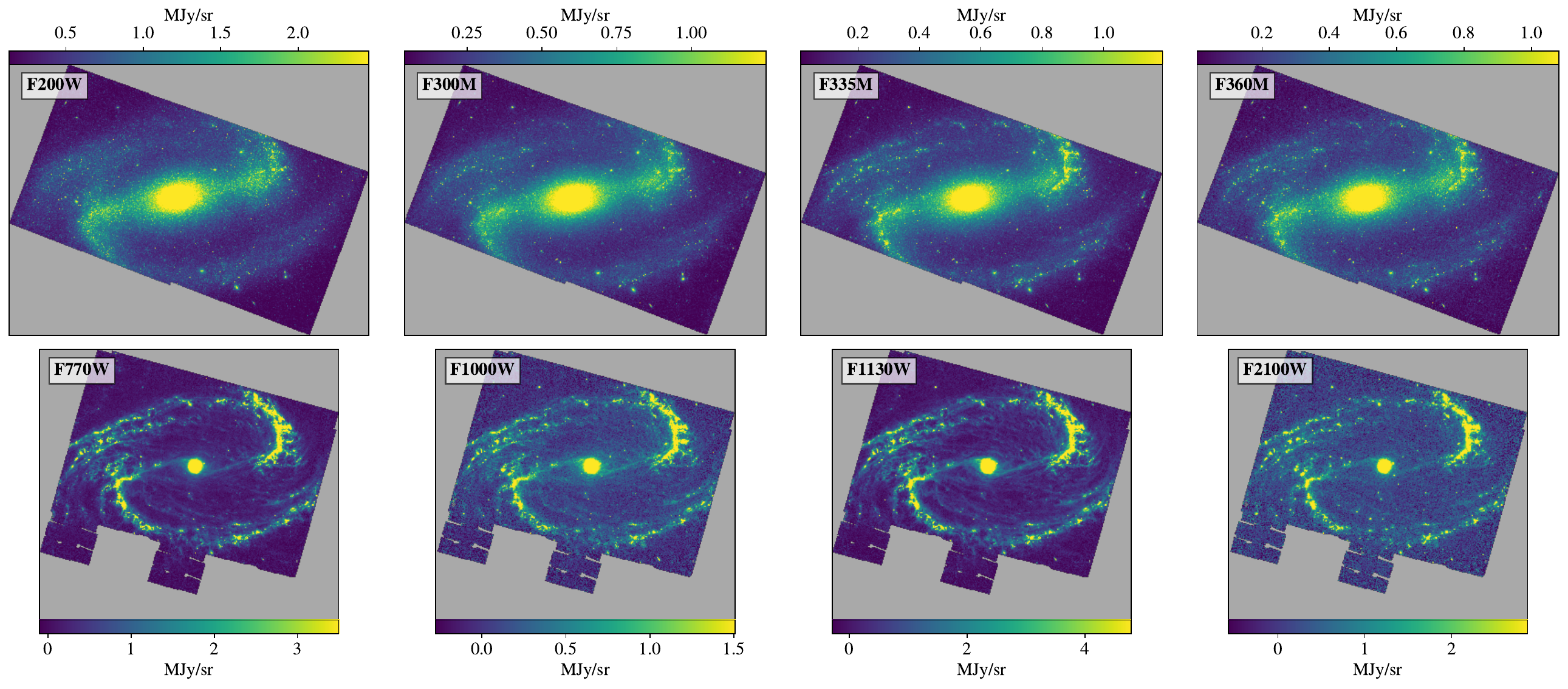}
    \caption{As Figure \ref{fig:final_mosaics}, but for NGC~1300.}
    \label{fig:ngc1300_mosaics}
\end{figure*}

\begin{figure*}
    \includegraphics[width=\textwidth]{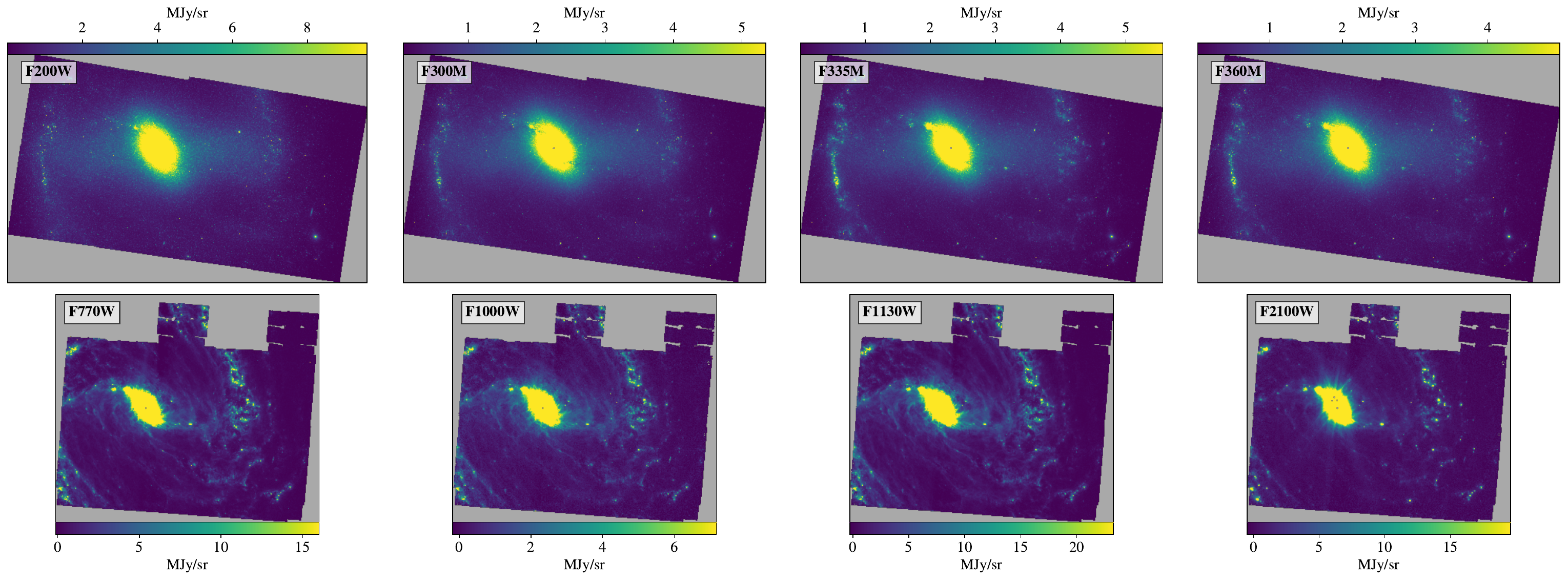}
    \caption{As Figure \ref{fig:final_mosaics}, but for NGC~1365.}
    \label{fig:ngc1365_mosaics}
\end{figure*}

\begin{figure*}
    \includegraphics[width=\textwidth]{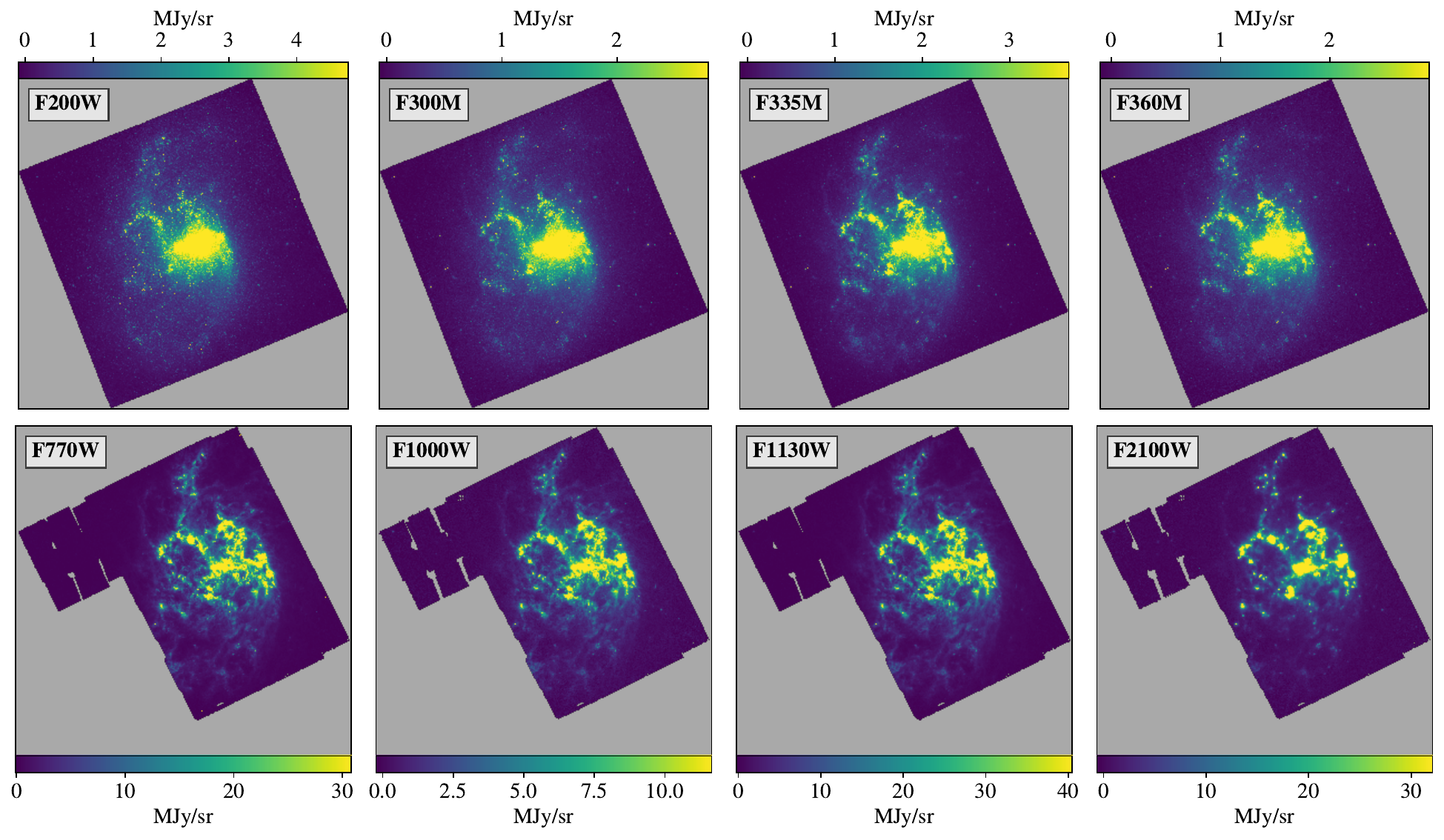}
    \caption{As Figure \ref{fig:final_mosaics}, but for NGC~1385.}
    \label{fig:ngc1385_mosaics}
\end{figure*}

\begin{figure*}
    \includegraphics[width=\textwidth]{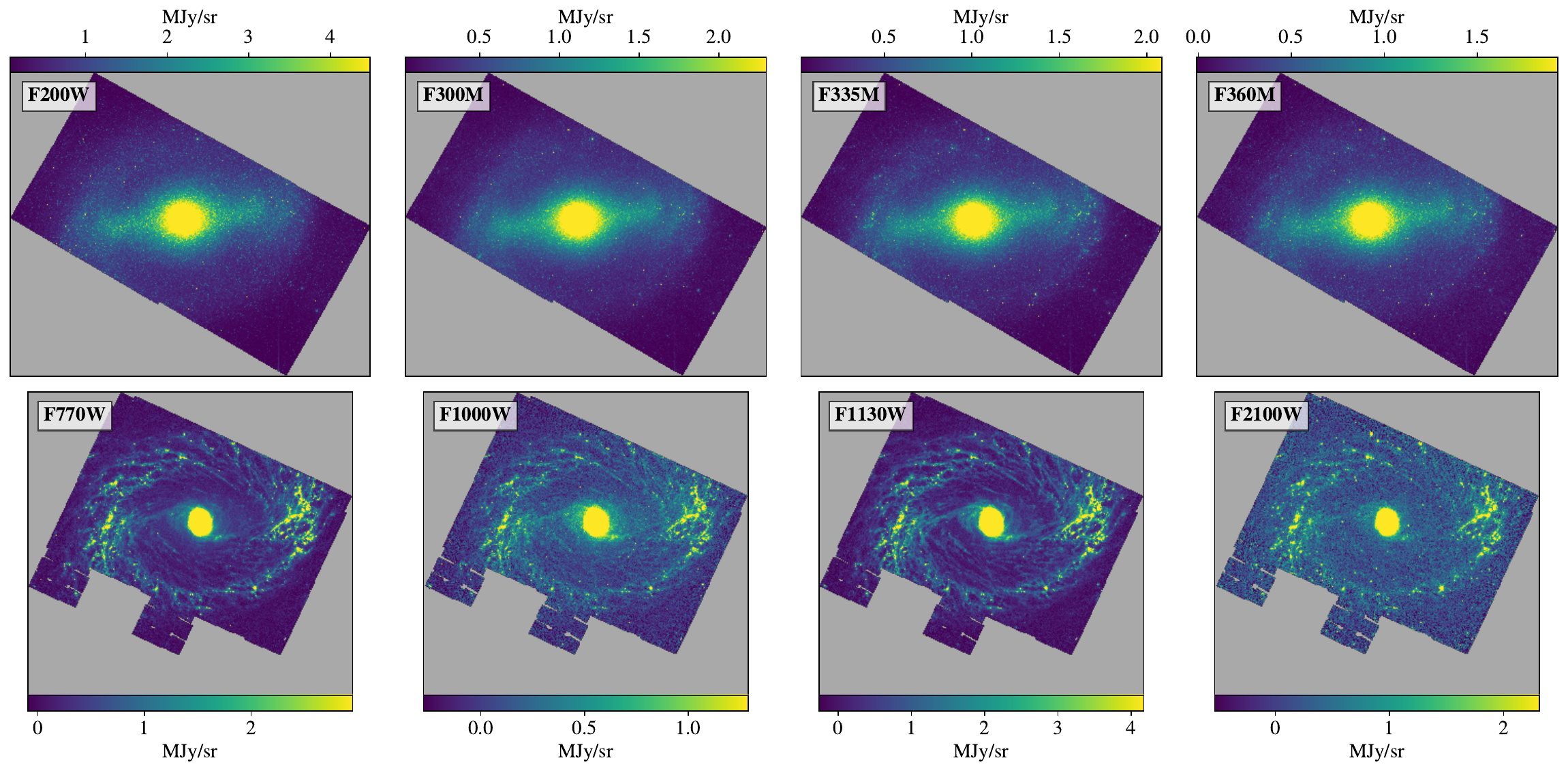}
    \caption{As Figure \ref{fig:final_mosaics}, but for NGC~1433.}
    \label{fig:ngc1433_mosaics}
\end{figure*}

\begin{figure*}
    \includegraphics[width=\textwidth]{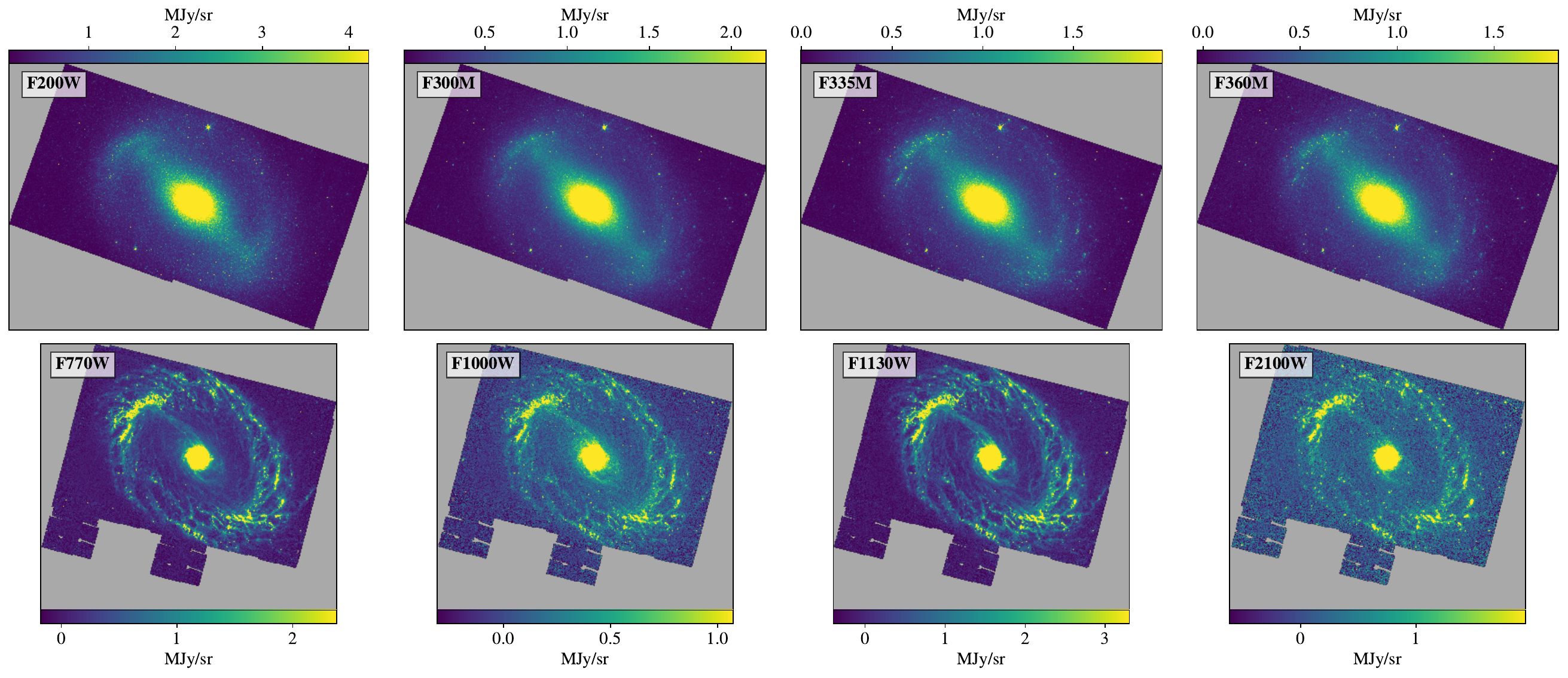}
    \caption{As Figure \ref{fig:final_mosaics}, but for NGC~1512.}
    \label{fig:ngc1512_mosaics}
\end{figure*}

\begin{figure*}
    \includegraphics[width=\textwidth]{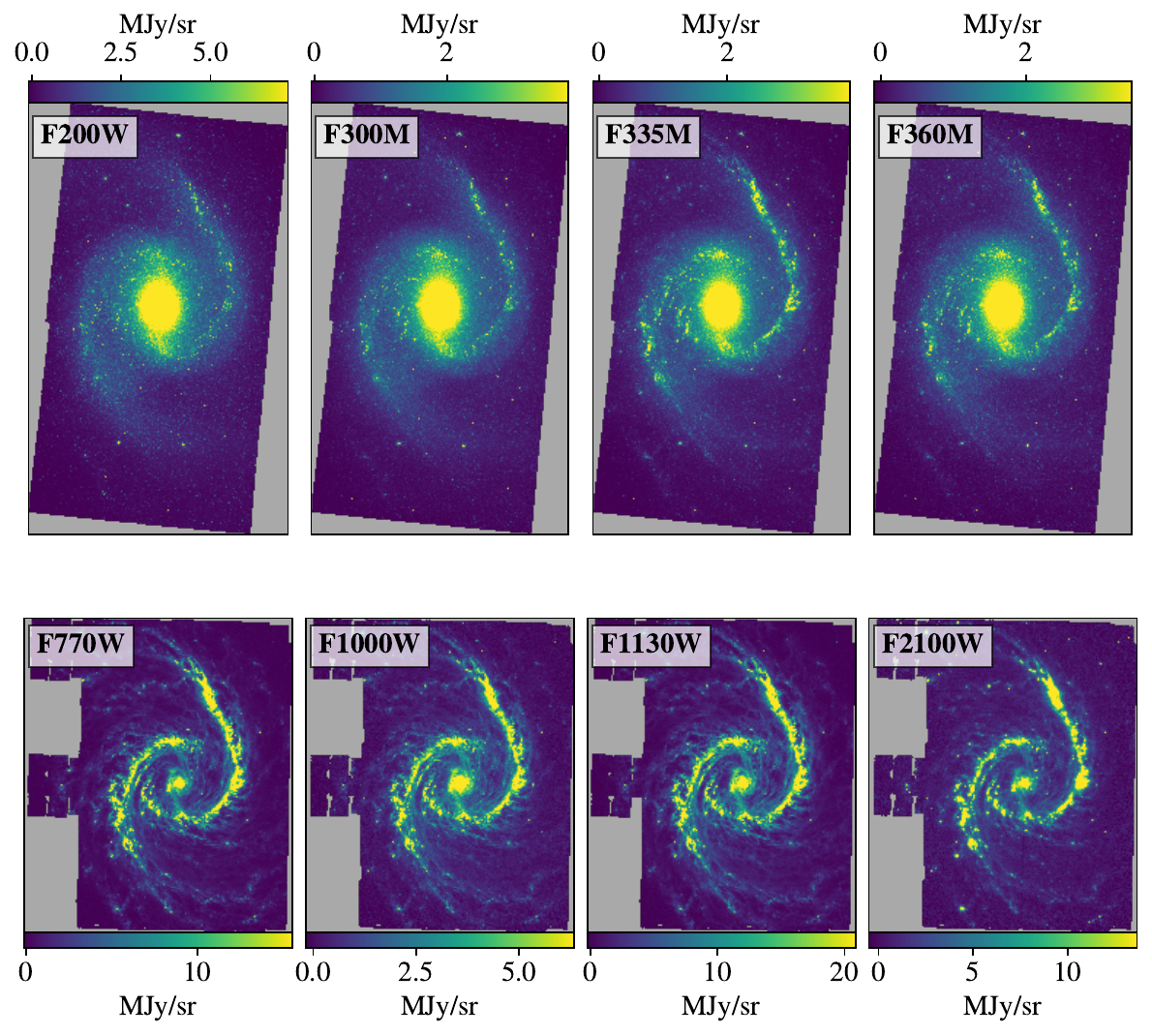}
    \caption{As Figure \ref{fig:final_mosaics}, but for NGC~1566.}
    \label{fig:ngc1566_mosaics}
\end{figure*}

\begin{figure*}
    \includegraphics[width=\textwidth]{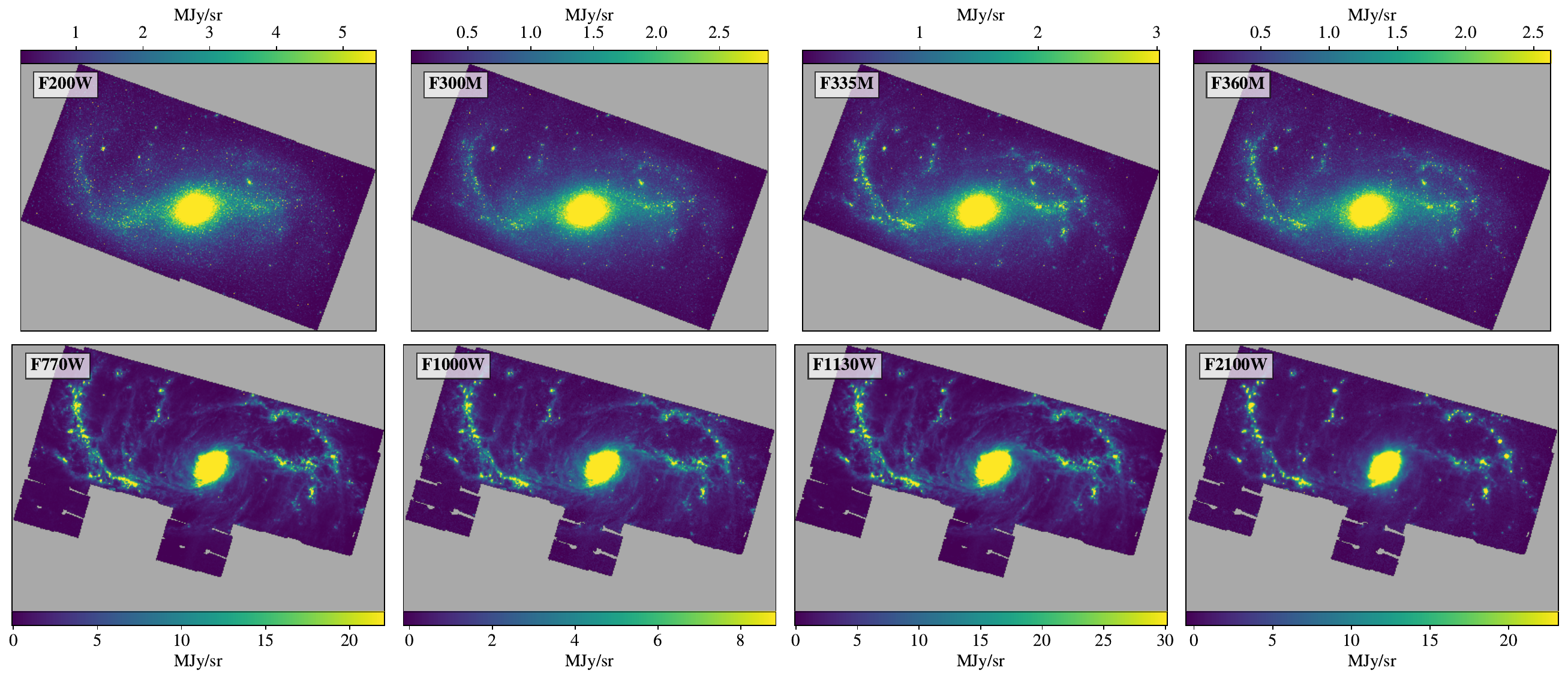}
    \caption{As Figure \ref{fig:final_mosaics}, but for NGC~1672.}
    \label{fig:ngc1672_mosaics}
\end{figure*}

\begin{figure*}
    \includegraphics[width=\textwidth]{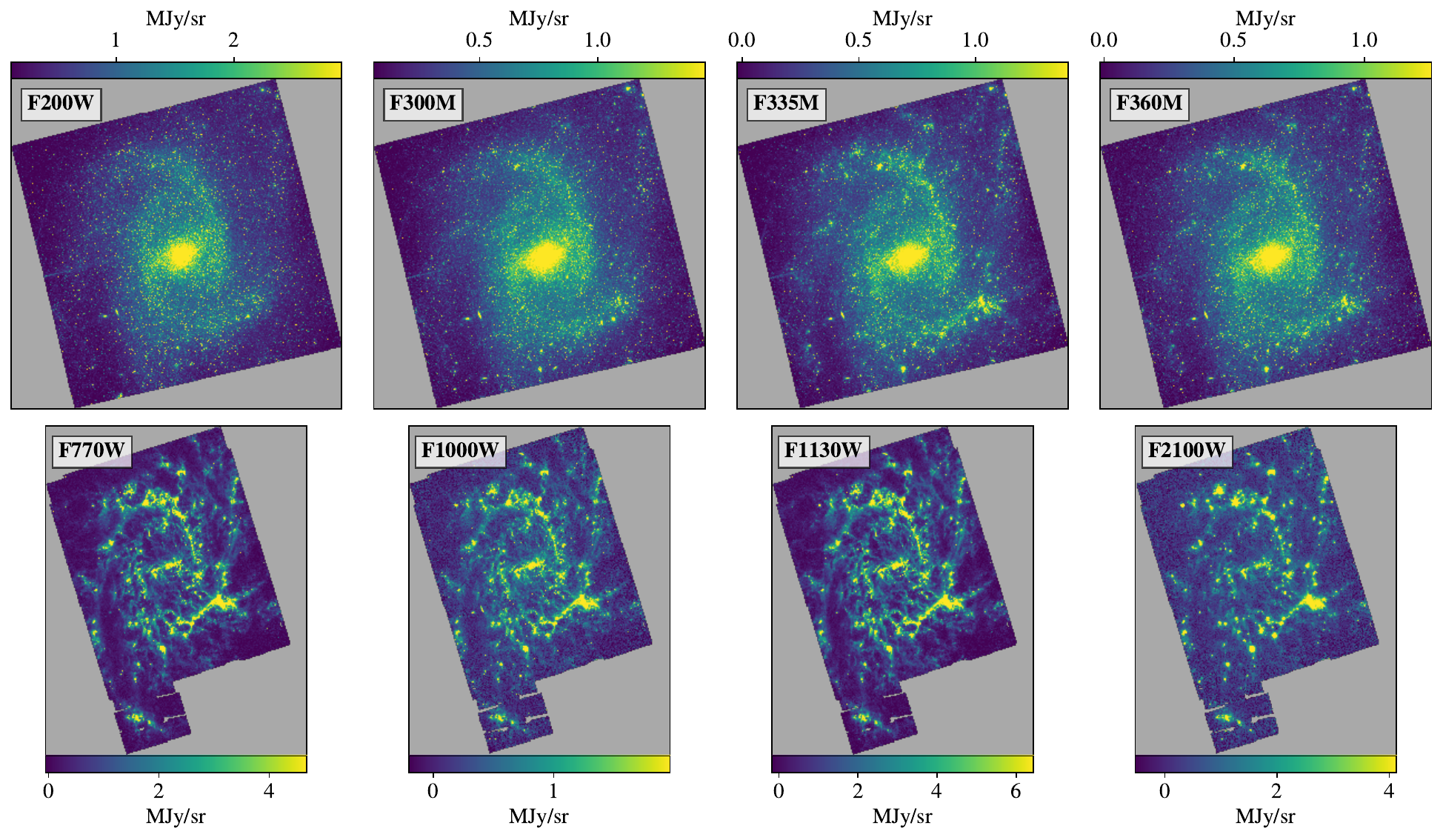}
    \caption{As Figure \ref{fig:final_mosaics}, but for NGC~2835.}
    \label{fig:ngc2835_mosaics}
\end{figure*}

\begin{figure*}
    \includegraphics[width=\textwidth]{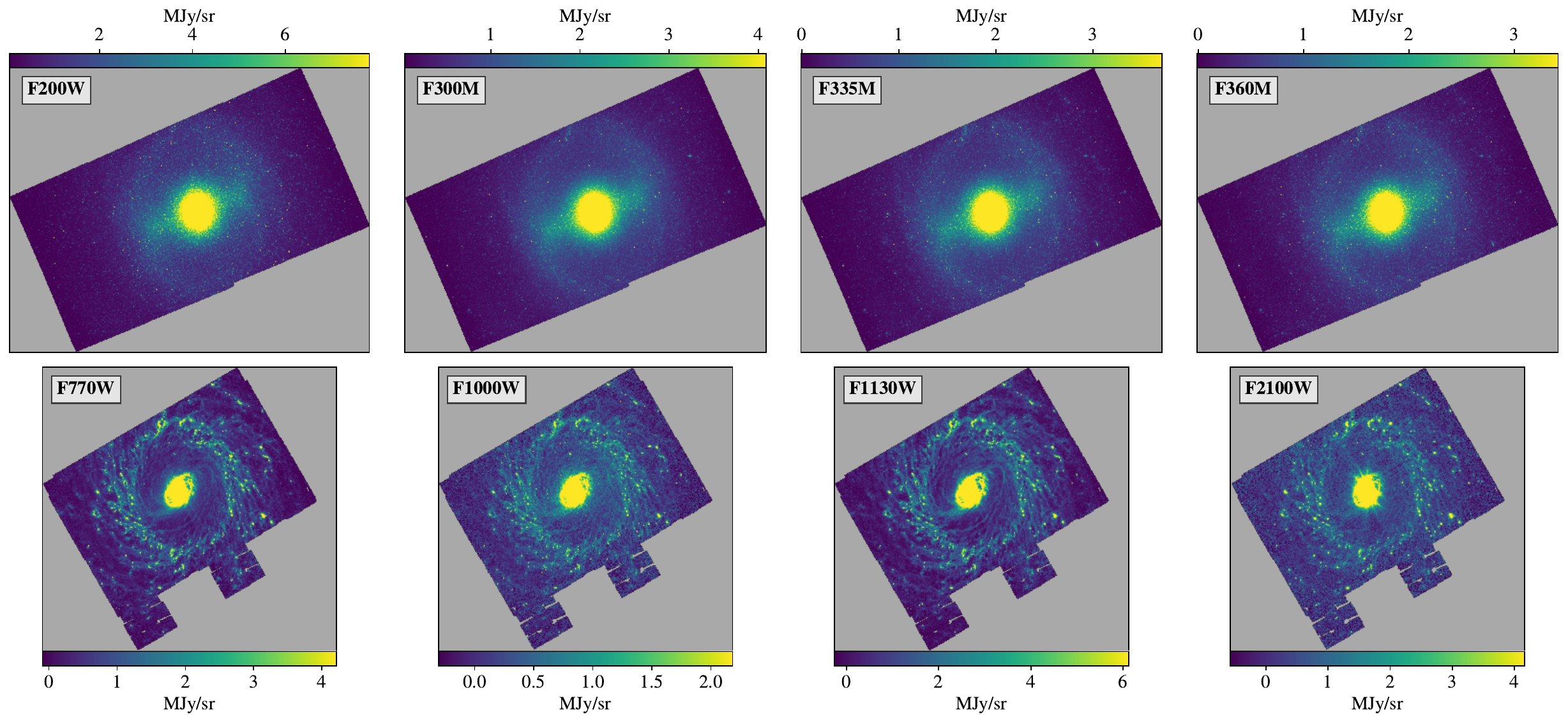}
    \caption{As Figure \ref{fig:final_mosaics}, but for NGC~3351.}
    \label{fig:ngc3351_mosaics}
\end{figure*}

\begin{figure*}
    \includegraphics[width=\textwidth]{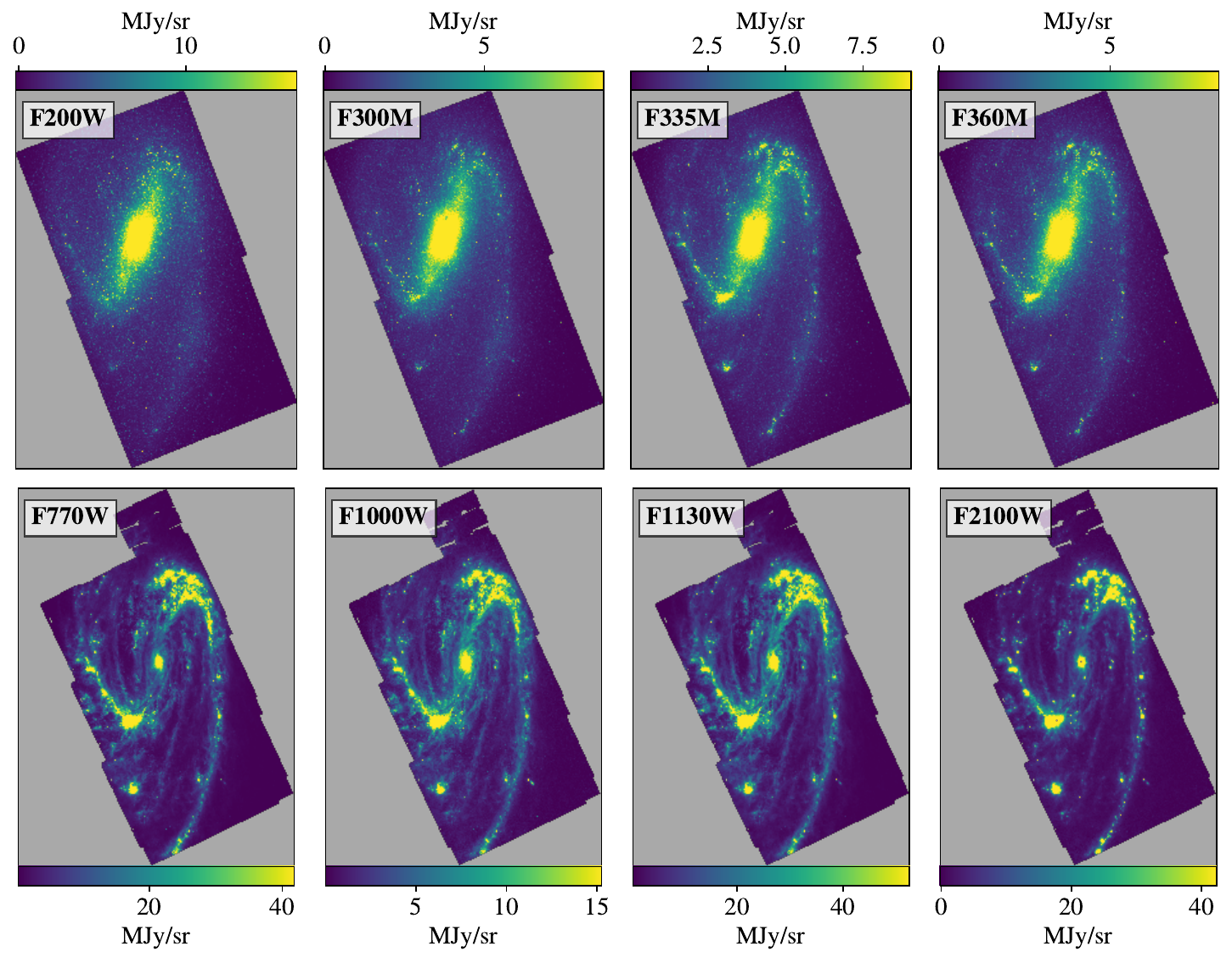}
    \caption{As Figure \ref{fig:final_mosaics}, but for NGC~3627.}
    \label{fig:ngc3627_mosaics}
\end{figure*}

\begin{figure*}
    \includegraphics[width=\textwidth]{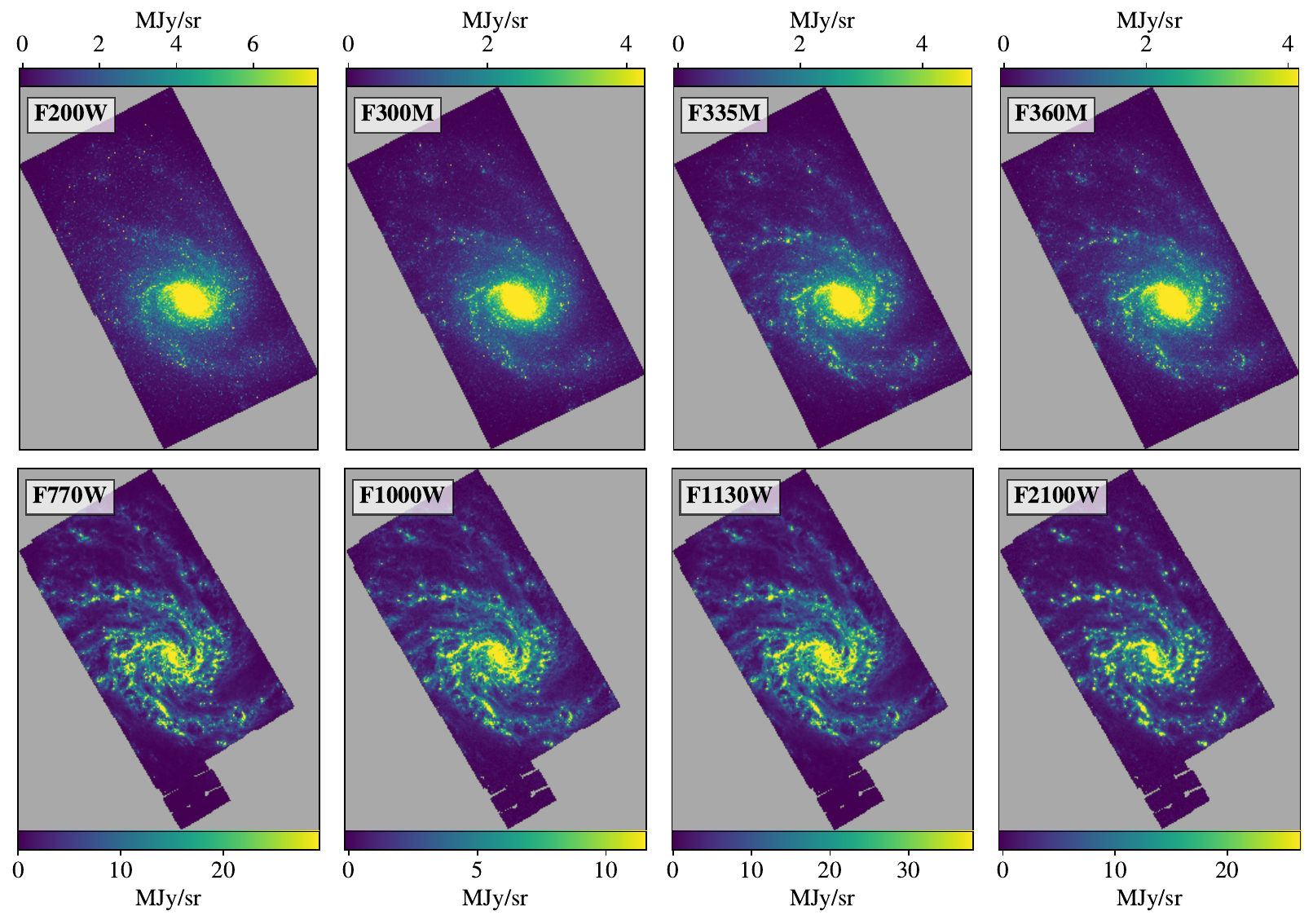}
    \caption{As Figure \ref{fig:final_mosaics}, but for NGC~4254.}
    \label{fig:ngc4254_mosaics}
\end{figure*}

\begin{figure*}
    \includegraphics[width=\textwidth]{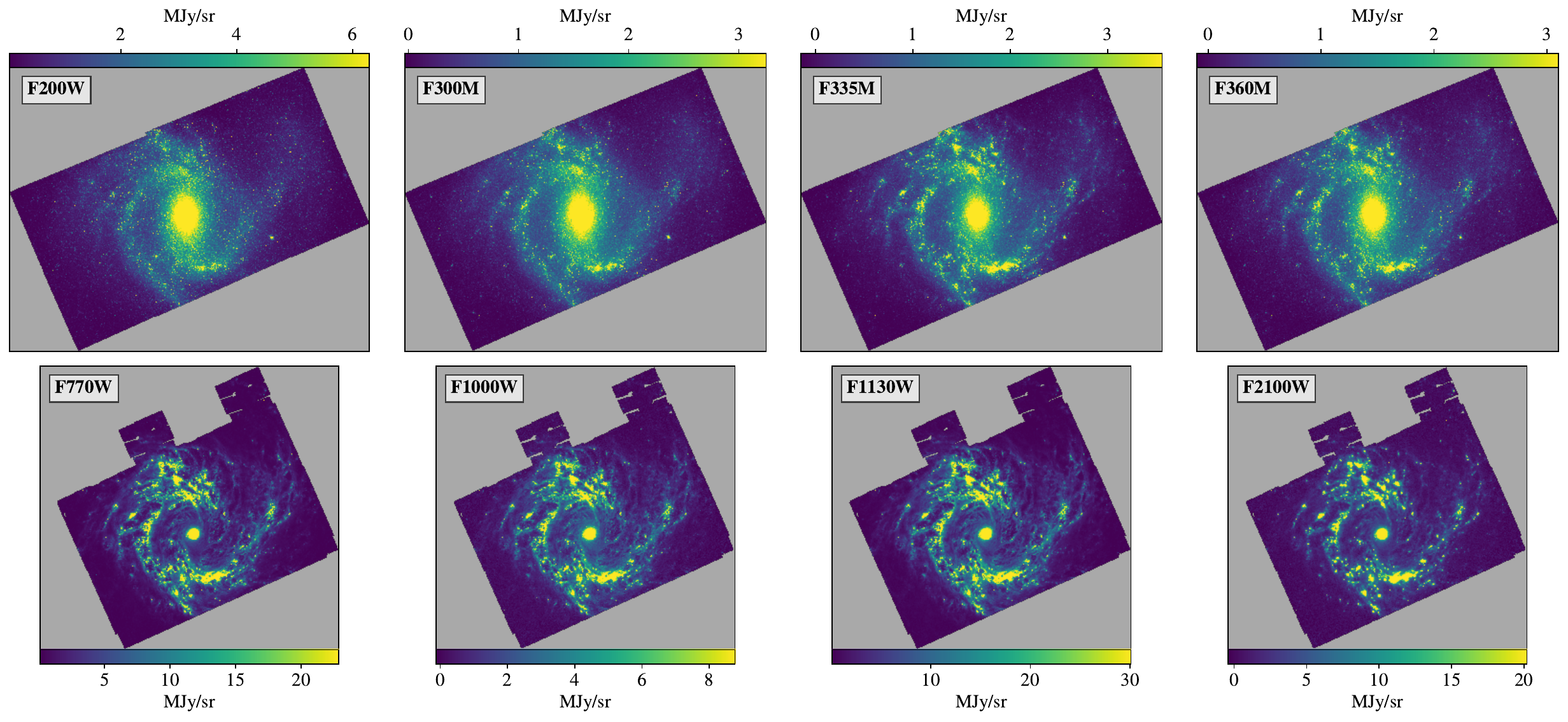}
    \caption{As Figure \ref{fig:final_mosaics}, but for NGC~4303.}
    \label{fig:ngc4303_mosaics}
\end{figure*}

\begin{figure*}
    \includegraphics[width=\textwidth]{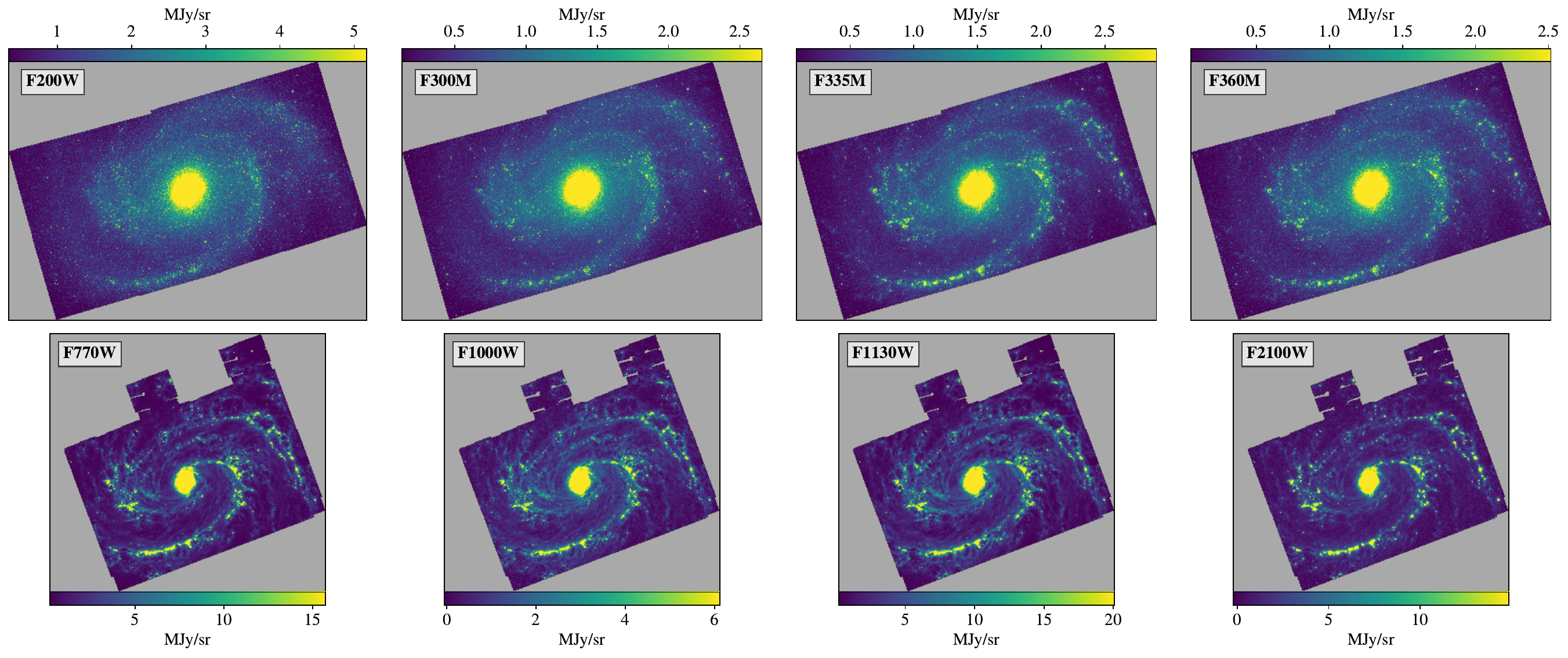}
    \caption{As Figure \ref{fig:final_mosaics}, but for NGC~4321.}
    \label{fig:ngc4321_mosaics}
\end{figure*}

\begin{figure*}
    \includegraphics[width=\textwidth]{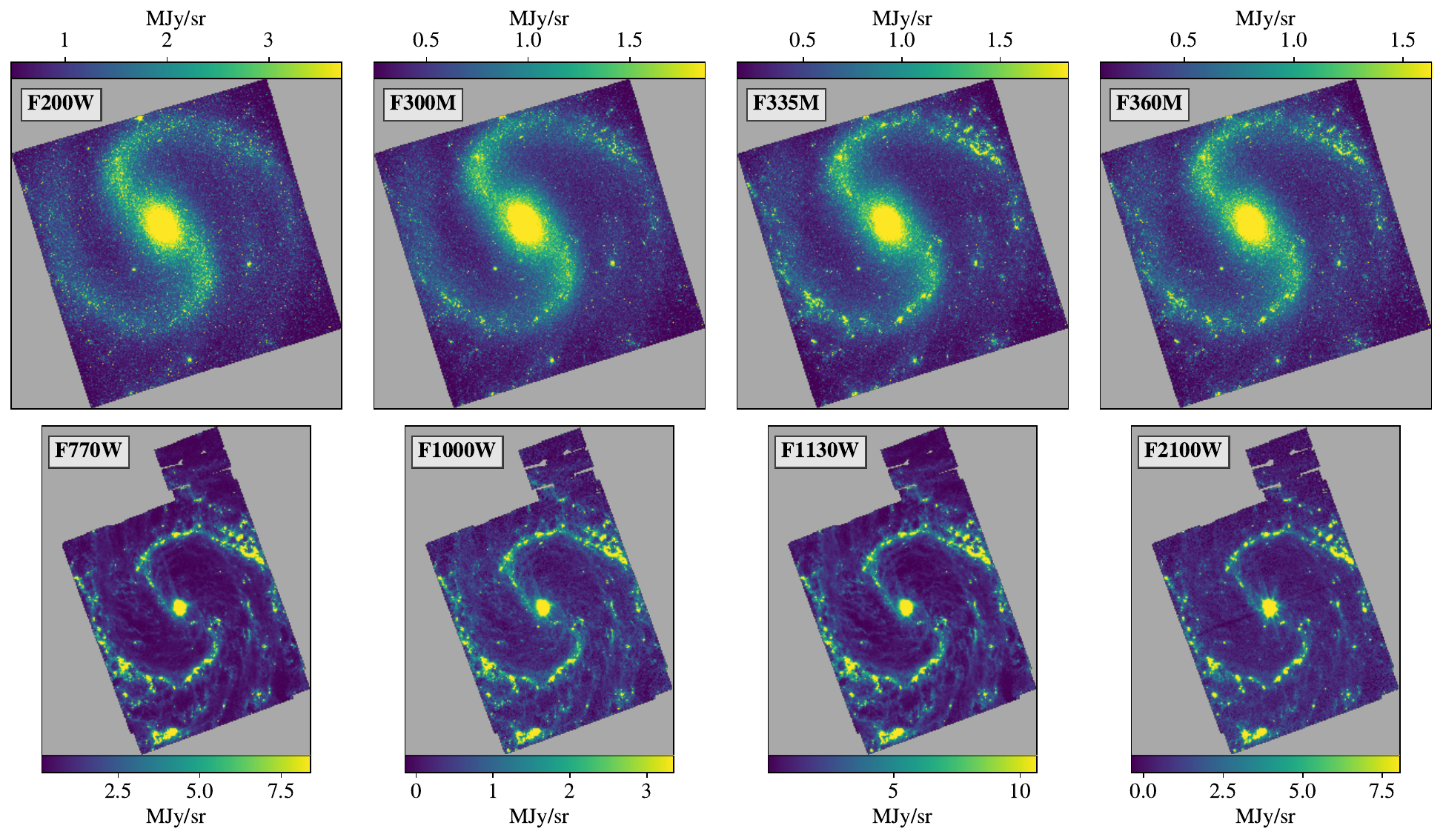}
    \caption{As Figure \ref{fig:final_mosaics}, but for NGC~4535.}
    \label{fig:ngc4535_mosaics}
\end{figure*}

\begin{figure*}
    \includegraphics[width=\textwidth]{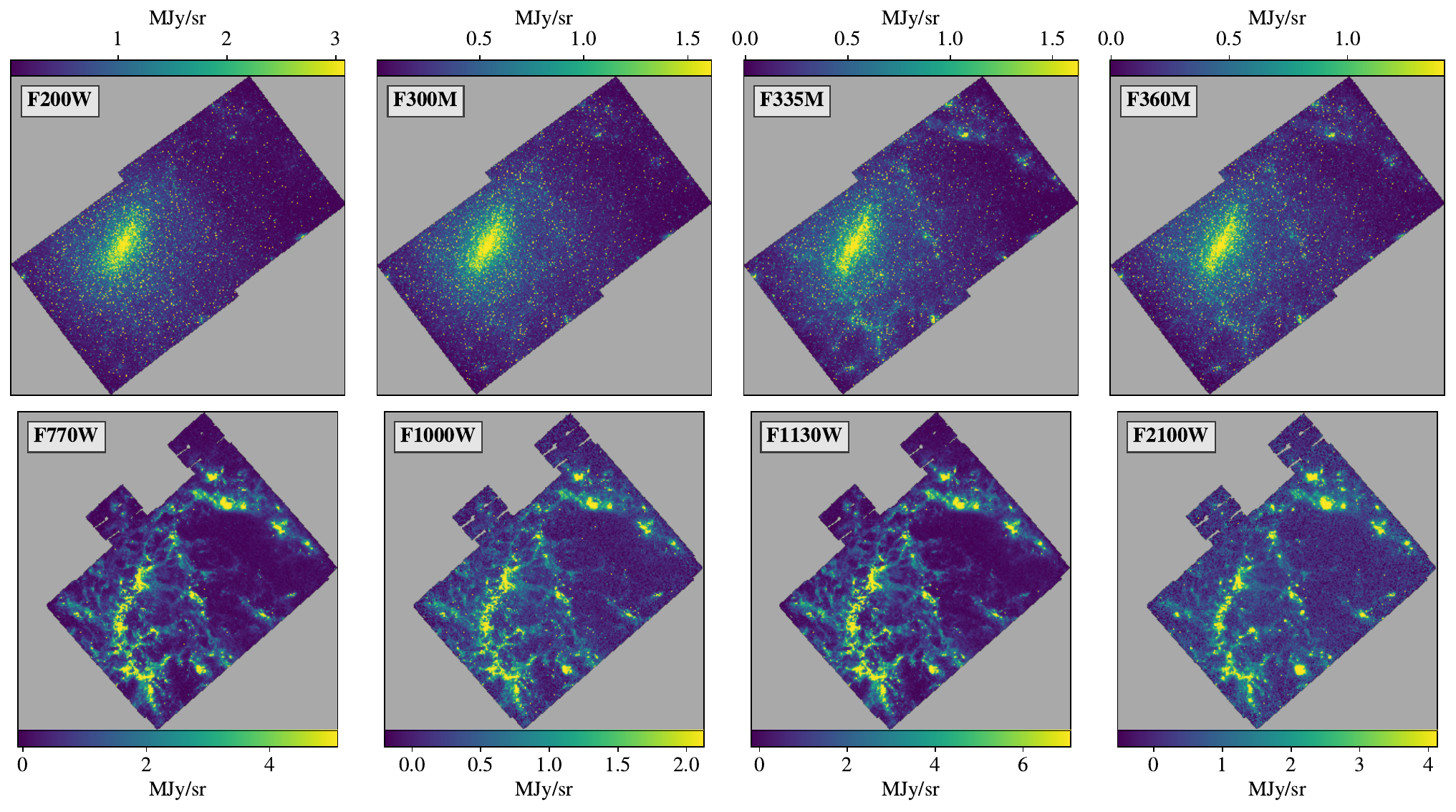}
    \caption{As Figure \ref{fig:final_mosaics}, but for NGC~5068.}
    \label{fig:ngc5068_mosaics}
\end{figure*}

\begin{figure*}
    \includegraphics[width=\textwidth]{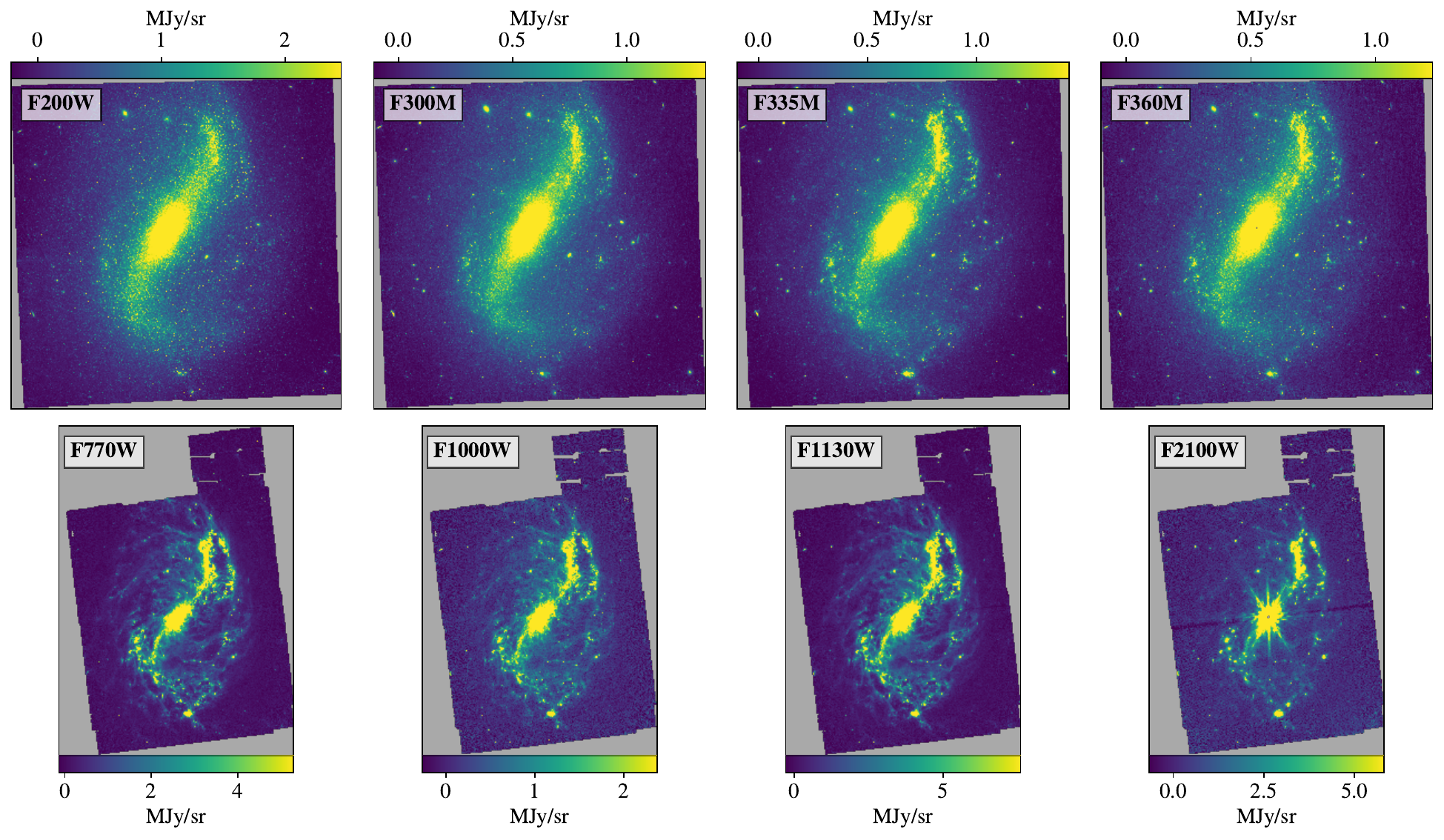}
    \caption{As Figure \ref{fig:final_mosaics}, but for NGC~7496.}
    \label{fig:ngc7496_mosaics}
\end{figure*}

\newpage

\section{Early Science Comparisons}\label{app:early_sci}

Here, we show equivalents to Figure \ref{fig:early_sci_comparison} for the other three galaxies that comprised our early science release.


\begin{figure*}
    \includegraphics[width=\textwidth]{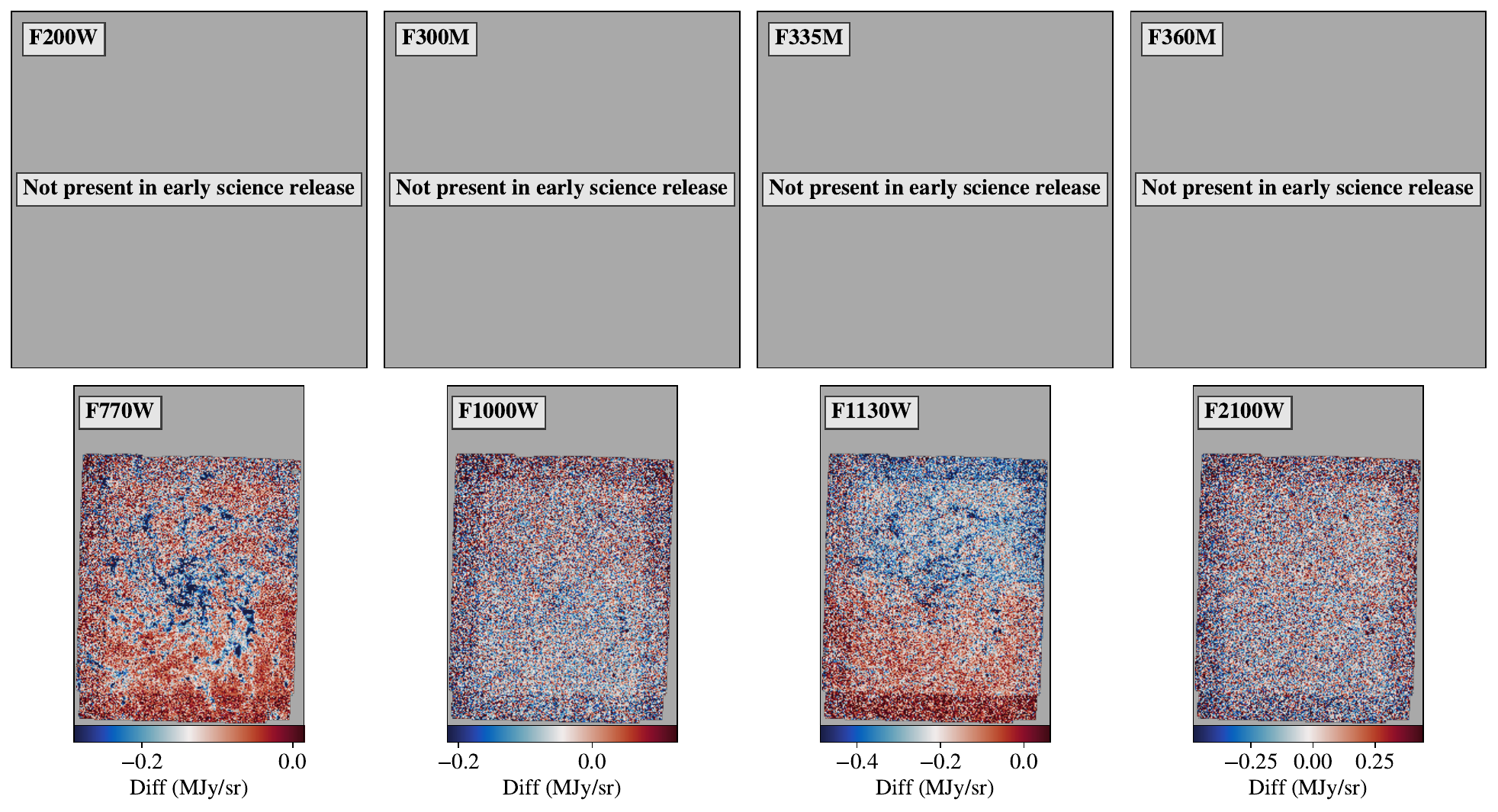}
    \caption{As Figure \ref{fig:early_sci_comparison}, but for IC~5332.}
    \label{fig:ic5332_early_sci_comparison}
\end{figure*}

\begin{figure*}
    \includegraphics[width=\textwidth]{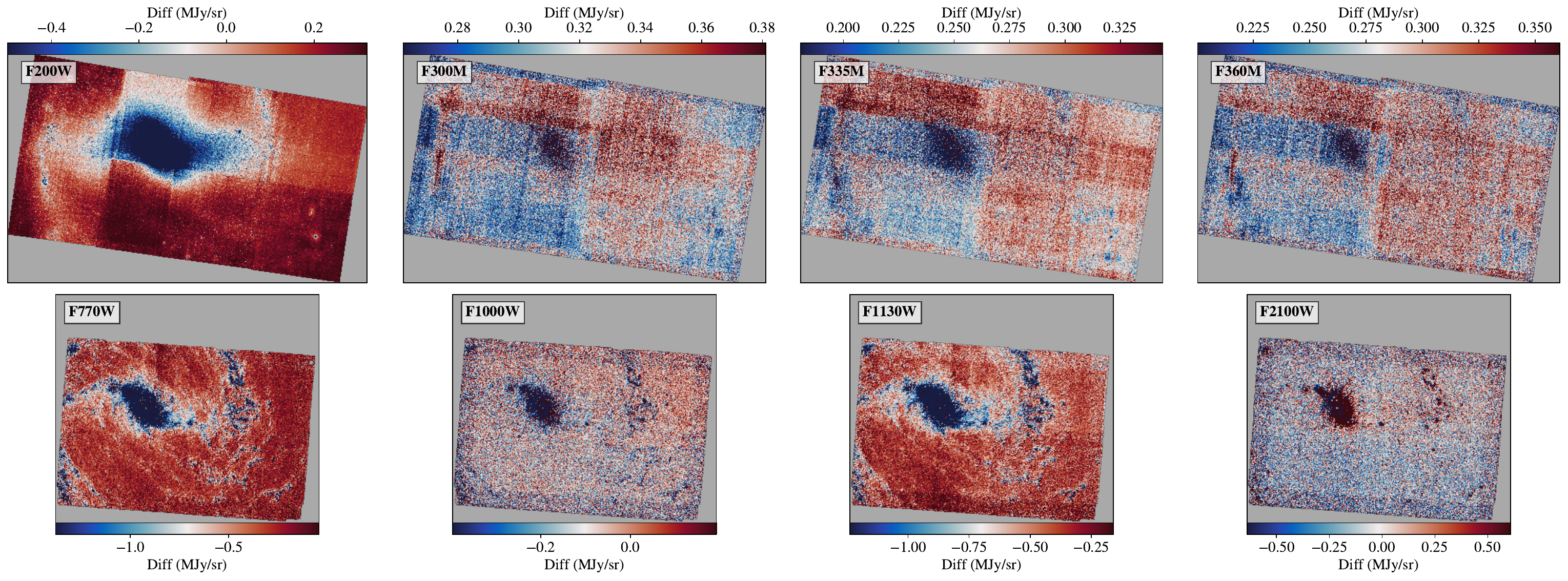}
    \caption{As Figure \ref{fig:early_sci_comparison}, but for NGC~1365.}
    \label{fig:ngc1365_early_sci_comparison}
\end{figure*}

\begin{figure*}
    \includegraphics[width=\textwidth]{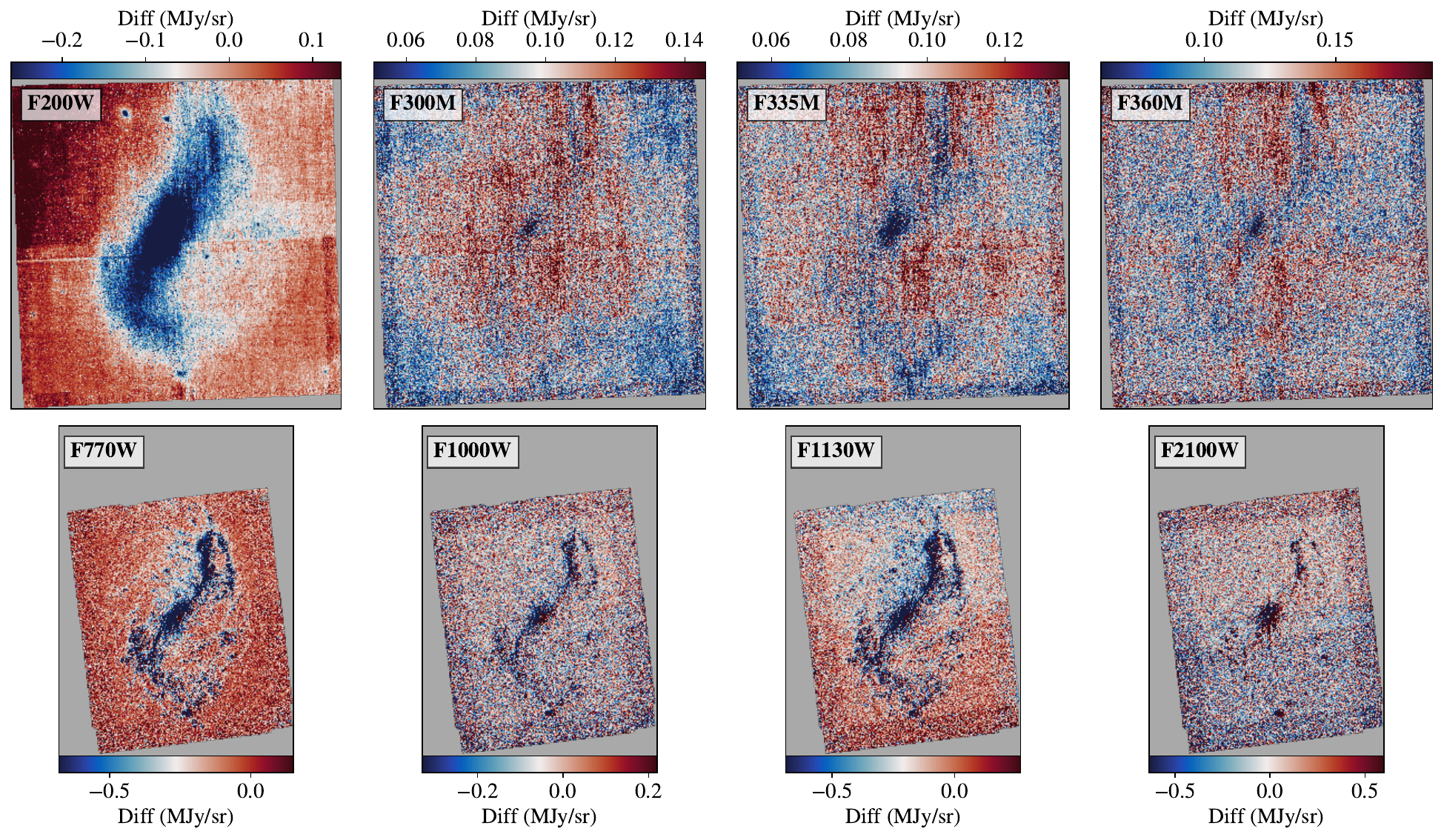}
    \caption{As Figure \ref{fig:early_sci_comparison}, but for NGC~7496.}
    \label{fig:ngc7496_early_sci_comparison}
\end{figure*}

\newpage

\section{MAST Comparisons}\label{app:mast}

Here, we show equivalents to Figure \ref{fig:mast_comparison} for the 18 other galaxies in our program.

\begin{figure*}
    \includegraphics[width=\textwidth]{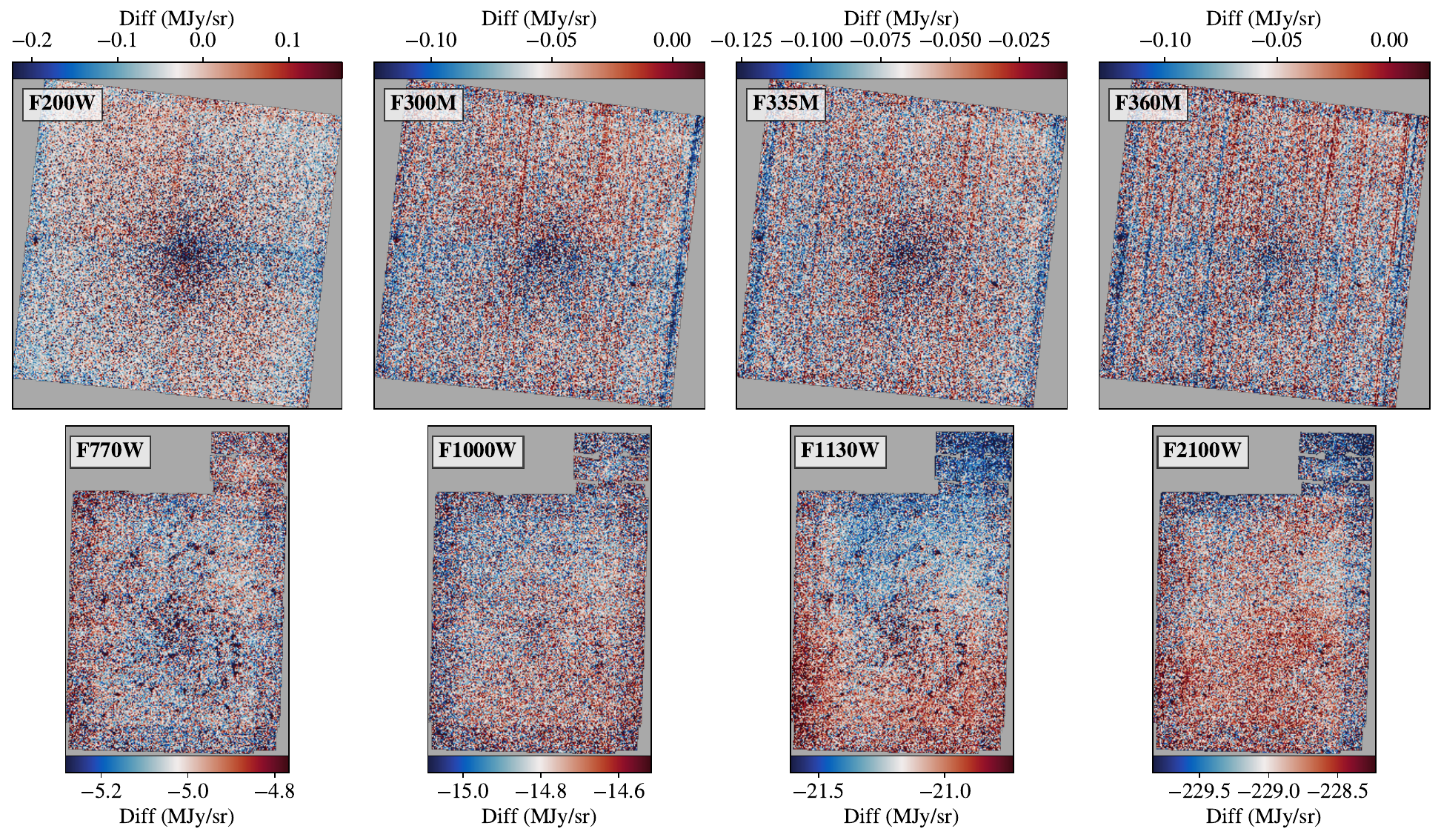}
    \caption{As Figure \ref{fig:mast_comparison}, but for IC~5332.}
    \label{fig:ic5332_mast_comparison}
\end{figure*}

\begin{figure*}
    \includegraphics[width=\textwidth]{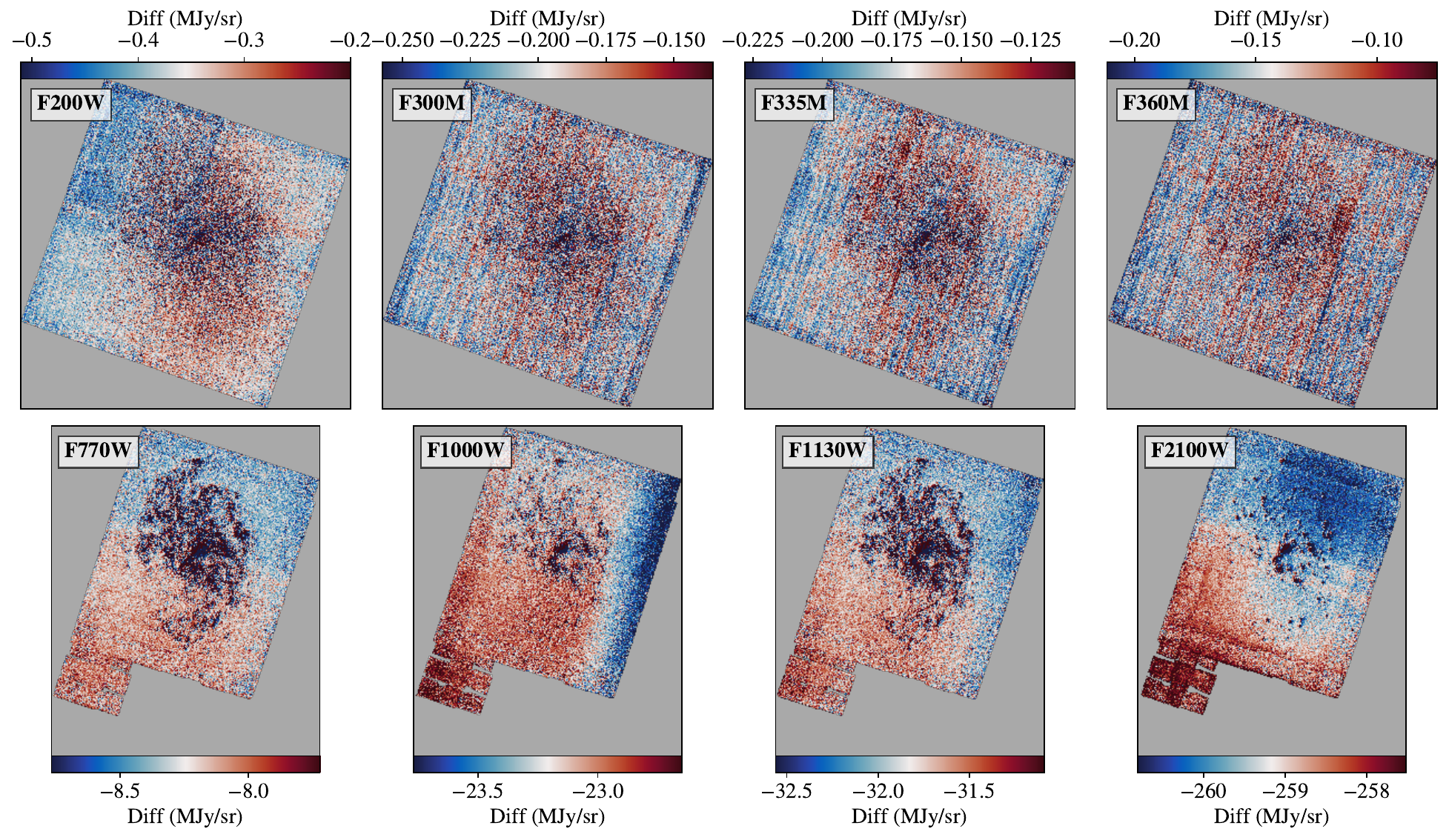}
    \caption{As Figure \ref{fig:mast_comparison}, but for NGC~1087.}
    \label{fig:ngc1087_mast_comparison}
\end{figure*}

\begin{figure*}
    \includegraphics[width=\textwidth]{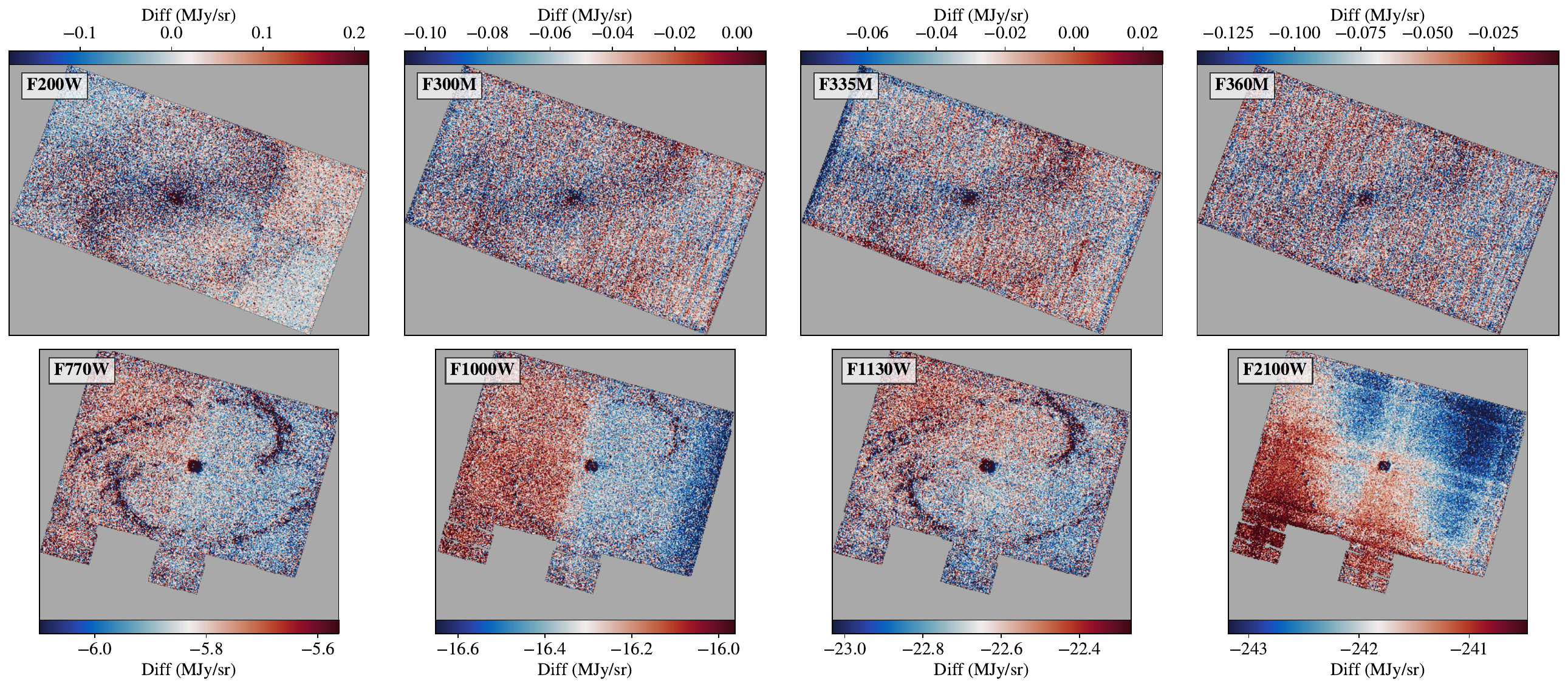}
    \caption{As Figure \ref{fig:mast_comparison}, but for NGC~1300.}
    \label{fig:ngc1300_mast_comparison}
\end{figure*}

\begin{figure*}
    \includegraphics[width=\textwidth]{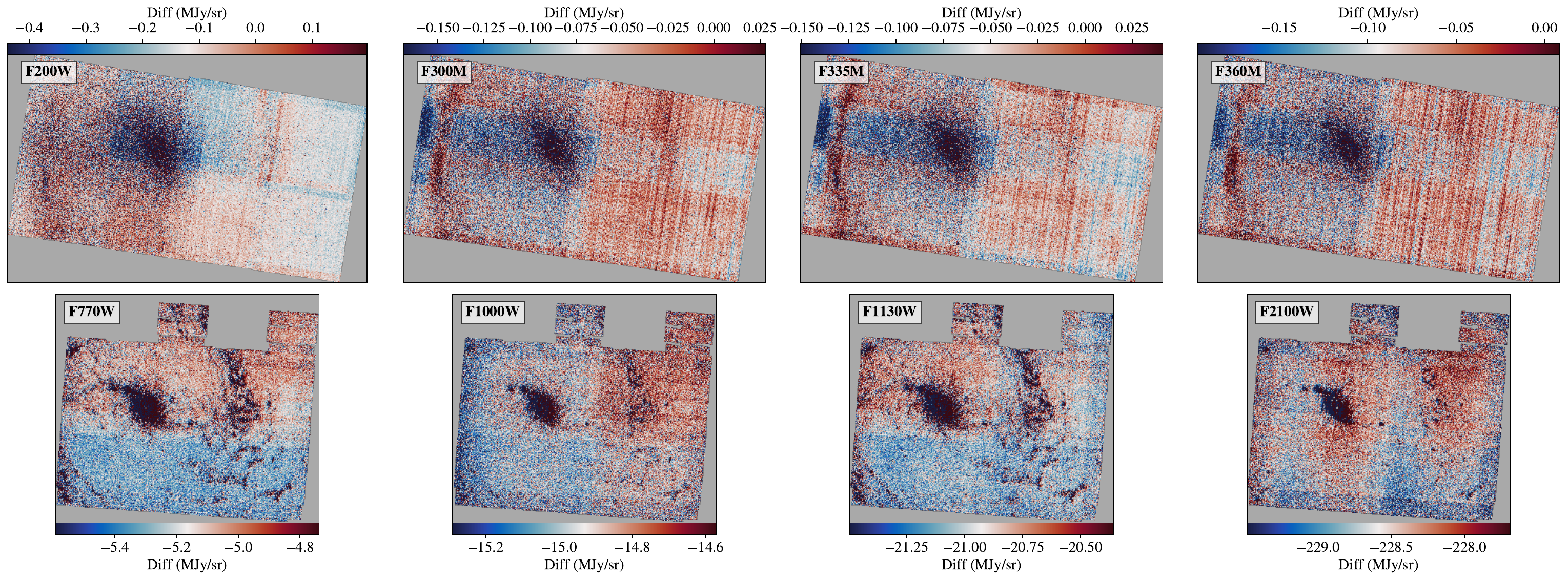}
    \caption{As Figure \ref{fig:mast_comparison}, but for NGC~1365.}
    \label{fig:ngc1365_mast_comparison}
\end{figure*}

\begin{figure*}
    \includegraphics[width=\textwidth]{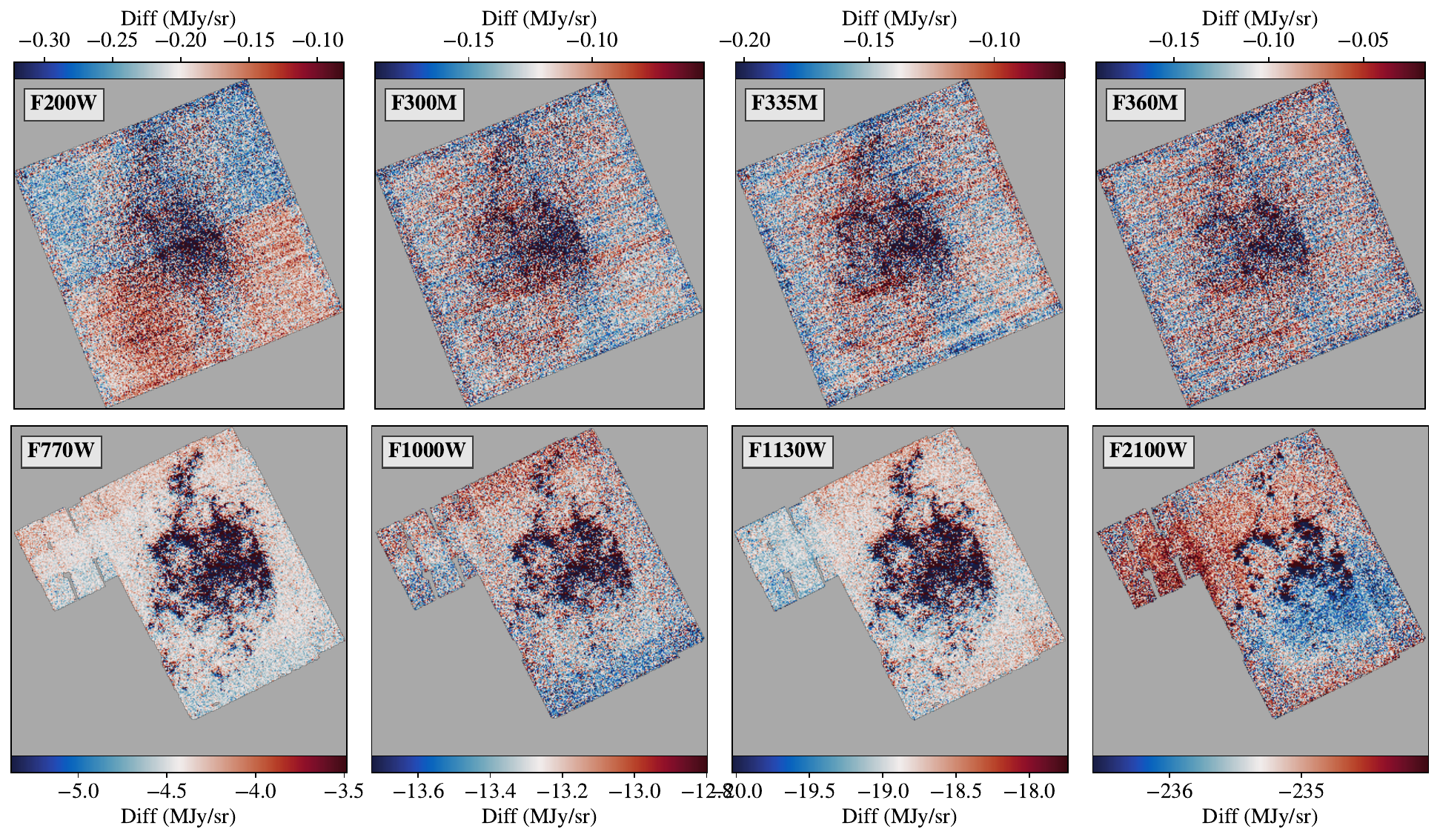}
    \caption{As Figure \ref{fig:mast_comparison}, but for NGC~1385.}
    \label{fig:ngc1385_mast_comparison}
\end{figure*}

\begin{figure*}
    \includegraphics[width=\textwidth]{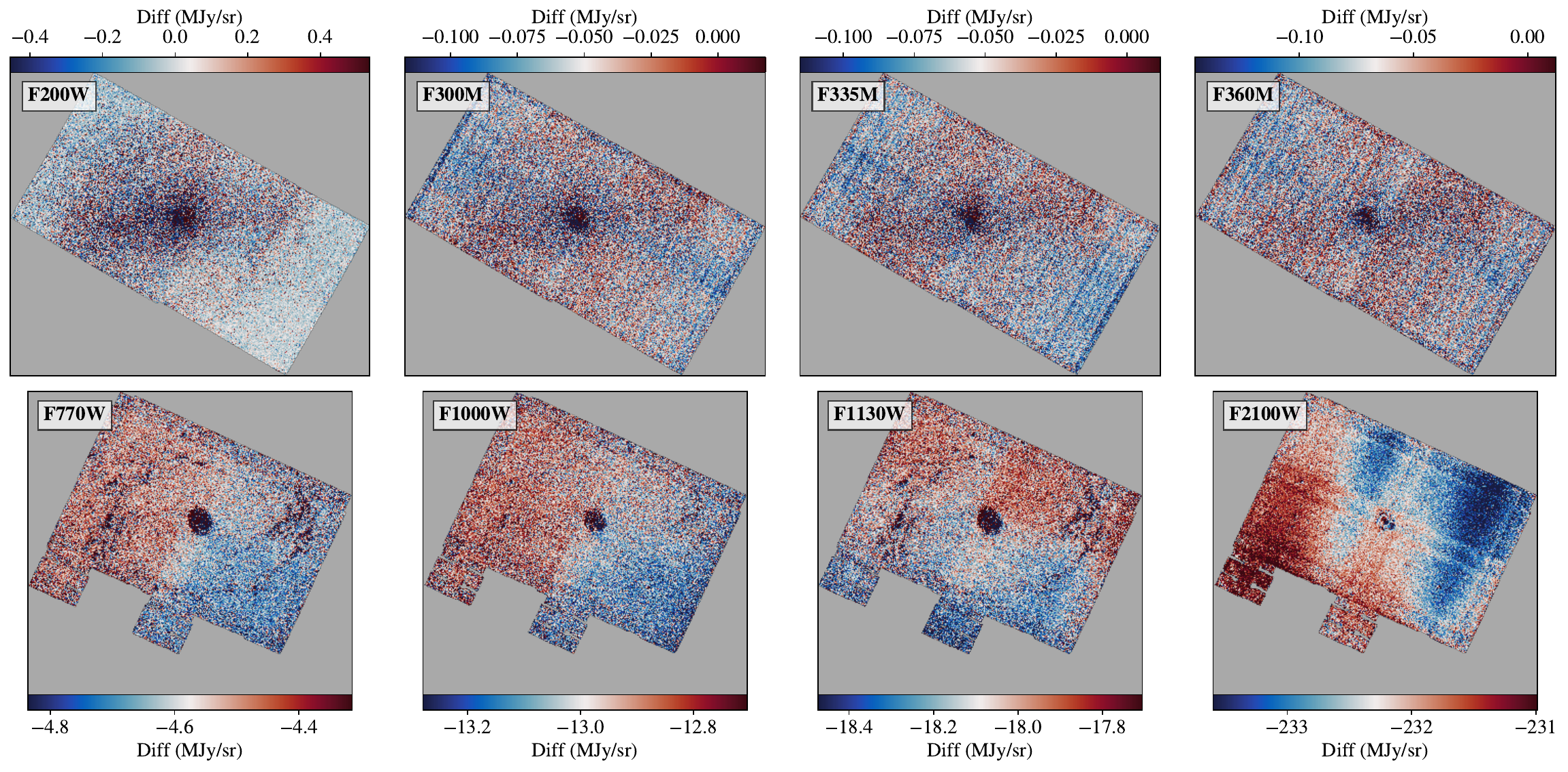}
    \caption{As Figure \ref{fig:mast_comparison}, but for NGC~1433.}
    \label{fig:ngc1433_mast_comparison}
\end{figure*}

\begin{figure*}
    \includegraphics[width=\textwidth]{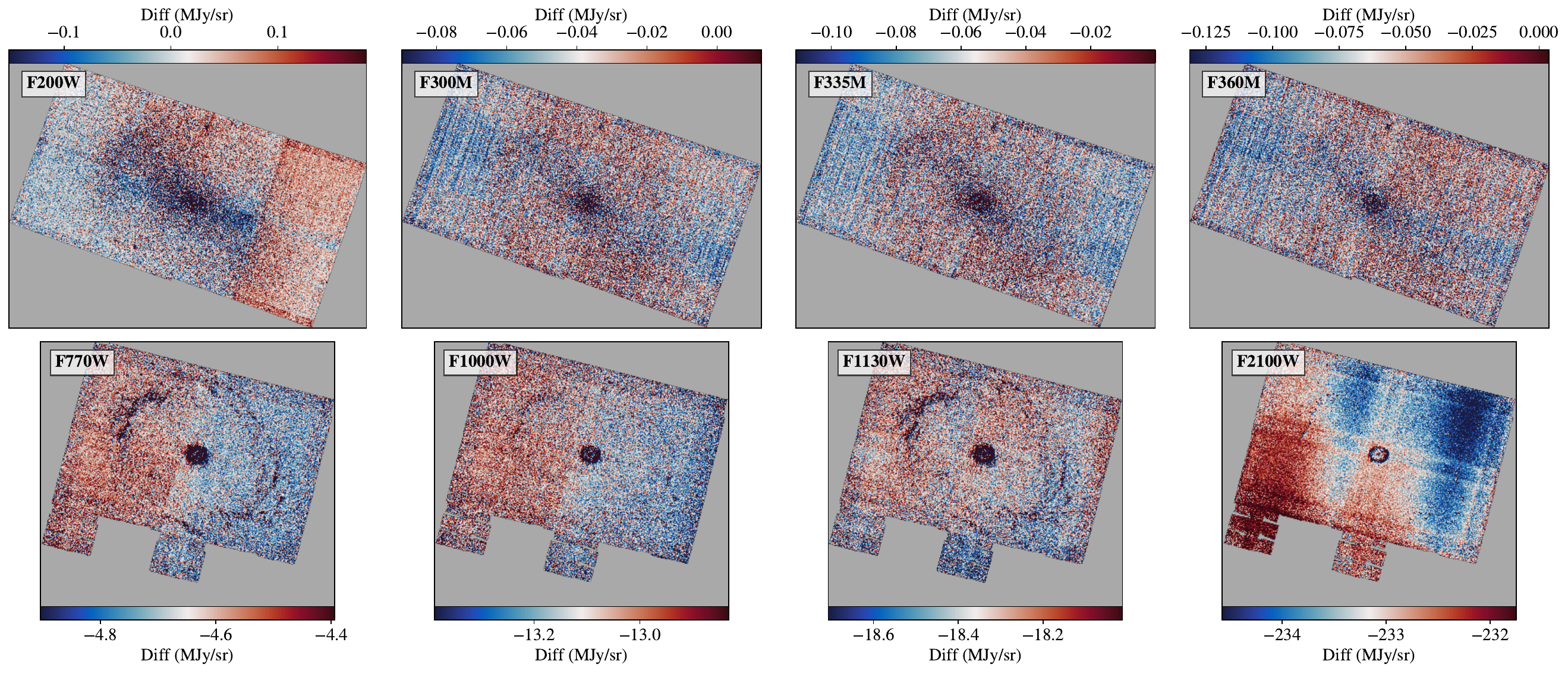}
    \caption{As Figure \ref{fig:mast_comparison}, but for NGC~1512.}
    \label{fig:ngc1512_mast_comparison}
\end{figure*}

\begin{figure*}
    \includegraphics[width=\textwidth]{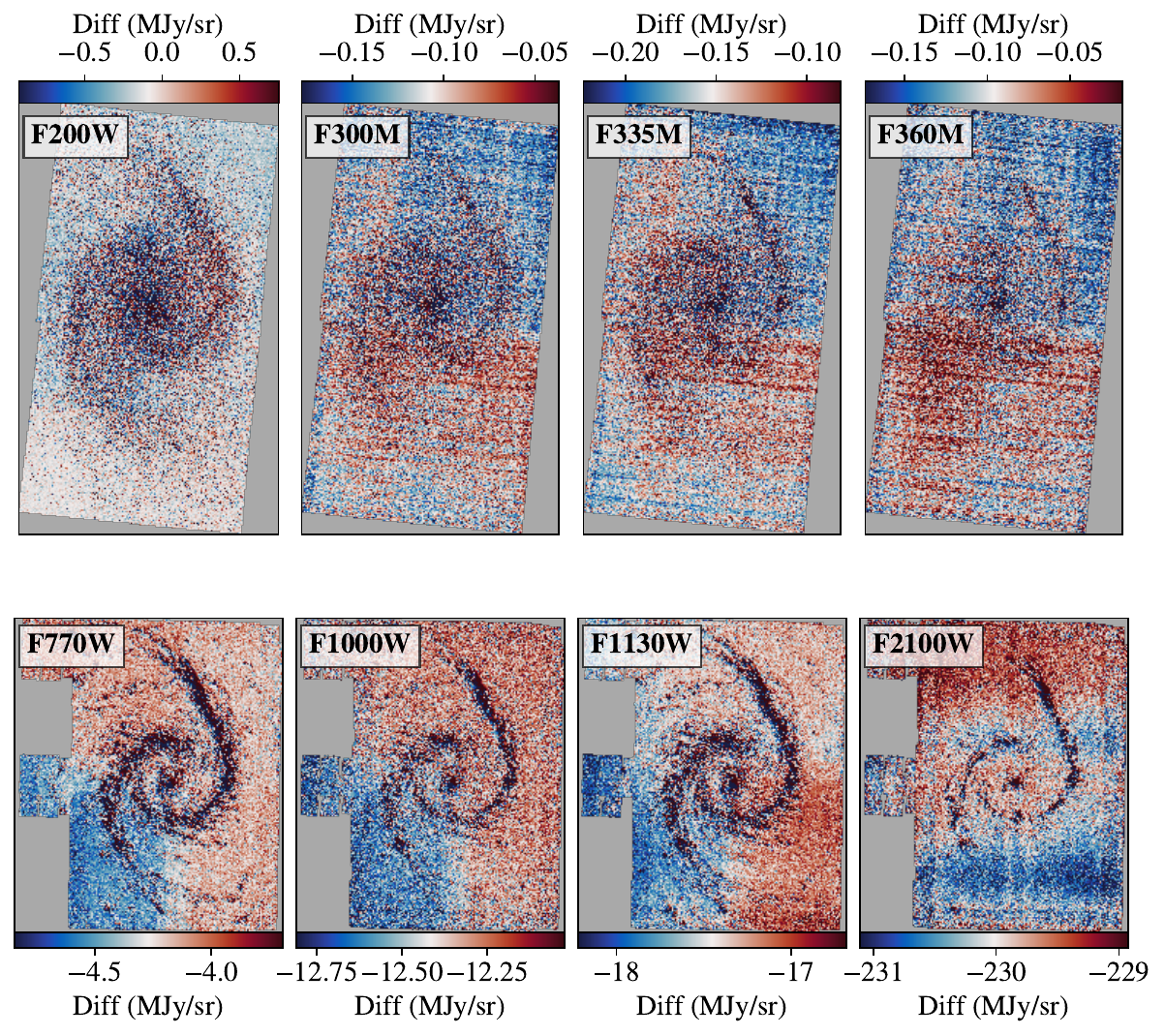}
    \caption{As Figure \ref{fig:mast_comparison}, but for NGC~1566.}
    \label{fig:ngc1566_mast_comparison}
\end{figure*}

\begin{figure*}
    \includegraphics[width=\textwidth]{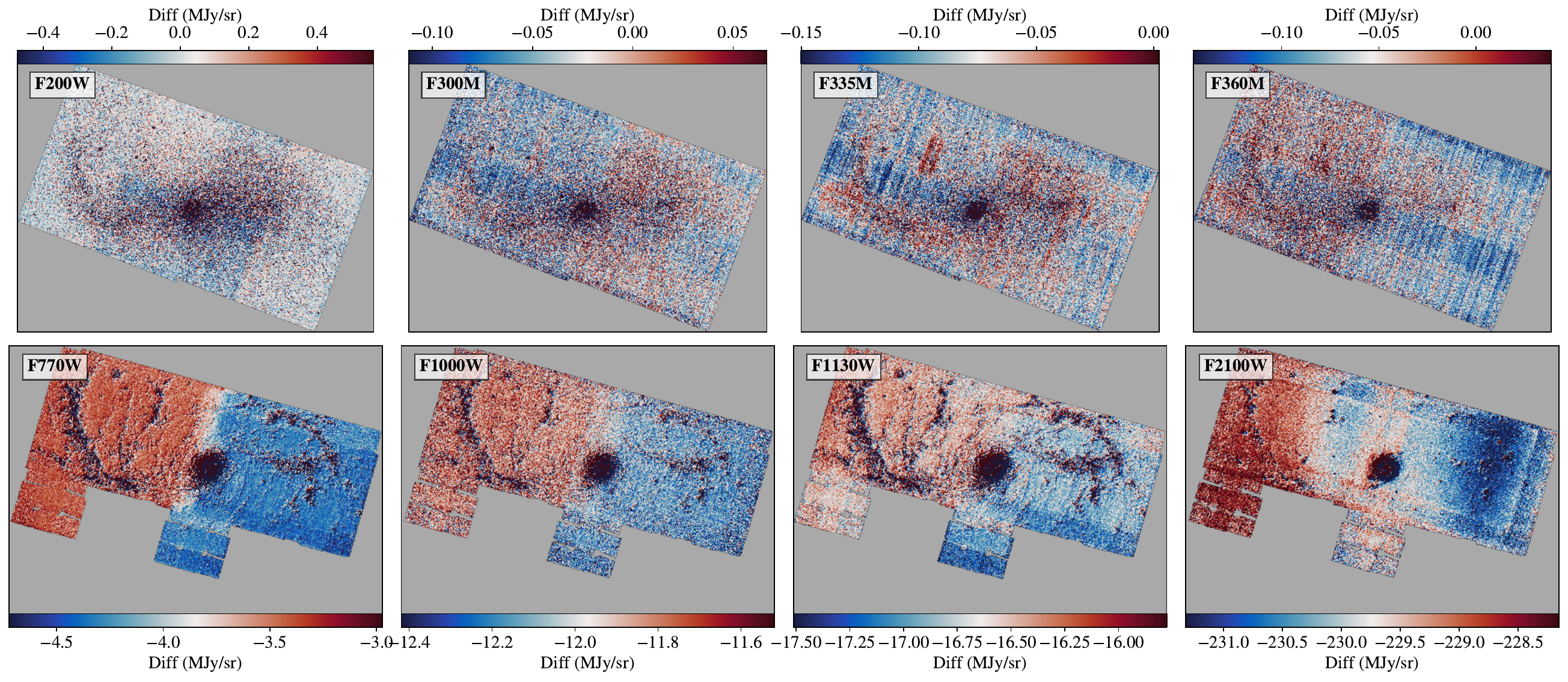}
    \caption{As Figure \ref{fig:mast_comparison}, but for NGC~1672.}
    \label{fig:ngc1672_mast_comparison}
\end{figure*}

\begin{figure*}
    \includegraphics[width=\textwidth]{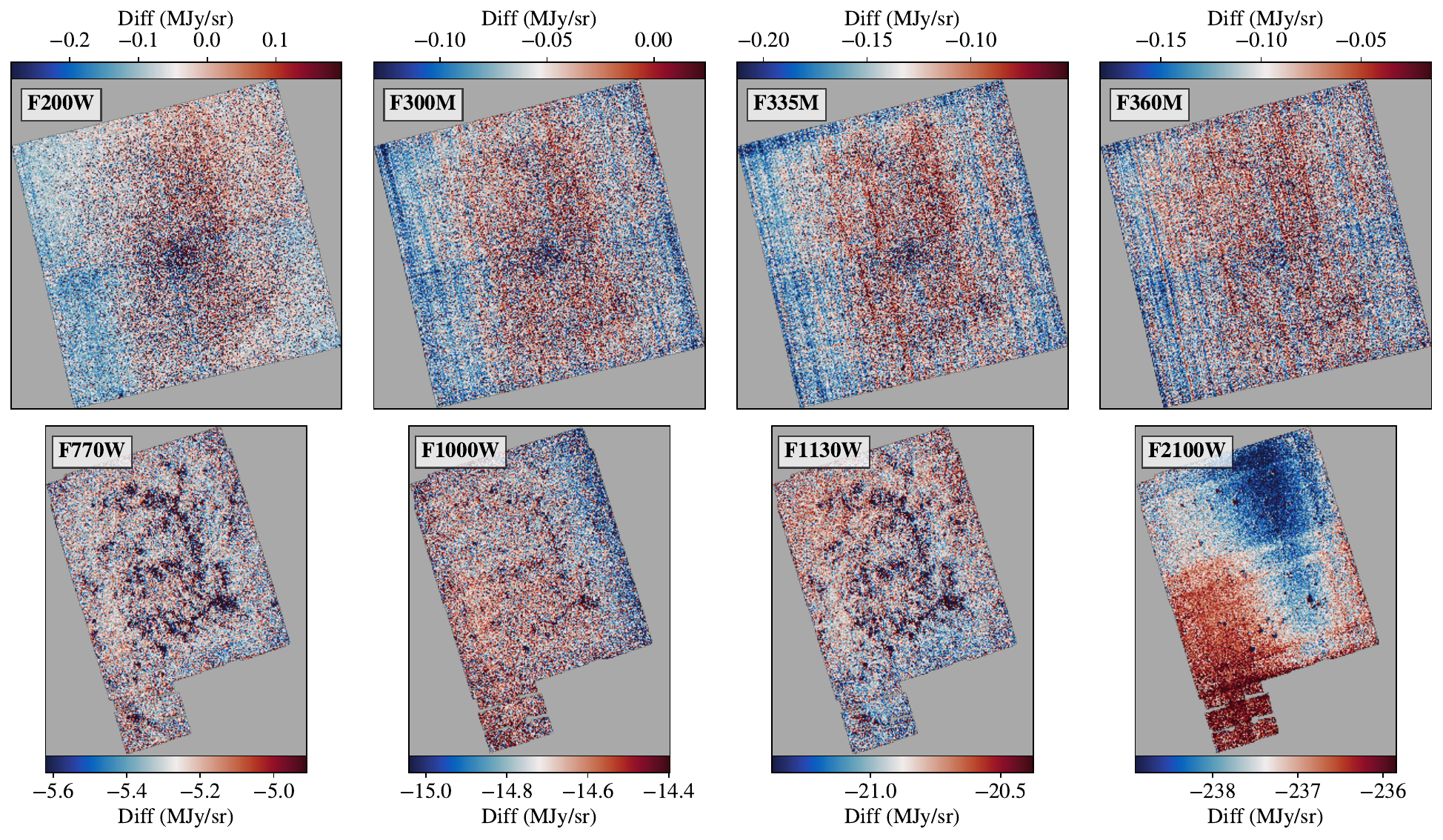}
    \caption{As Figure \ref{fig:mast_comparison}, but for NGC~2835.}
    \label{fig:ngc2835_mast_comparison}
\end{figure*}

\begin{figure*}
    \includegraphics[width=\textwidth]{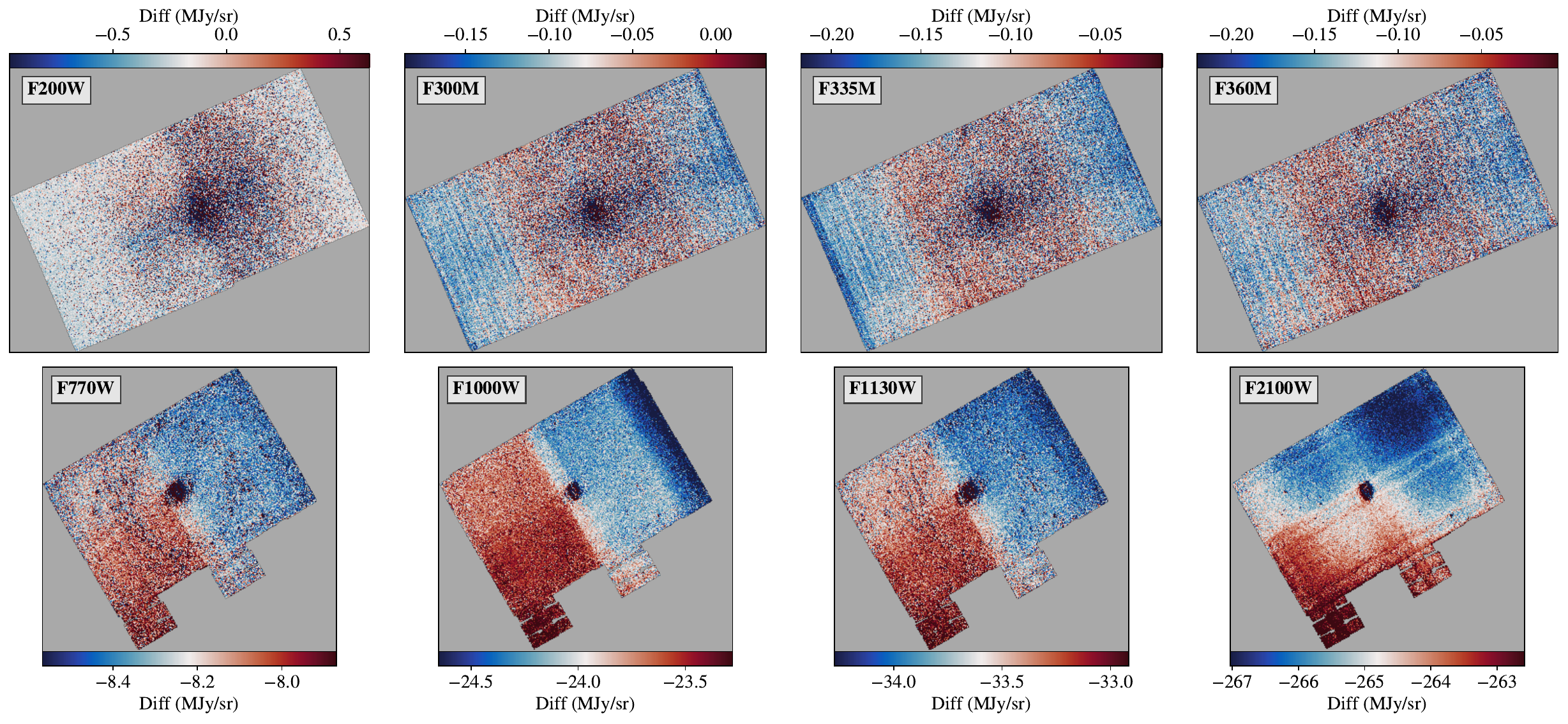}
    \caption{As Figure \ref{fig:mast_comparison}, but for NGC~3351.}
    \label{fig:ngc3351_mast_comparison}
\end{figure*}

\begin{figure*}
    \includegraphics[width=\textwidth]{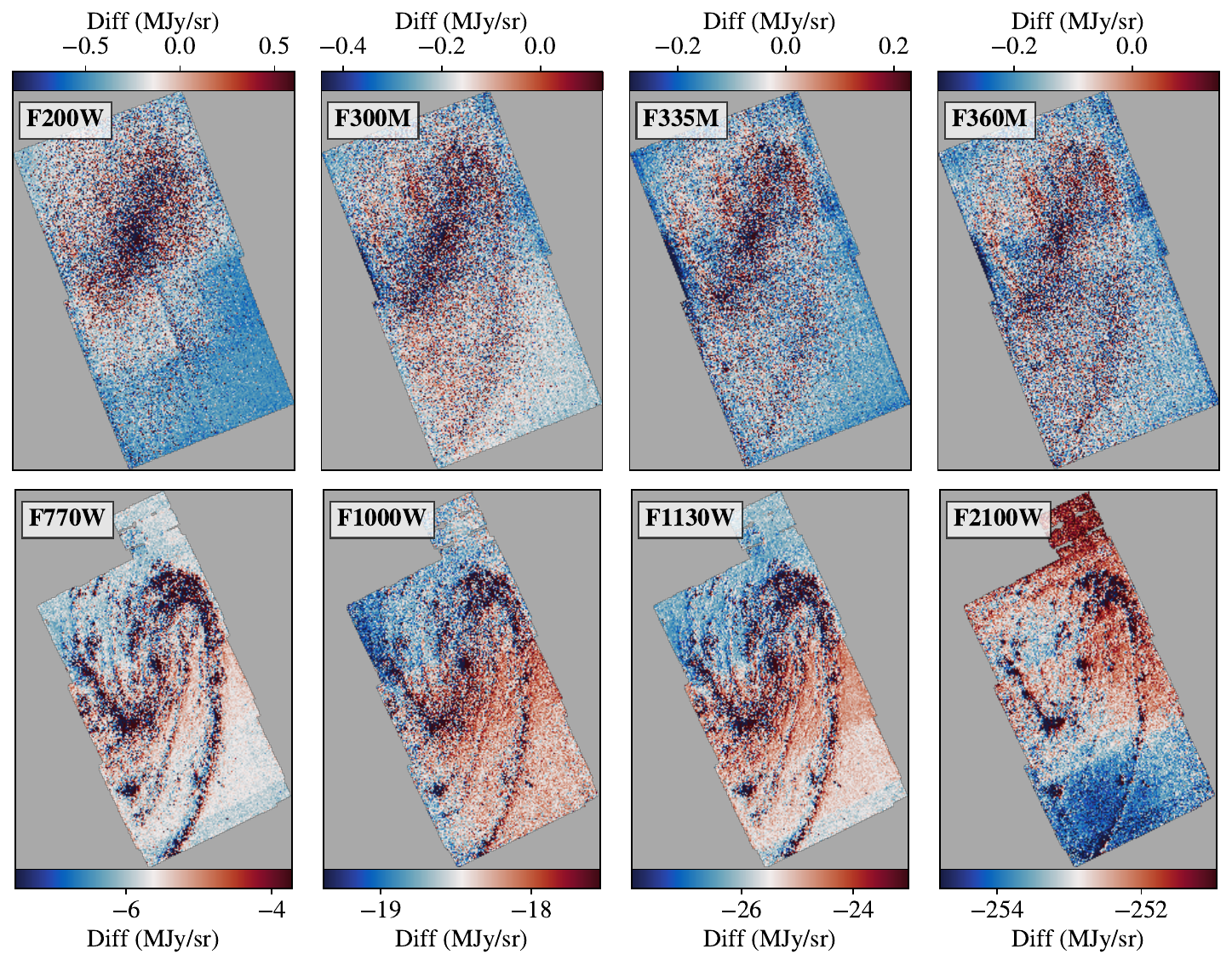}
    \caption{As Figure \ref{fig:mast_comparison}, but for NGC~3627.}
    \label{fig:ngc3627_mast_comparison}
\end{figure*}

\begin{figure*}
    \includegraphics[width=\textwidth]{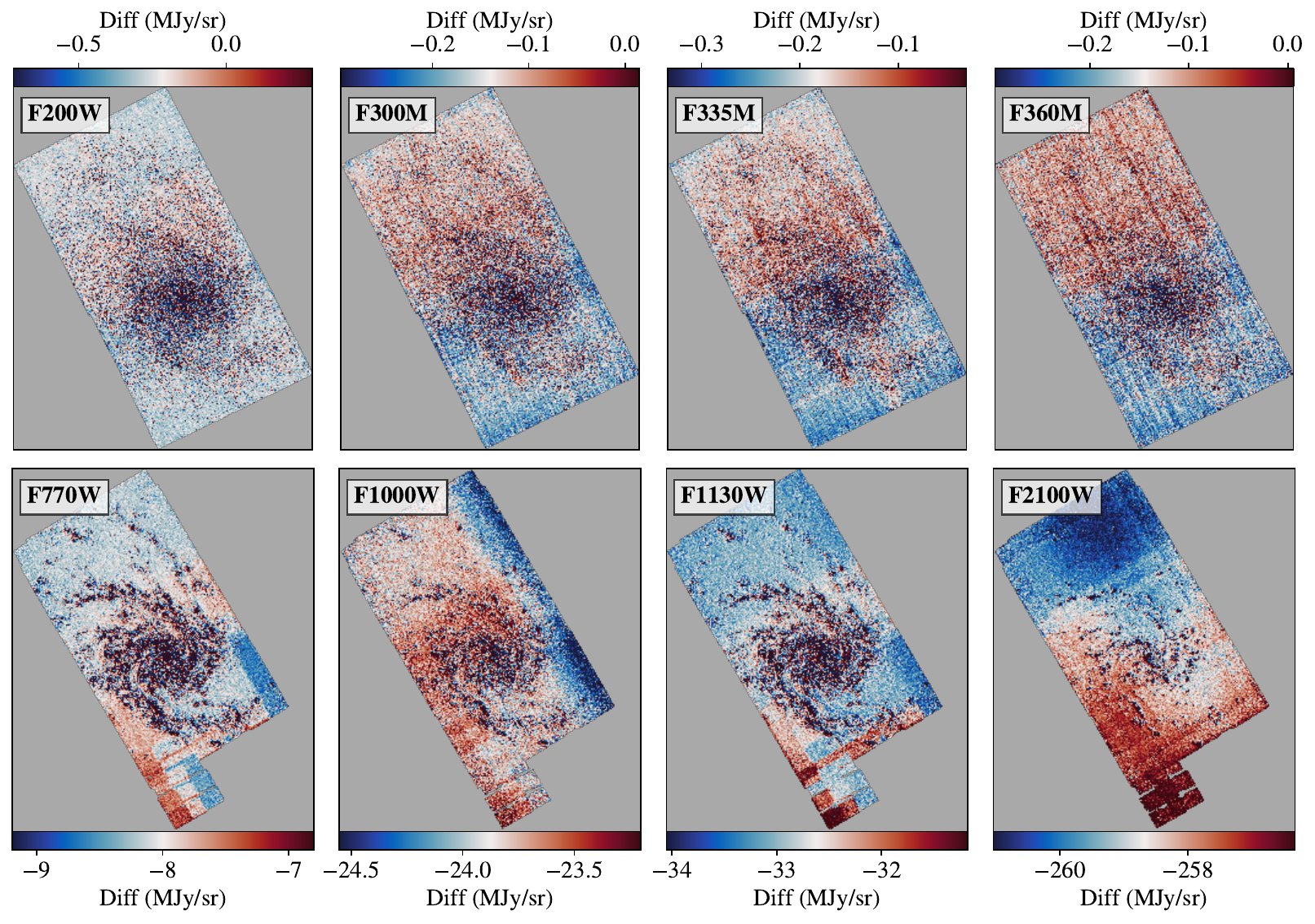}
    \caption{As Figure \ref{fig:mast_comparison}, but for NGC~4254.}
    \label{fig:ngc4254_mast_comparison}
\end{figure*}

\begin{figure*}
    \includegraphics[width=\textwidth]{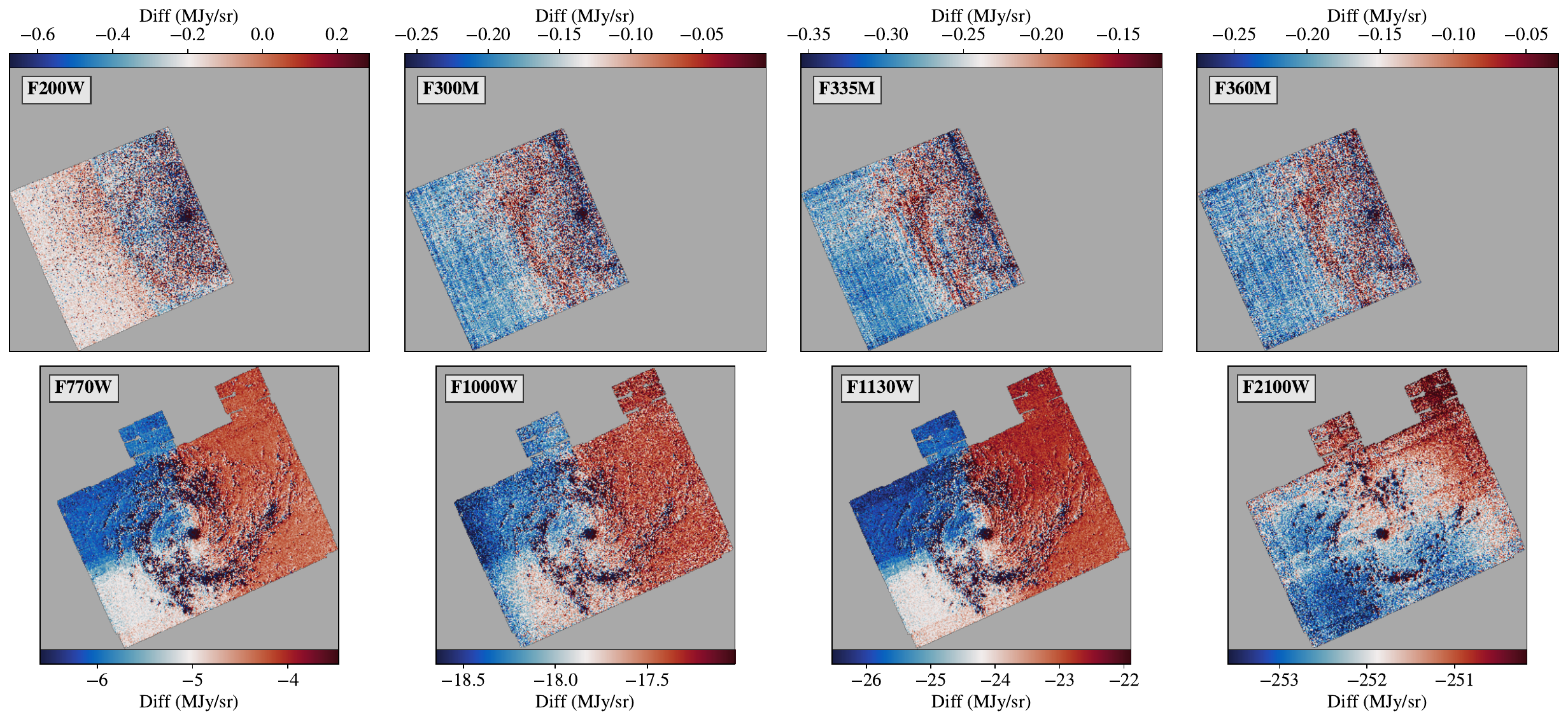}
    \caption{As Figure \ref{fig:mast_comparison}, but for NGC~4303.}
    \label{fig:ngc4303_mast_comparison}
\end{figure*}

\begin{figure*}
    \includegraphics[width=\textwidth]{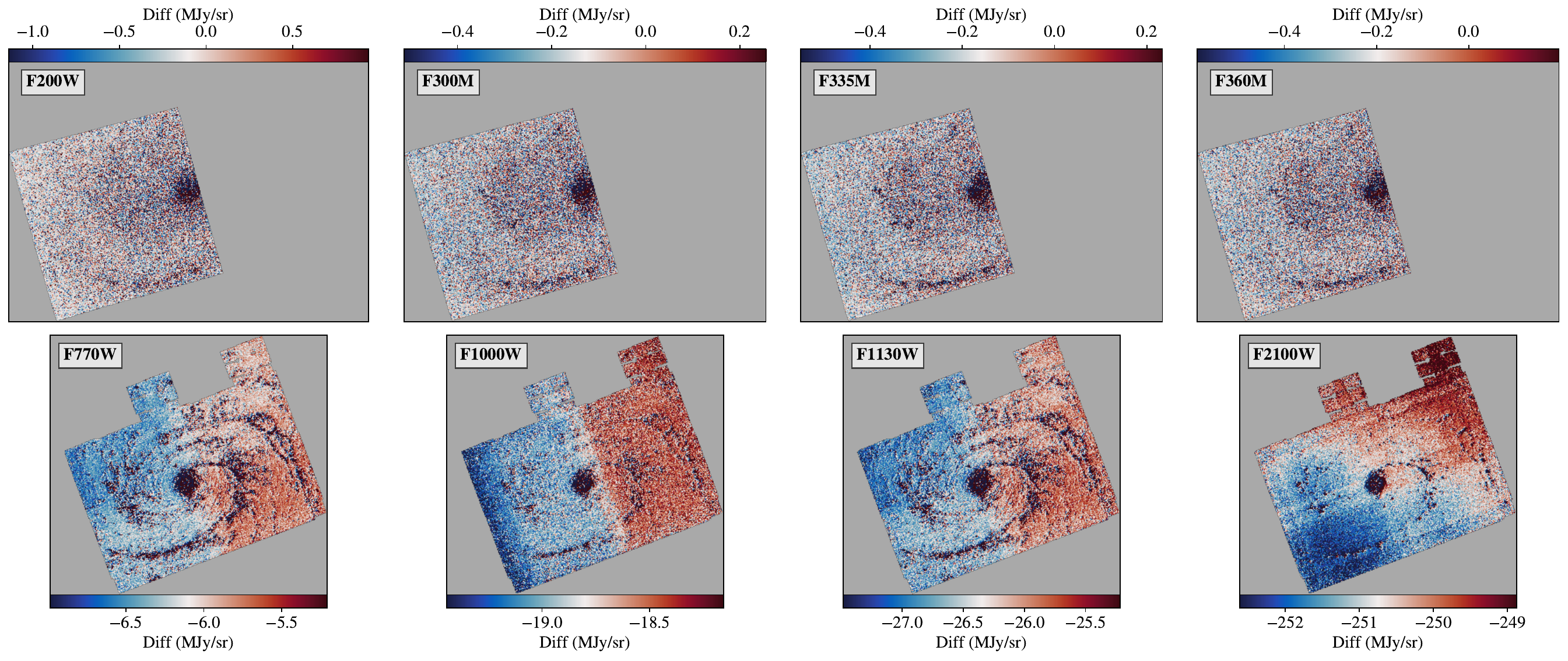}
    \caption{As Figure \ref{fig:mast_comparison}, but for NGC~4321.}
    \label{fig:ngc4321_mast_comparison}
\end{figure*}

\begin{figure*}
    \includegraphics[width=\textwidth]{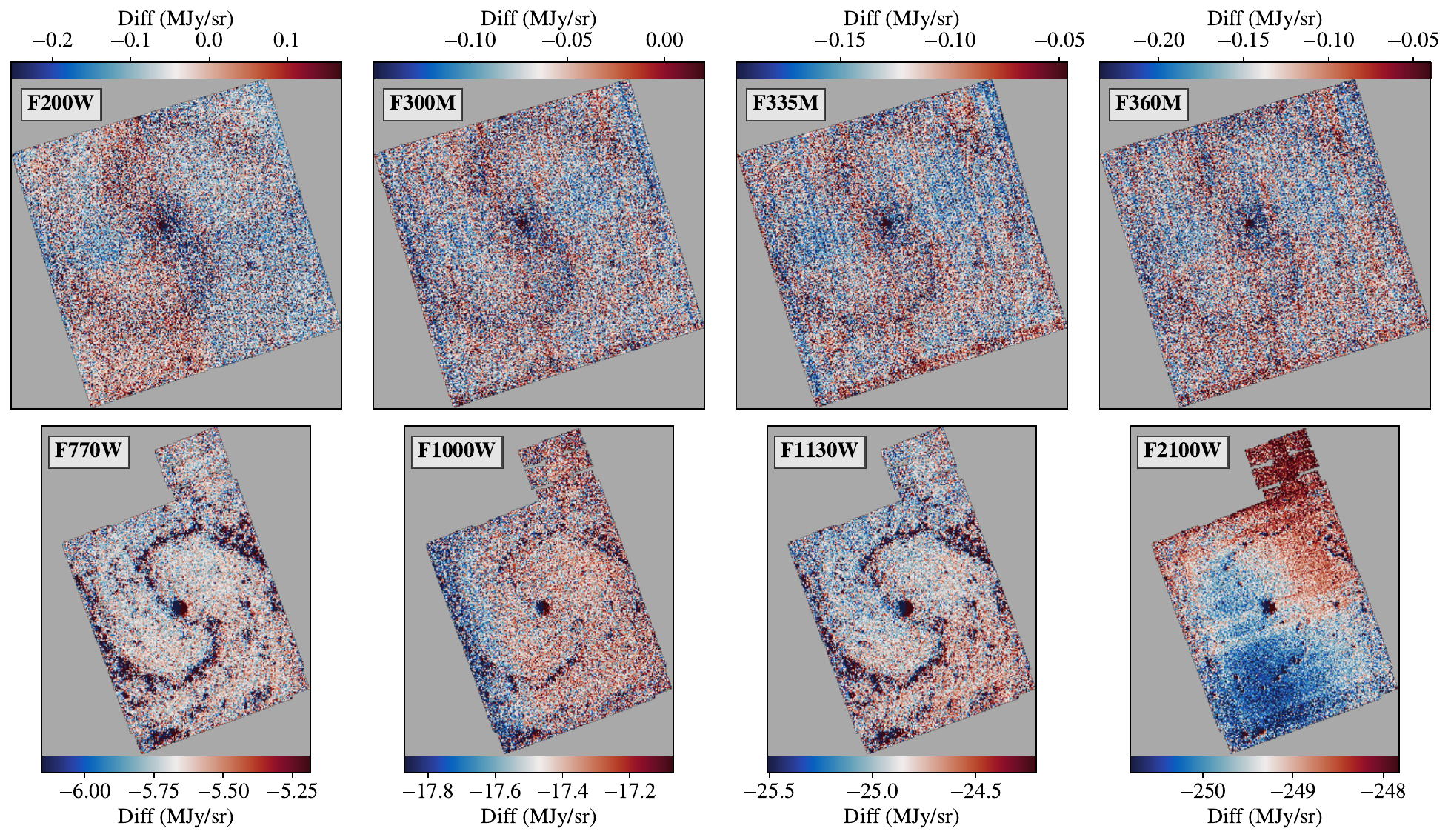}
    \caption{As Figure \ref{fig:mast_comparison}, but for NGC~4535.}
    \label{fig:ngc4535_mast_comparison}
\end{figure*}

\begin{figure*}
    \includegraphics[width=\textwidth]{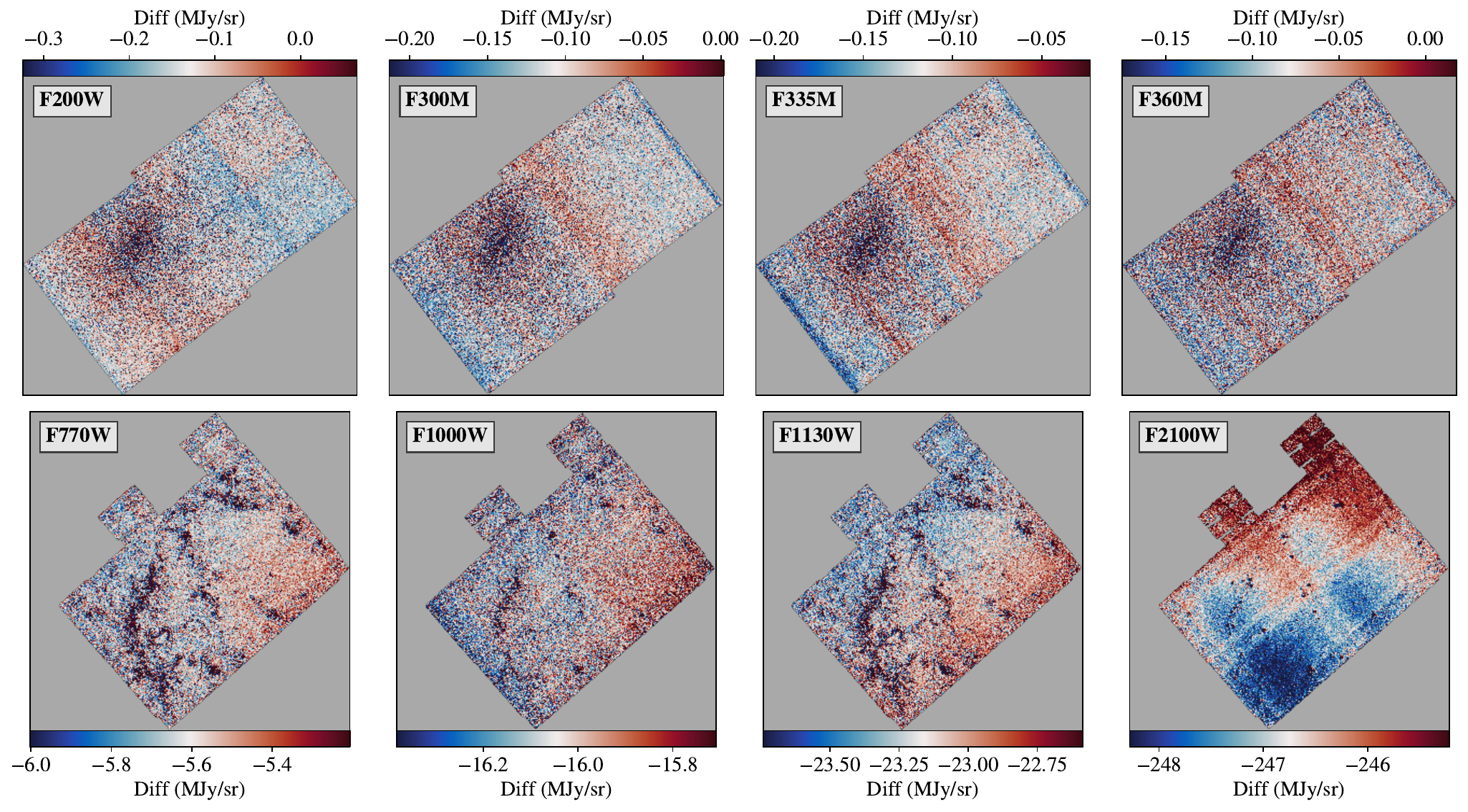}
    \caption{As Figure \ref{fig:mast_comparison}, but for NGC~5068.}
    \label{fig:ngc5068_mast_comparison}
\end{figure*}

\begin{figure*}
    \includegraphics[width=\textwidth]{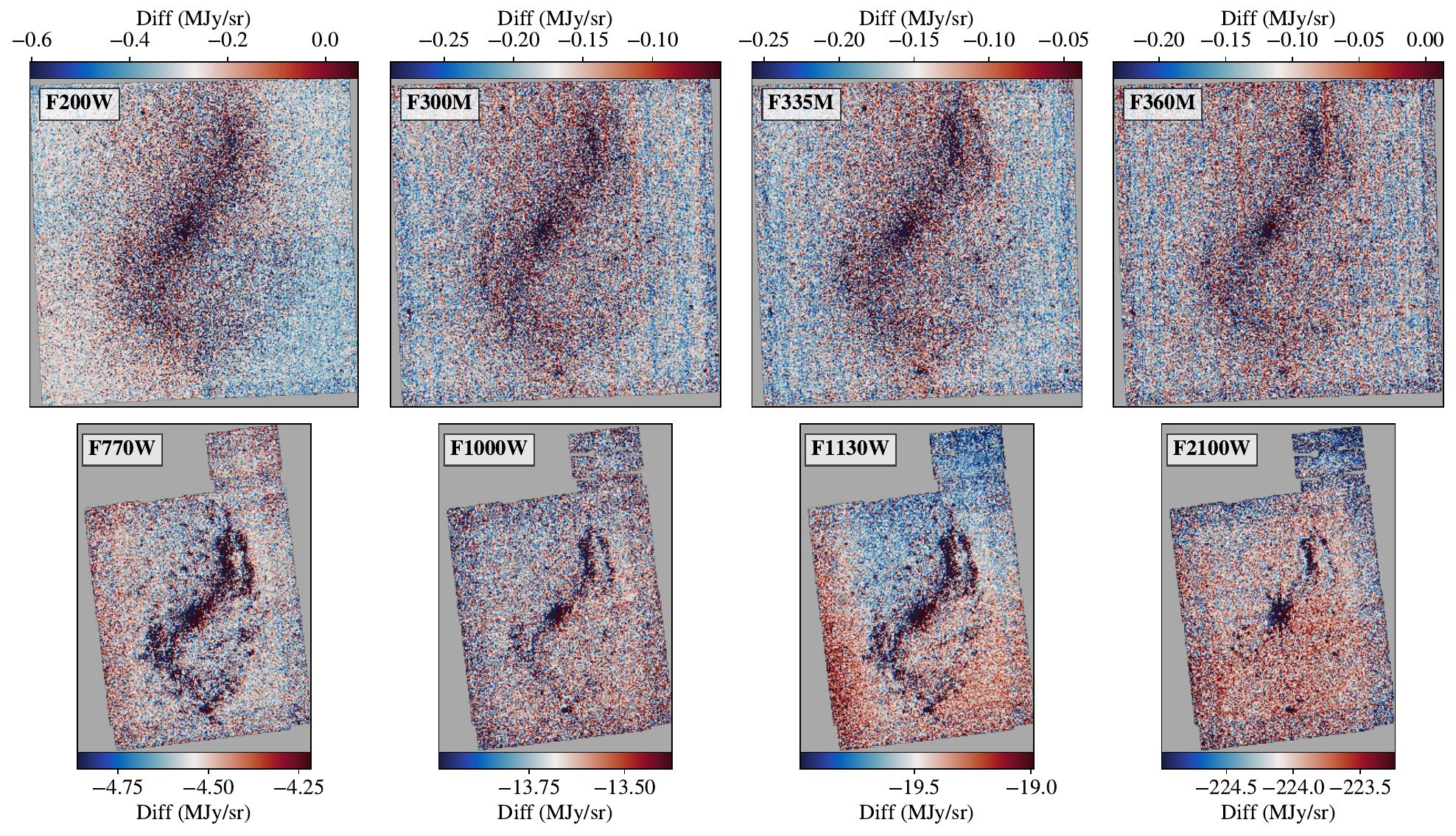}
    \caption{As Figure \ref{fig:mast_comparison}, but for NGC~7496.}
    \label{fig:ngc7496_mast_comparison}
\end{figure*}

\bibliography{bibliography}{}
\bibliographystyle{aasjournal}

\suppressAffiliationsfalse
\allauthors


\end{document}